\documentclass[a4paper,11pt]{article}
\pdfoutput=1 

\usepackage{jcappub} 

\usepackage[T1]{fontenc} 

\usepackage{amsmath}
\usepackage{booktabs}
\usepackage{amsmath}
\usepackage{caption} 
\usepackage{array}
\usepackage{siunitx}
\usepackage{amssymb}
\usepackage{graphicx}
\usepackage{caption}
\usepackage{lipsum}
\usepackage{subcaption}  
\usepackage{hyperref}
\usepackage{mathtools}
\usepackage{placeins}
\usepackage{cleveref}
\usepackage{physics}
\usepackage{xcolor,ulem}

\title{\boldmath Probing Lorentz-violating effects via precession and accretion disk images in a rotating bumblebee spacetime}

\author{Qing Ou,}
\author{Zhen-Bo Wu,}
\author[1]{Qian Wan\note{Corresponding author.}}
\author[2]{and Peng-Cheng Li\note{Corresponding author.}}
\affiliation{School of Physics and Optoelectronics, South China University of Technology,\\ Guangzhou 510641, People's Republic of China}

\emailAdd{wanqian@scut.edu.cn, pchli2021@scut.edu.cn}

\abstract{We investigate kinematic and optical signatures of Lorentz violation in the strong-field region of a rotating bumblebee spacetime generated by a scalar-gradient bumblebee field. 
For generic nonextremal rotating configurations with \(al\neq0\), the surface \(r=r_+\) is curvature singular and is therefore modeled phenomenologically as a perfectly absorbing inner boundary in the ray-tracing calculation. 
By analyzing the spin precession of test gyroscopes and equatorial timelike geodesics, we find that Lorentz violation suppresses the Lense--Thirring precession of static observers, enhances geodetic precession in the static, spherically symmetric limit, and increases the periastron-precession frequency of bound circular orbits. 
Images of a geometrically thin accretion disk further show that the Lorentz violation has a negligible impact on the critical curve, while shrinking the inner-shadow-like feature associated with the absorbing boundary and enhancing the lensed ring. 
These results suggest that selected precession observables and exterior imaging features may provide complementary diagnostics of Lorentz-violating effects in strong-field gravity, whereas the central dark feature remains dependent on the adopted inner-boundary prescription.
}

\begin{document}

\maketitle
\flushbottom

\section{Introduction}

General Relativity (GR) remains the cornerstone of our classical understanding of gravitation, having passed numerous precision tests in the Solar System and beyond \cite{Will:2001mx,Turyshev:2008dr,LIGOScientific:2016aoc,LIGOScientific:2019fpa}. 
Nevertheless, constructing a self-consistent theory of quantum gravity that unifies GR with the Standard Model of particle physics remains one of the most profound open problems in fundamental physics \cite{Crivellin:2023zui,Carlip:2015asa}. 
It is widely expected that quantum effects of gravity become relevant near the Planck scale, far beyond the reach of current particle accelerators. 
However, several candidate frameworks for quantum gravity may leave observable low-energy imprints in the form of Lorentz symmetry violation \cite{Collins:2004bp,Kostelecky:2008ts,MAGIC:2020egb}. 
In this sense, the possible detection of Lorentz-violating effects may offer an important observational window into the underlying structure of quantum gravity.

The Standard Model Extension (SME) provides a systematic effective field theory framework for studying gravity coupled to the Standard Model at low energies \cite{Kostelecky:1994rn,Colladay:1996iz,Colladay:1998fq,Kostelecky:2003fs,Tasson:2016xib,Bluhm:2019ato}. 
Within its gravitational sector, bumblebee theory offers a simple and widely studied mechanism for spontaneous Lorentz symmetry breaking \cite{Maluf:2014dpa,Casana:2017jkc}. 
In this framework, a vector field, known as the bumblebee field, acquires a nonzero vacuum expectation value, which selects a preferred direction in spacetime and thereby spontaneously breaks Lorentz symmetry \cite{Bluhm:2004ep,Bluhm:2007xzd}. 
To date, black hole solutions in bumblebee gravity have been investigated mainly in static, spherically symmetric settings \cite{Casana:2017jkc}, together with several extensions \cite{Ding:2019mal,Maluf:2020kgf,Xu:2022frb,Filho:2022yrk,Liu:2024axg,AraujoFilho:2024ykw,Li:2025rjv,Zhu:2025fiy,Liu:2025oho,Ovcharenko:2026rvj,Xu:2026zgd}. 
These studies have explored a variety of phenomenological implications, including the thermodynamic properties of black holes \cite{Mai:2023ggs,AraujoFilho:2024iox,Liu:2025bpp,AraujoFilho:2025hkm}, gravitational lensing and black hole imaging \cite{2011.14916,2106.14602,Izmailov:2022jon,Islam:2024sph,Kumar:2025bim,AraujoFilho:2025zaj,Sui:2026pyf,QiQi:2026zwp}, and gravitational wave signatures \cite{2102.06303,Liu:2022dcn,Liu:2024oeq,Mai:2024lgk,deOliveira:2025yeo,Chen:2023cjd,Li:2025itp,Liu:2025swi,AraujoFilho:2025nmc,Shi:2025plr,Shi:2026zxx}. 
However, since astrophysical black holes are expected to carry angular momentum, it is essential to investigate genuinely rotating spacetimes in bumblebee gravity and their potential observational consequences. 
In this direction, Ref.~\cite{Poulis:2021nqh} obtained an exact, though special, rotating bumblebee spacetime by taking the bumblebee field to be a scalar-gradient field at its vacuum expectation value. 
For generic nonextremal configurations with \(al\neq0\), this rotating configuration does not possess a regular horizon of the physical metric \cite{Guo:2026sfg}; nevertheless, its exterior geometry provides a useful and tractable framework for exploring Lorentz-violating effects in the strong-field region.

In experimental observations, the rapid advancement of technology has provided complementary windows for testing Lorentz violation. 
For instance, the Event Horizon Telescope (EHT) has achieved event horizon-scale imaging of supermassive black holes M87* and Sgr A* \cite{EventHorizonTelescope:2019dse, EventHorizonTelescope:2019ggy, EventHorizonTelescope:2020qrl, EventHorizonTelescope:2022wkp}, while the GRAVITY collaboration has conducted precise measurements of stellar orbits near the galactic center black hole Sgr A* \cite{Genzel:2024vou,Islam:2021dyk,Cunha:2018gql}, marking a new era in strong gravitational field detection. 
Dynamical features near black holes---such as spin precession and orbital precession---directly encode spacetime curvature and frame-dragging effects \cite{2509.26270,Wang:2021gtd,Zhen:2025nah,QiQi:2024dwc,Wu:2023wld,Li:2023djs,Bambhaniya:2021jum,Wang:2025zis,2505.20736,astro-ph/9811198,1709.08935,Kraniotis:2005zm,2402.01192}, whereas black hole shadows and photon rings reflect near-horizon geometry through strong gravitational lensing \cite{2410.13661,Jha:2022bpv,Kuang:2022xjp,2507.03981,Ayzenberg:2018jip,Alhamzawi:2017iyn,2004.08857,Hou:2022gge,2206.05878,Vagnozzi:2022moj,Zhang:2023cuw,Zhang:2024lsf,2409.06218,Zhang:2025vyx}. 
Additionally, the potential correlation between orbital precession frequencies and high-frequency quasi-periodic oscillations offers supplementary pathways for dynamical detection \cite{Ingram:2019mna,Motta:2016vwf,vanderKlis:2000ca,Bambi:2013fea}. 
It is therefore natural to study precession effects and shadow features within the same rotating spacetime, in order to extract complementary information about Lorentz-violating corrections from different observational channels.

In this work, we study the rotating bumblebee spacetime of \cite{Poulis:2021nqh} and examine how Lorentz violation modifies timelike and null motion in the strong-field regime. 
Through detailed derivation of test gyroscope spin precession frequencies and equatorial timelike motion, coupled with numerical imaging of accretion disk features, we find that increasing the Lorentz violation parameter $l$ suppresses Lense--Thirring (LT) precession of test gyroscopes while enhancing periastron precession frequencies of bound circular orbits. 
Meanwhile, although the critical curve in the image is nearly unaffected by $l$, the size of the inner shadow decreases significantly as the Lorentz-violating effect becomes stronger within the adopted absorbing-boundary and emission prescriptions. 
These results suggest that a combined analysis of orbital precession and inner-shadow measurements may provide an effective probe of Lorentz violation in strong-field gravity.

The paper is organized as follows. In Sec.~\ref{sec2}, we  briefly review the rotating bumblebee spacetime obtained in \cite{Poulis:2021nqh}. 
Subsequently, in Sec.~\ref{sec3}, we study the geodesic motion and orbital precession of equatorial timelike particles. 
Sec.~\ref{sec4} focuses on analyzing the influence of Lorentz violation on the spin precession frequency of a test gyroscope. 
The appearance of a rotating bumblebee spacetime illuminated by a geometrically thin accretion disk is discussed in Sec.~\ref{sec5}, where the effects of Lorentz violation on the main observables in the images are also analyzed. 
Finally, in Sec.~\ref{sec6}, we present our summary and discussion. Throughout this paper, we adopt geometric units with $G = c = 1$.

\section{A rotating solution in bumblebee gravity}
\label{sec2}

In the Einstein--bumblebee theory, the Lorentz symmetry is spontaneously broken by the bumblebee vector field $B_\mu$ acquiring a nonzero vacuum expectation value. 
The action of the theory can be written as (for the convenience of discussion, we have ignored the matter field part here) \cite{Poulis:2021nqh}
\begin{equation}
	\label{action}
	\begin{aligned}
		S&=\int\sqrt{-g}\dd^4x\left(\frac{R}{2\kappa}+\mathcal{L}_B\right), \\
		\mathcal{L}_B&=\frac{\xi}{2\kappa}B^\mu B^\nu R_{\mu\nu}-\frac{1}{4}B_{\mu\nu}B^{\mu\nu}-V,
	\end{aligned}
\end{equation}
where $\kappa=8\pi$ is the Einstein gravitational constant, $R_{\mu\nu}$ and $R$ are the Ricci tensor and scalar, and $B_{\mu\nu}=\partial_\mu B_\nu-\partial_\nu B_\mu$ is the field strength tensor of the bumblebee field. 
The interaction between the bumblebee field and gravity is characterized by the real parameter $\xi$. 
The specific form of the potential $V$ in the theory depends on the model, but to ensure that the $B_\mu$ field acquires a nonzero vacuum expectation value, it is required that 
$V$ has the general form $V=V(B^2\pm b^2)$, where $B^2=g^{\mu\nu}B_\mu B_\nu$ and $b$ is a real constant that fixes the symmetry-breaking scale.

Varying the above action with respect to the metric $g_{\mu\nu}$ and the bumblebee field $B_\mu$, respectively, yields the equations of motion for the gravitational field and the bumblebee field
\begin{equation}
	\begin{aligned}
		&R_{\mu\nu}-\frac{1}{2}g_{\mu\nu}R=\kappa T_{\mu\nu}^B,\\
		&\nabla^\mu B_{\mu\nu}=2V'B_\nu-\frac{\xi}{\kappa}B^\mu R_{\mu\nu},
	\end{aligned}
\end{equation}
where $T_{\mu\nu}^B$ is the energy-momentum tensor of the bumblebee field, and $V'=\partial V/\partial x$ denotes the derivative with respect to its argument $x=B^2\pm b^2$. 
It is evident that $g_{\mu\nu}$ and $B_\mu$ are coupled to each other in the equations of motion, making an exact solution very difficult to obtain. 
However, it was discovered in \cite{Poulis:2021nqh} that when the bumblebee field is at its VEV and can be expressed as the gradient of a scalar field $\lambda$, i.e.,
\begin{equation}
	\label{condi}
	B^2=\mp b^2,\quad B_\mu=\partial_\mu\lambda,
\end{equation}
the coupled equations of motion simplify to a single equation of motion. 
In this case, both $B_{\mu\nu}$ and $V$ vanish in the action, which can therefore be further simplified to
\begin{equation}
	S=\int\dd^4x\frac{\sqrt{-g}}{2\kappa}\left(g^{\mu\nu}+\xi B^\mu B^\nu\right)R_{\mu\nu}.
\end{equation}
If we define the effective (disformally related) metric and its inverse as
\begin{equation}
	\label{gtilde}
	\tilde{g}^{\mu\nu}=g^{\mu\nu}+\xi B^\mu B^\nu,\quad
	\tilde{g}_{\mu\nu}=g_{\mu\nu}-\frac{\xi}{1+\xi B^2}B_\mu B_\nu,
\end{equation}
then the action can be expressed in a more compact form \cite{Poulis:2021nqh}
\begin{equation}
	S=\int\dd^4x\frac{\sqrt{-\tilde{g}}}{2\tilde{\kappa}}\tilde{R},
\end{equation}
where $\tilde{\kappa}=\kappa/\sqrt{1+\xi B^2}$ and $\tilde{g}=\det\,(\tilde{g}_{\mu\nu})$ is the determinant of the background metric, $\tilde{R}_{\mu\nu}$ and $\tilde{R}$ 
are the Ricci tensor and scalar completely expressed in terms
of the background metric $\tilde{g}_{\mu\nu}$. 
This shows that $\tilde{g}_{\mu\nu}$ satisfies the vacuum Einstein field equations
\begin{equation}
	\tilde{R}_{\mu\nu}=0,
\end{equation}
It is important to emphasize that this reduction does not correspond to a test-field approximation, but rather defines a restricted sector of the full Einstein--bumblebee theory in which back reaction effects of bumblebee field are fully encoded in an effective metric structure. 
The condition \eqref{condi} for the bumblebee field can also be rewritten in terms of the effective metric as
\begin{equation}
	\label{condi2}
	\tilde{B}^2=\tilde{g}^{\mu\nu}B_\mu B_\nu=B^2(1+\xi B^2),\quad\partial_\mu B_\nu=\partial_\nu B_\mu.
\end{equation}
In this framework, one may first solve the vacuum Einstein equations for the effective metric $\tilde{g}_{\mu\nu}$, and then reconstruct the physical metric $g_{\mu\nu}$ once a compatible bumblebee configuration is specified.

As an explicit example, consider stationary, axisymmetric vacuum solutions of general relativity. 
The Kerr metric,
\begin{equation}
	\label{kerr}
	\begin{aligned}
		\dd s^2&=-\left(1-\frac{2Mr}{\Sigma}\right)\dd t^2-\frac{4aMr}{\Sigma}\sin^2\theta\dd t\dd\phi+\frac{\Sigma}{\Delta}\dd r^2 \\
		&\quad+\Sigma\dd\theta^2+\left(r^2+a^2+\frac{2a^2Mr}{\Sigma}\sin^2\theta\right)\sin^2\theta\dd\phi^2,
	\end{aligned}
\end{equation}
with
\begin{equation}
	\Sigma=r^2+a^2\cos^2\theta,\quad \Delta=(r-r_+)(r-r_-),\quad r_\pm=M\pm\sqrt{M^2-a^2}
\end{equation}
is a vacuum solution of $\tilde{R}_{\mu\nu}=0$, and therefore may be used as the effective metric $\tilde{g}_{\mu\nu}$. 
Importantly, this does not mean that Kerr is assumed as a fixed background in a test-field sense. Instead, Kerr arises as a solution of the effective vacuum Einstein equations in the constrained Einstein--bumblebee sector. 

In Ref. \cite{Poulis:2021nqh}, a class of spacelike bumblebee configurations compatible with the Kerr effective geometry (i.e., satisfying the conditions in \eqref{condi2}) was constructed,
\begin{equation}
	B_\mu=b\sqrt{1+\xi b^2}\left(0,\sigma_r r/\sqrt{\Delta},\sigma_\theta a\cos\theta,0\right),
\end{equation}
where $\sigma_r,\,\sigma_\theta=\pm 1$ and we will set $\sigma_r=\sigma_\theta=1$ in subsequent discussions for the sake of simplicity. 
It is evident that only $B_r$ and $B_\theta$ are non-zero in the given $B_{\mu}$. 
Therefore, from \eqref{gtilde}, it can be seen that its back reaction to the Kerr metric will only appear in the $rr$, $\theta\theta$, and $r\theta$ components. 
Specifically, we have
\begin{equation}
	\label{metric}
	\begin{aligned}
		\dd s^2&=-\left(1-\frac{2Mr}{\Sigma}\right)\dd t^2-\frac{4aMr}{\Sigma}\sin^2\theta\dd t\dd\phi+\frac{\Sigma+lr^2}{\Delta}\dd r^2+\frac{2lar\cos\theta}{\sqrt{\Delta}}\dd r\dd\theta \\
		&\quad+(\Sigma+la^2\cos^2\theta)\dd\theta^2+\left(r^2+a^2+\frac{2a^2Mr}{\Sigma}\sin^2\theta\right)\sin^2\theta\dd\phi^2,
	\end{aligned}
\end{equation}
where $l=\xi b^2$ is the Lorentz violation parameter. 
It can be observed that the corrections due to the bumblebee field are all proportional to the parameter $l$. 
Consequently, when $l=0$, the above solution reduces to the Kerr spacetime; and when $a=0$, the solution simplifies to the spherically symmetric bumblebee spacetime \cite{Casana:2017jkc}. 
The modified solution obtained here describes a stationary and axisymmetric rotating spacetime in the bumblebee gravity theory.

It is evident that no perturbative expansion in the parameter $l$ is performed in the present construction. 
The resulting spacetime corresponds to a closed-form solution within a specific consistent sector of the Einstein--bumblebee theory, rather than a perturbation around the Kerr geometry, and therefore does not require $l$ to be small at the level of formal derivation. 
However, weak-field tests of Lorentz symmetry violation, such as Solar System experiments, typically impose stringent constraints corresponding to $l\ll 1$ \cite{Casana:2017jkc}. 
These bounds are derived in perturbative and weak-field regimes and may not directly constrain the strong-field region near compact objects, where nonlinear effects become important. 
In this work, we therefore treat $l$ as an effective phenomenological parameter and explore relatively large values to highlight its impact on strong-field observables. 
The goal is not to provide observational constraints on $l$, but to identify characteristic signatures of Lorentz violation in a rotating strong-field spacetime.

It should be noted that the aforementioned method imposes very strict constraints on the bumblebee field. 
Therefore, one should not expect the resulting modified solution to fully preserve the spacetime properties of the effective metric. 
For instance, the solution derived from the Kerr metric inevitably introduces a cross term $g_{r\theta}\sim 1/\sqrt{\Delta}$. 
Since $\sqrt{\Delta}=\sqrt{r-r_+)(r-r_-)}$ becomes imaginary in the interval $r_-<r<r_+$, this metric \eqref{metric} does not provide a real continuation from the exterior region across $r=r_+$. 
More importantly, for generic nonextremal rotating configurations
with $al\neq0$, the inner surface $r=r_+$ is curvature singular
rather than a regular horizon. As shown in Appendix~\ref{app:ricci},
the Ricci scalar behaves generically as
$R\propto 1/\sqrt{\Delta}$ when $r\to r_+^+$. 
We therefore restrict the following analysis to $r>r_+$ and refer to $r_+$ as the inner surface of the rotating bumblebee spacetime. 
Despite this limitation, the domain $r>r_+$, away from the symmetry axis, defines a regular exterior region that can still be used as a phenomenological framework for studying the effects of Lorentz violation on exterior particle motion and photon propagation.  
In the following, we will take the rotating bumblebee metric derived from the Kerr spacetime as an example to study the impact of Lorentz violation on orbital precession, spin precession and accretion disk imaging.

\section{Geodesic motion and orbital precession of equatorial timelike particles}
\label{sec3}
In this section, we investigate the geodesic motion of timelike particles on the equatorial plane of the rotating bumblebee spacetime and analyze the associated orbital precession effects for bound circular orbits. Our focus is on equatorial motion because it provides a clean and astrophysically relevant setup for extracting the fundamental orbital frequencies in the strong-field region. These frequencies are not only useful for describing matter motion in a thin accretion flow, but may also serve as theoretical templates for other timelike observables near supermassive black holes, such as stellar-orbit precession and frequency relations appearing in relativistic interpretations of quasi-periodic oscillations.

\subsection{Timelike geodesics on the equatorial plane}
In the following, we investigate the properties of timelike geodesics on the equatorial plane in the rotating bumblebee spacetime. 
The dynamics of test particles is completely described by the Lagrangian $\mathcal{L}=g_{\mu\nu}\dot{x}^\mu \dot{x}^\nu/2$, where $\dot{x}^\mu=\dd x^\mu/\dd\tau$ is the four-velocity of the particle and $\tau$ is a proper time along the geodesic. 
Due to the stationary and axisymmetric nature of the rotating bumblebee spacetime, the energy $E=-p_t$ and angular momentum $L=p_\phi$ are conserved quantities. 
The $t$- and $\phi$-components of the four-velocity $\dot{x}^\mu$ can thus be expressed as  
\begin{equation}
	\label{ttdot}
	\dot{t} = \frac{E g_{\phi\phi} + L g_{t\phi}}{g_{t\phi}^2 - g_{tt} g_{\phi\phi}}, \qquad  
	\dot{\phi} = -\frac{E g_{t\phi} + L g_{tt}}{g_{t\phi}^2 - g_{tt} g_{\phi\phi}}.
\end{equation}
Hence, the equation of motion for the particle can be written as
\begin{equation}
	\label{veff}
	g_{rr}\dot{r}^2+2g_{r\theta}\dot{r}\dot{\theta}+g_{\theta\theta}\dot{\theta}^2=V_\text{eff}(r,\theta).
\end{equation}
where
\begin{equation}
	V_\text{eff}(r,\theta)=\frac{E^2g_{\phi\phi}+2ELg_{t\phi}+L^2g_{tt}}{g_{t\phi}^2-g_{tt}g_{\phi\phi}}-1,
\end{equation}
denotes the effective potential. 

For timelike particles confined to the equatorial plane, the polar velocity satisfies $\dot{\theta} = 0$. 
Therefore, from Eq.~\eqref{veff} it can be obtained 
\begin{equation}
	\label{rdot}
	\dot{r}=-\sqrt{V_\text{eff}(r,\pi/2)/g_{rr}}.
\end{equation}
Here we have chosen infalling timelike geodesics ($\dot{r}<0$), as they are more physically plausible for constructing accretion disk models in the subsequent analysis. 
In our model, we are interested in bound circular orbital motion, for which the conditions $V_{\text{eff}}(r,\pi/2)=0$ and $\partial_rV_{\text{eff}}(r,\pi/2)=0$ yield the orbital angular frequency as
\begin{equation}
	\label{omegakep}
	\Omega_\phi=\frac{\dd{\phi}}{\dd{t}}=\frac{-\partial_r g_{t\phi}\pm\sqrt{(\partial_r g_{t\phi})^2-\partial_rg_{tt}\partial_rg_{\phi\phi}}}{\partial_rg_{\phi\phi}}
\end{equation}
where the signs $\pm$ correspond to prograde and retrograde circular orbits, respectively. 
In the subsequent discussion, we will only consider prograde circular orbits.
Furthermore, since the radial effective potential $V_{\text{eff}}(r,\theta)$ in the rotating bumblebee spacetime is identical to that in the Kerr spacetime, they also share the same innermost stable circular orbit (ISCO), which is given by \cite{Bardeen:1972fi}
\begin{equation}
	\begin{aligned}
		&r_{\text{ISCO}}=M[3+Z_2\mp\sqrt{(3-Z_1)(3+Z_1+2Z_2)}], \\
		&Z_1=1+(1-a_*^2)^{1/3}[(1+a_*)^{1/3}+(1-a_*)^{1/3}], \\
		&Z_2=\sqrt{3a_*^2+Z_1^2}.
	\end{aligned}
\end{equation}
where $a_*=a/M$ is the dimensionless spin parameter, and $\mp$ corresponds to prograde and retrograde orbits, respectively. 

\subsection{Precession of bound circular orbits}

We now discuss the orbital precession associated with bound circular timelike orbits on the equatorial plane. Although such orbits naturally arise in thin accretion flows, the frequencies derived here are more broadly characteristic of the strong-field timelike dynamics of the rotating bumblebee spacetime.
Assuming that the deviations $\delta r$ and $\delta\theta$ from a reference circular orbit $(r_0, \pi/2)$ are small, we expand the right-hand side of Eq.~\eqref{veff} to second order in the perturbations. 
Noting that $g_{r\theta}(\theta=\pi/2)=0$, and using the circular orbit conditions as well as the reflection symmetry of $V_{\text{eff}}(r,\theta)$ about the equatorial plane, we obtain
\begin{equation}
	\left(\frac{\dd}{\dd t}\delta r\right)^2-\frac{\partial_r^2V_\text{eff}}{2g_{rr}\dot{t}^2}(\delta r)^2+\left(\frac{\dd}{\dd t}\delta\theta\right)^2-\frac{\partial_\theta^2V_\text{eff}}{2g_{\theta\theta}\dot{t}^2}(\delta\theta)^2=0,
\end{equation}
where the metric and derivatives of $V_\text{eff}(r,\theta)$ are all evaluated at $\theta=\pi/2$. 
Evidently, the above expressions indicate that the motions in $r$ and $\theta$ directions are equivalent to two independent harmonic oscillations, whose eigenfrequencies are given by \cite{PhysRevD.52.5707,PhysRevD.90.044004} 
\begin{equation}
	\Omega_i=\left.\sqrt{-\frac{\partial_i^2V_\text{eff}}{2g_{ii}\dot{t}^2}}\right|_{\theta=\pi/2}
\end{equation}
where $i=r,\theta$ denotes radial and polar oscillations, respectively.

Perturbations in the $\theta$-direction of a circular orbit cause the orbital plane of the accretion disk to precess, known as nodal precession. Its frequency is given by \cite{Motta:2013wga}
\begin{equation}
	\Omega_{\text{nod}} = \Omega_\phi - \Omega_\theta.
\end{equation}
Clearly, for the rotating bumblebee spacetime, since the angular frequency $\Omega_\phi$ and the effective potential $V_\text{eff}$ are identical to those in the Kerr spacetime, the nodal precession frequency differs from that of Kerr only through the $g_{\theta\theta}$ component appearing in the denominator of $\Omega_\theta$. 
However, for bound circular orbits on the equatorial plane, $g_{\theta\theta}(\theta=\pi/2)=r^2$, which is identical to the Kerr case. 
Consequently, the nodal precession frequencies in the two spacetimes coincide.

\begin{figure*}
	\centering
	\includegraphics[width=4.5in]{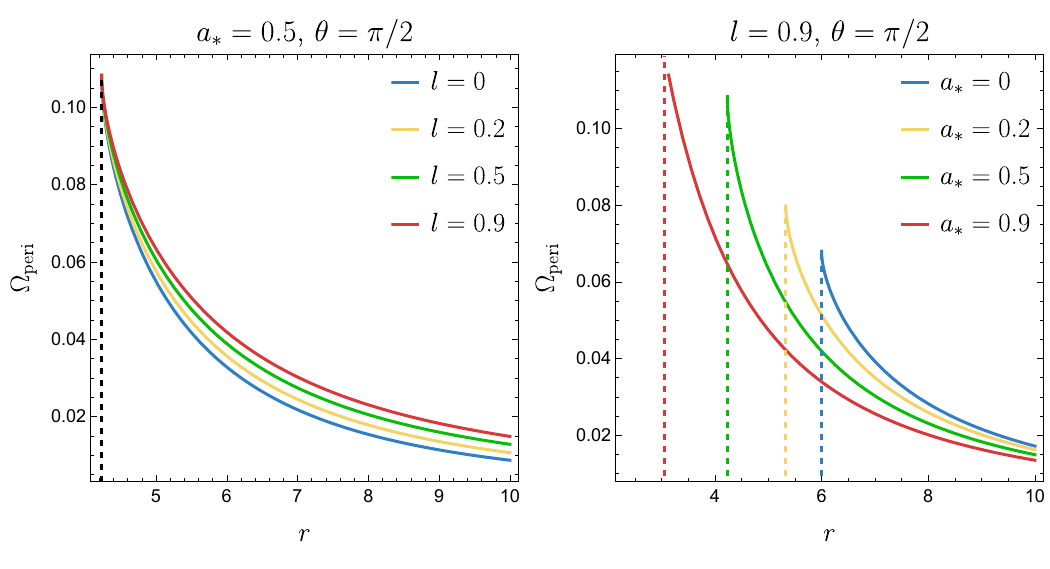}
	\caption{The variation of the periastron precession frequency $\Omega_\text{peri}$ (in $M^{-1}$) with $r$ (in $M$) for different $l$ and $a_*$, the vertical lines indicate the location of the ISCO.} 
	\label{fig:pre}
\end{figure*}

As for perturbations in the radial direction, they induce precession of the orbit's periastron. The corresponding precession frequency is \cite{Motta:2013wga}
\begin{equation}
	\Omega_{\text{peri}} = \Omega_\phi - \Omega_r.
\end{equation}
In Fig.~\ref{fig:pre}, we display the behavior of the periastron precession frequency as a function of the orbital radius $r$ for different parameters. 

From the figure, it can be seen that as the orbit approaches the ISCO, the periastron precession frequency increases monotonically. 
Moreover, an increase in the Lorentz-violating parameter $l$ leads to a larger periastron precession frequency. 
This can be understood from the relation $\Omega_r \sim 1/\sqrt{g_{rr}} \sim 1/\sqrt{l+1}$: an increase in $l$ reduces the radial eigenfrequency, which in turn results in a larger periastron precession frequency. 
It is worth noting that the periastron precession frequency at the ISCO remains unchanged for different values of $l$. 
This is because at the ISCO, $\Omega_r^2\sim\partial_r^2 V_{\text{eff}} = 0$; hence, $\Omega_{\text{peri}} = \Omega_{\phi}$ depends only on the orbital radius of ISCO. 
Since $l$ does not alter the ISCO, $\Omega_{\text{peri}}$ remains unchanged.
Finally, as can be seen from the right panel of Fig.~\ref{fig:pre}, the influence of the spin parameter $a$ on the periastron precession frequency is similar to that for a Kerr black hole. 
Specifically, for circular orbits at the same radius, the periastron precession frequency decreases as $a$ increases. However, considering that increasing the spin brings the ISCO closer to the inner surface, the periastron precession frequency at the ISCO consequently becomes larger.

\section{Spin precession of test gyroscope}
\label{sec4}

\subsection{Spin precession frequency}
We now investigate the spin precession of a test gyroscope in the rotating bumblebee spacetime. 
For a rotating spacetime, frame-dragging prevents observers from remaining fixed at a point in space without changing all three spatial coordinates inside the ergosphere. 
Therefore, a more suitable approach is to attach the test gyroscope to a stationary observer, whose four-velocity is given by
\begin{equation}
	\label{equ}
	u=K/\sqrt{-K^2},
\end{equation}
where $K=\partial_t+\Omega\partial_\phi$ is the most general Killing vector in a stationary axisymmetric spacetime, and $\Omega$ is the angular velocity of the observer. 
Since we are interested in timelike observers, the allowed range for $\Omega$ is
\begin{equation}
    \Omega_-(r,\theta)<\Omega(r,\theta)<\Omega_+(r,\theta),
\end{equation}
with $\Omega_\pm=(-g_{t\phi}\pm\sqrt{g^2_{t\phi}-g_{tt}g_{\phi\phi}})/g_{\phi\phi}$. 
By introducing a parameter $0 < k < 1$, the angular velocity $\Omega$ can be expressed in a general form as
\begin{equation}
	\label{omegak}
	\Omega =k\Omega _{+}+(1-k)\Omega _{-}=\frac{(2k-1)\sqrt{g^2_{t\phi}-g_{tt}g_{\phi\phi}}-g_{t\phi}}{g_{\phi\phi}}.
\end{equation}
It is evident that the parameter $k$ covers the entire range of $\Omega$ from $\Omega_-$ to $\Omega_+$. 
When $k = 1/2$, we have $\Omega = -g_{t\phi}/g_{\phi\phi}$, and the  specific angular momentum of a stationary observer (relative to a static observer at infinity) is $u_\phi=g_{\phi\mu}u^\mu=0$, meaning that it is a zero angular momentum observer (ZAMO).

Consider a test gyroscope attached to the stationary observer. 
Its spin vector undergoes Fermi-Walker transported along Eq.~\eqref{equ}, and the spin precession frequency can be expressed as \cite{straumann2004general}
\begin{equation}
	\tilde{\Omega}_p=\frac{1}{2K^2}*(\tilde{K}\wedge\dd\tilde{K}),
\end{equation}
where $\tilde{K}=g_{\mu\nu}K^\mu\dd x^\nu$ is the covector of the Killing vector $K$, $*$ and $\wedge$ denote the Hodge dual and wedge product. 
For the rotating bumblebee spacetime, this can be further written as
\begin{equation}
	\begin{aligned}
		\tilde{\Omega}_p&=\frac{\epsilon_{3ij}C_ig_{j\mu}}{2\sqrt{-g}\left(1+2\Omega\frac{g_{t\phi}}{g_{tt}}+\Omega^2\frac{g_{\phi\phi}}{g_{tt}}\right)}\dd x^\mu,
	\end{aligned}
\end{equation}
where
\begin{equation}
	\begin{aligned}
		C_i=\left(\partial_i g_{t\phi}-\frac{g_{t\phi}}{g_{tt}}\partial_i g_{tt}\right)+\Omega\left(\partial_i g_{\phi\phi}-\frac{g_{\phi\phi}}{g_{tt}}\partial_i g_{tt}\right)+\Omega^2\left(\frac{g_{t\phi}}{g_{tt}}\partial_i g_{\phi\phi}-\frac{g_{\phi\phi}}{g_{tt}}\partial_i g_{t\phi}\right)
	\end{aligned}
\end{equation}
In the above, $\epsilon_{3ij}$ is the Levi-Civita symbol, and the indices $i,\,j$ denote spatial indices.

\begin{figure*}
	\centering
	\includegraphics[width=6.2in]{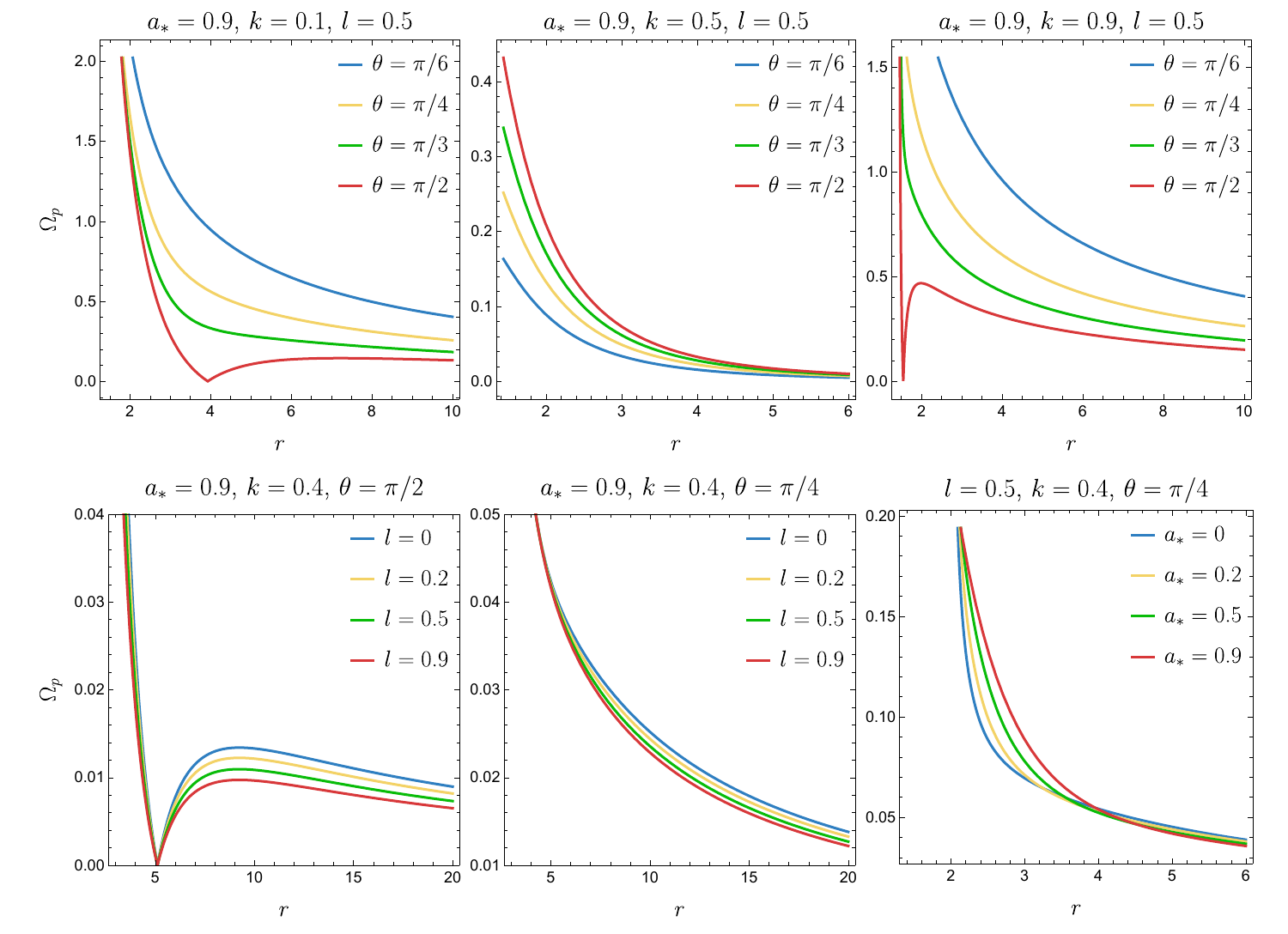}
	\caption{The modulus of spin precession frequency $\Omega_p$ (in $M^{-1}$) as a function of radial coordinate $r$ (in $M$) for the rotating bumblebee spacetime. In the top row, we adopt different values of parameter $k$ and focus on the influence of the angle $\theta$ on the precession frequency. In the bottom row, we concentrate on the effect of the Lorentz-violating parameter $l$ and spin parameters $a_*$ on the precession frequency.}
	\label{spin_gen}
\end{figure*}

We now discuss the precession properties of a gyroscope in the rotating bumblebee spacetime. 
Using $\Omega$ defined in Eq.~\eqref{omegak}, the magnitude of the spin precession frequency can be expressed as
\begin{equation}
	\label{omega_p}
	\Omega_p=|\tilde{\Omega}_p|=\frac{g_{tt}g_{\phi\phi}\sqrt{g_{rr}C_2^2-2g_{r\theta}C_1C_2+g_{\theta\theta}C_1^2}}{8\sqrt{-g}(k-1)k(g_{t\phi}^2-g_{tt}g_{\phi\phi})},
\end{equation}
In Fig.~\ref{spin_gen}, we illustrate the behavior of the spin precession frequency as a function of the radial coordinate $r$. 
The top panels display the spin precession frequency for different values of $k$ (i.e., for different stationary observers) and the angle $\theta$. 
It can be seen from the figures that, except for the ZAMO with $k=1/2$, the spin precession frequency $\Omega_p$ of the test gyroscope carried by an observer diverges as the observer approaches the surface $r=r_+$ from the exterior. 
From Eq.~\eqref{omega_p}, this divergence arises because the term $g_{t\phi}^2-g_{tt}g_{\phi\phi}=\Delta\sin^2\theta$ in the denominator vanishes at the inner surface. 
However, when $k=1/2$, the factor $g_{rr}C_2^2-2g_{r\theta}C_1C_2+g_{\theta\theta}C_1^2\sim (g_{t\phi}^2-g_{tt}g_{\phi\phi})^2$ appearing in the numerator cancels the divergent term in the denominator, thereby keeping the spin precession frequency finite.

The bottom panels of Fig.~\ref{spin_gen} show the variation of the spin precession frequency with the radial coordinate $r$ for different values of the spin parameter $a_*$ and the Lorentz-violating parameter $l$. 
The lower left and middle panels show that Lorentz violation reduces $\Omega_p$. 
Moreover, when the test gyroscope is located in the equatorial plane, $\Omega_p$ first increases and then decreases to zero as the radial coordinate $r$ decreases, finally rising again and diverging at $r\to r_+$. 
It is noteworthy that the position of $\Omega_p=0$ is independent of the Lorentz-violating parameter $l$, because at $\theta = \pi/2$ one has $g_{rr}C_2^2-2g_{r\theta}C_1C_2+g_{\theta\theta}C_1^2=r^2C_1^2$, and $C_1$ depends only on $a_*$ and $k$. 
The lower right panel further indicates that the dependence of $\Omega_p$ on the spin parameter $a_*$ remains consistent with that of the Kerr black hole \cite{Chakraborty:2016mhx}.

\subsection{LT precession and geodetic precession}
The frequency $\Omega_p$ discussed in the previous subsection includes the overall contribution to the spin precession. 
To gain a clearer understanding of the precession arising from different causes, we consider two limiting cases: $\Omega = 0$ and $a = 0$. These correspond, respectively, to the LT precession induced by the rotation of the spacetime and the geodetic precession induced by the spacetime curvature.

\begin{figure*}
	\centering
	\includegraphics[width=6in]{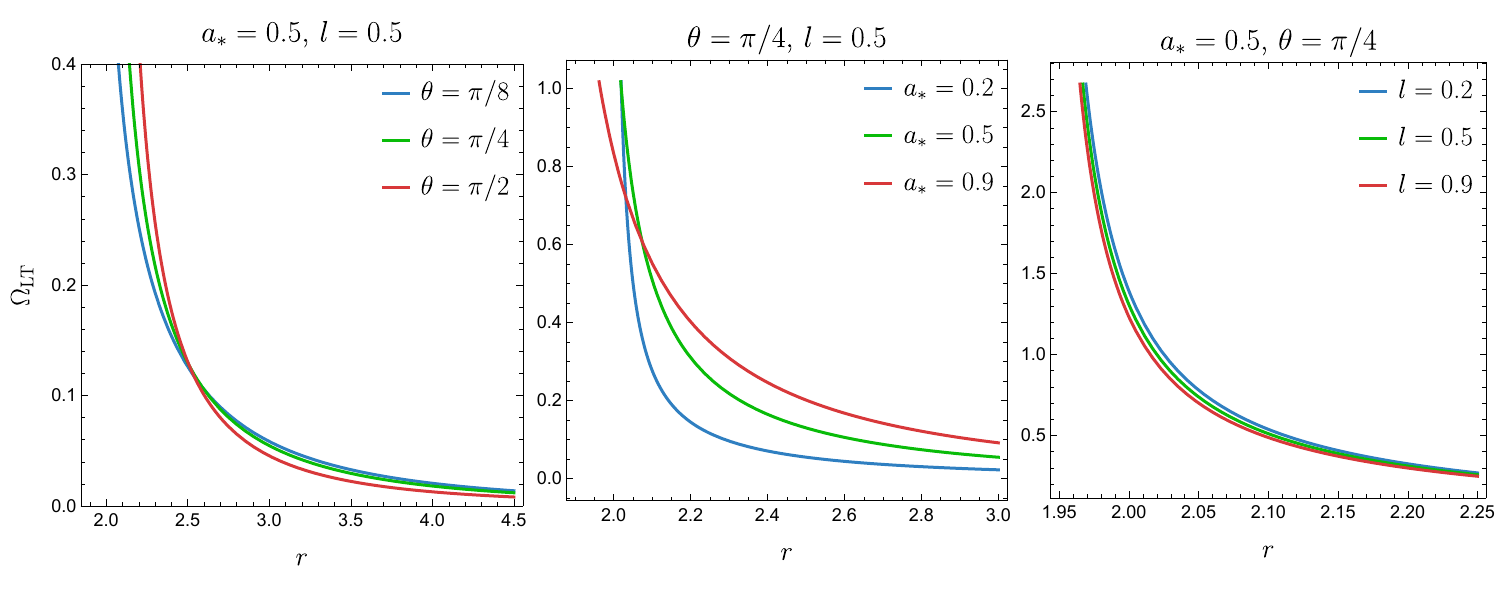}
	\caption{LT precession frequency $\Omega_\text{LT}$ (in $M^{-1}$) of a test gyroscope as a function of the radial coordinate $r$ (in $M$) in the rotating bumblebee spacetime.}
	\label{fig:3a}
\end{figure*}

When the angular velocity $\Omega$ vanishes, the stationary observer reduces to a static observer \footnote{It should be noted that static observers can exist only outside the ergosphere. The rotating bumblebee spacetime discussed in this paper has the same $r_+$ as the Kerr black hole, and its metric component $g_{tt}$ is also independent of the Lorentz-violating parameter. As a result, its ergosphere is exactly identical to that of the Kerr spacetime.}, and the spin precession frequency simplifies to the LT precession frequency. 
In Fig.~\ref{fig:3a}, we illustrate the behavior of the magnitude of the LT precession frequency as a function of the radial coordinate $r$ in the rotating bumblebee spacetime. 
It is evident that its magnitude increases monotonically as $r$ decreases and exhibits divergent behavior near the ergosurface. 
When the test gyroscope is far away from the ergosurface, the differences in the magnitude of the LT precession frequency for various parameters gradually diminish. 

Furthermore, as can be seen from the left and middle panels of Fig.~\ref{fig:3a}, the dependence of the LT precession frequency on the polar angle $\theta$ and the spin parameter $a$ is similar to that in the Kerr black hole \cite{Chakraborty:2016ipk}. 
Specifically, both a decrease in $a_*$ and the approach of the test gyroscope toward the equatorial plane tend to suppress $\Omega_{\mathrm{LT}}$. 
However, in these cases the ergosurface (or infinite redshift surface) radius $r_{\infty}=M(1+\sqrt{1-a_*^2\cos^2\theta})$ also shifts outward, so that $\Omega_{\mathrm{LT}}$ diverges more rapidly.
From the right panel, it can be seen that an increase in the Lorentz-violating parameter $l$ leads to a decrease in the LT precession frequency.

When $a = 0$, the rotating bumblebee solution, Eq.~\eqref{metric}, reduces to the spherically symmetric bumblebee black hole solution. 
In this case, the spacetime structure is static and spherically symmetric. 
For computational convenience, the test gyroscope can be placed on the equatorial plane ($\theta = \pi/2$). 
From Eq.~\eqref{omega_p}, the spin precession frequency becomes
\begin{equation}
	\Omega_p|_{a=0,\theta=\pi/2}=\frac{(r-3M)\Omega}{\sqrt{1+l}(r-2M-r^3\Omega^2)}.
\end{equation}
Clearly, this frequency is non-zero \footnote{Circular orbits on the equatorial plane exist only for radii larger than the photon orbit radius, which is $r_c=3M$ for the spherically symmetric bumblebee spacetime discussed here.}, indicating that the spin of a test gyroscope also precesses in a static and spherically symmetric spacetime, which arises from the intrinsic curvature of the spacetime.

Consider a test gyroscope moving along a timelike circular geodesic, whose angular velocity is given by Eq.~\eqref{omegakep} $\Omega=\sqrt{M/r^3}$ . 
In this scenario, the spin precession frequency is
\begin{equation}
\label{eq32}
	\Omega_p=\frac{1}{\sqrt{1+l}}\sqrt{\frac{M}{r^3}}.
\end{equation}
It is worth mentioning that, unlike in the Schwarzschild spacetime where $\Omega_p=\Omega$ \cite{PhysRevD.19.2280}, these two frequencies are not equal in the spherically symmetric bumblebee spacetime. 
This discrepancy arises because the Lorentz-violating effect alters the spacetime structure. 
The frequency \eqref{eq32} is calculated with respect to the proper time $\tau$ measured in the Copernican frame. 
Using $\dd\tau=\sqrt{1-3M/r}\dd t$, the precession frequency with respect to the coordinate time $t$ can be expressed as
\begin{equation}
	\Omega_p'=\sqrt{1-\frac{3M}{r}}\Omega_p
\end{equation}
Consequently, the geodetic precession frequency of the spin vector of the test gyroscope is the difference of $\Omega_p'$ and $\Omega$, and the result is \cite{Chakraborty:2016mhx,10.1119/1.1604390}
\begin{equation}
	\Omega_{\mathrm{geo}}=\left(1-\frac{1}{\sqrt{1+l}}\sqrt{1-\frac{3M}{r}}\right)\sqrt{\frac{M}{r^3}}.
\end{equation}
Clearly, when $l=0$, the result above is consistent with that in the Schwarzschild spacetime \cite{PhysRevD.19.2280}.

\begin{figure}
	\centering
	\includegraphics[width=2.5in]{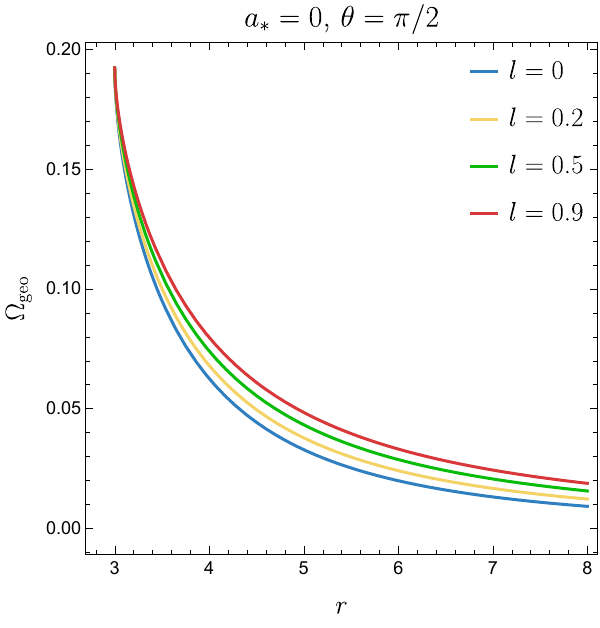}
	\caption{The geodetic precession $\Omega_\text{geo}$ (in $M^{-1}$) as a function of radial coordinate $r$ (in $M$) for the spherically symmetric bumblebee spacetime.}
	\label{fig:5b}
\end{figure}

In Fig.~\ref{fig:5b}, we show the variation of the geodetic precession frequency with the radial coordinate $r$. 
It can be seen that the magnitude of the precession frequency increases monotonically as $r$ decreases, reaching a maximum value of $\Omega_\mathrm{geo}=(3\sqrt{3}M)^{-1}$ at $r=3M$. Furthermore, an increase in the Lorentz-violating parameter $l$ enhances the geodetic precession frequency.

\section{Accretion disk images in a rotating bumblebee spacetime}
\label{sec5}

In addition to affecting the orbital precession of timelike particles, the Lorentz-violating effects may also alter the trajectories of photons in the strong-field region via strong gravitational lensing, thereby leaving potential observable imprints in accretion disk images. 
In the following, we take a geometrically thin accretion disk lying on the equatorial plane as an example to study its imaging characteristics and analyze the influence of Lorentz-violating effects.

For convenience, we consider a prograde geometrically thin accretion disk extending from the inner boundary $r_+$ to $r=60\,r_+$, composed of electrically neutral plasma. 
Outside the ISCO, the accretion flow follows circular motion, whereas inside the ISCO, the accretion flow can no longer maintain stable circular orbits and will plunge rapidly toward, and terminate at, $r=r_+$ while conserving its energy $E_\text{ISCO}$ and angular momentum $L_\text{ISCO}$. 
The four-velocity of the particles is described by Eqs.~\eqref{ttdot} and \eqref{rdot}. 
In the ray-tracing calculation, $r=r_+$ is treated phenomenologically as a perfectly absorbing inner boundary \footnote{This absorption condition is imposed as part of the imaging prescription and is not inferred from the local properties of the curvature-singular surface.}, and photons reaching it are removed from the integration.
It is worth noting that although the accretion disk model introduced here is quite simple, numerous studies \cite{Gralla:2020srx, Hou:2022eev, Zhang:2023bzv, Zeng:2025pch, Zeng:2025tji, Zeng:2025kqw, Meng:2025ivb,Wan:2026xzs} have shown that it can still capture the essential features of optical images with very modest computational resources.

\subsection{Imaging scheme}

Imaging an accretion disk around a rotating bumblebee spacetime requires propagating the photons emitted by the disk to a distant observer. 
Owing to the symmetries of the spacetime in the $t$ and $\phi$ directions, we consider a ZAMO located at $(t_o,r_o,\theta_o,\phi_o)$, whose orthonormal tetrad can be written as
\begin{equation}
	\begin{aligned}
		& e_{(0)}=\zeta\left(1,0,0,-\frac{g_{t\phi}}{g_{\phi\phi}}\right),\quad e_{(1)}=\left(0,\frac{1}{\sqrt{g_{rr}}},0,0\right), \\
		& e_{(2)}=\eta\left(0,-\frac{g_{r\theta}}{g_{rr}},1,0\right),\quad e_{(3)}=\left(0,0,0,\frac{1}{\sqrt{g_{\phi\phi}}}\right),
	\end{aligned}
\end{equation}
with  
\begin{equation}
	\zeta=\sqrt{\frac{g_{\phi\phi}}{g_{t\phi}^2-g_{tt}g_{\phi\phi}}},\quad \eta=\sqrt{\frac{g_{rr}}{g_{rr}g_{\theta\theta}-g_{r\theta}^2}}.
\end{equation}
Note that the presence of a non-zero $g_{r\theta}$ term in the metric causes the form of $e_{(2)}$ to differ slightly from the corresponding tetrad in the Kerr spacetime.

To save computational resources, we actually employ the backward ray-tracing method in the numerical calculations. 
Starting from the observer’s position, we reverse the directions of $e_{(1)}$ and $e_{(3)}$ in the ZAMO frame and integrate the null geodesic equations backward to trace the photon trajectories. 
Technical details can be found in \cite{Hu:2020usx}.
Once the ray paths are obtained, the change of specific intensity $I_{\nu}$ along a geodesic is computed using the general relativistic radiative transfer (GRRT) equation \cite{LINDQUIST1966487}  
\begin{equation}
	\label{grrt}
	\frac{\dd}{\dd\lambda}\left(\frac{I_\nu}{\nu^3}\right)=\frac{j_\nu-\alpha_\nu I_\nu}{\nu^2}.
\end{equation}
Here $\lambda$ is affine parameter, $j_\nu$ and $\alpha_\nu$ denote the emission and absorption coefficients at frequency $\nu$, respectively. 
For a geometrically thin accretion disk lying on the equatorial plane, each time a backward traced ray crosses the equatorial plane it undergoes one emission and absorption process, and the absorption and emission coefficients can be treated as constants during that crossing. 
Hence the solution of Eq.~\eqref{grrt} can be written as  
\begin{equation}
	\label{inten}
	I_{\nu_o}=\sum_{n=1}^{N_\text{max}}f_ng_n^3J_n,
\end{equation}
where the subscript $n$ indicates the $n$-th crossing of the equatorial plane, $g_n=\nu_o/\nu_n$ is the redshift factor, and $f_n$ is a factor related to the absorption effect when crossing the accretion disk. 
In the following calculations, we consider the extreme optically-thin limit with $\alpha_\nu\to 0$, which gives $f_n=\nu_n\Delta\lambda_n$ and we will take $f_n=1$ for simplicity \cite{Hou:2022eev}. 
For the emissivity we adopt  
\begin{equation}
	\label{emit}
	J_n=\exp(-\frac{1}{2}z_n^2-2z_n),\quad z_n=\log(r_n/r_+),
\end{equation}
which has also been used to fit the time-averaged 230 GHz black hole image observed by the EHT \cite{Chael:2021rjo}.

\subsection{Intensity maps}

\begin{figure*}[htbp]
	\centering
	\includegraphics[width=6in]{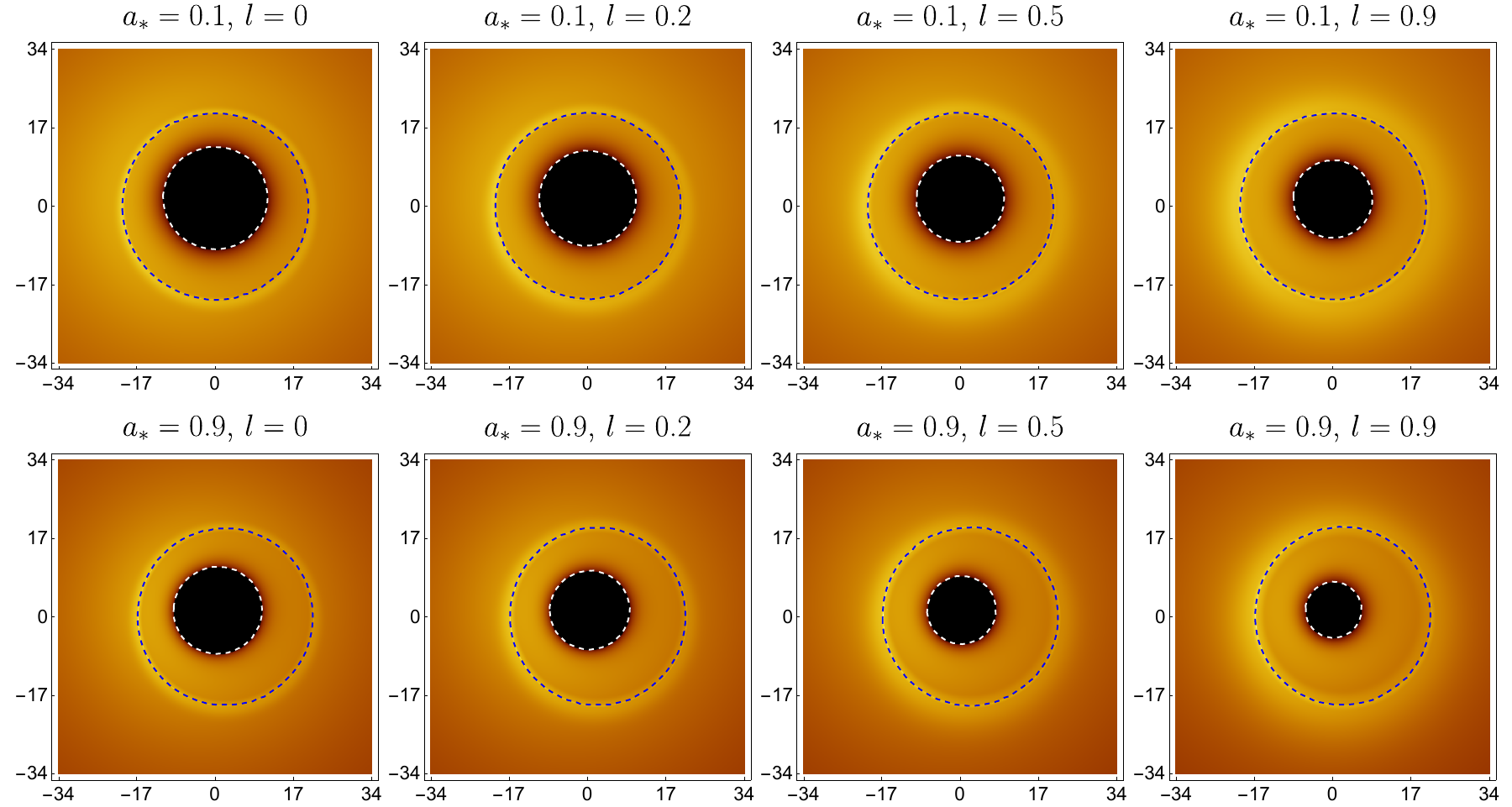}\\
	\includegraphics[width=6in]{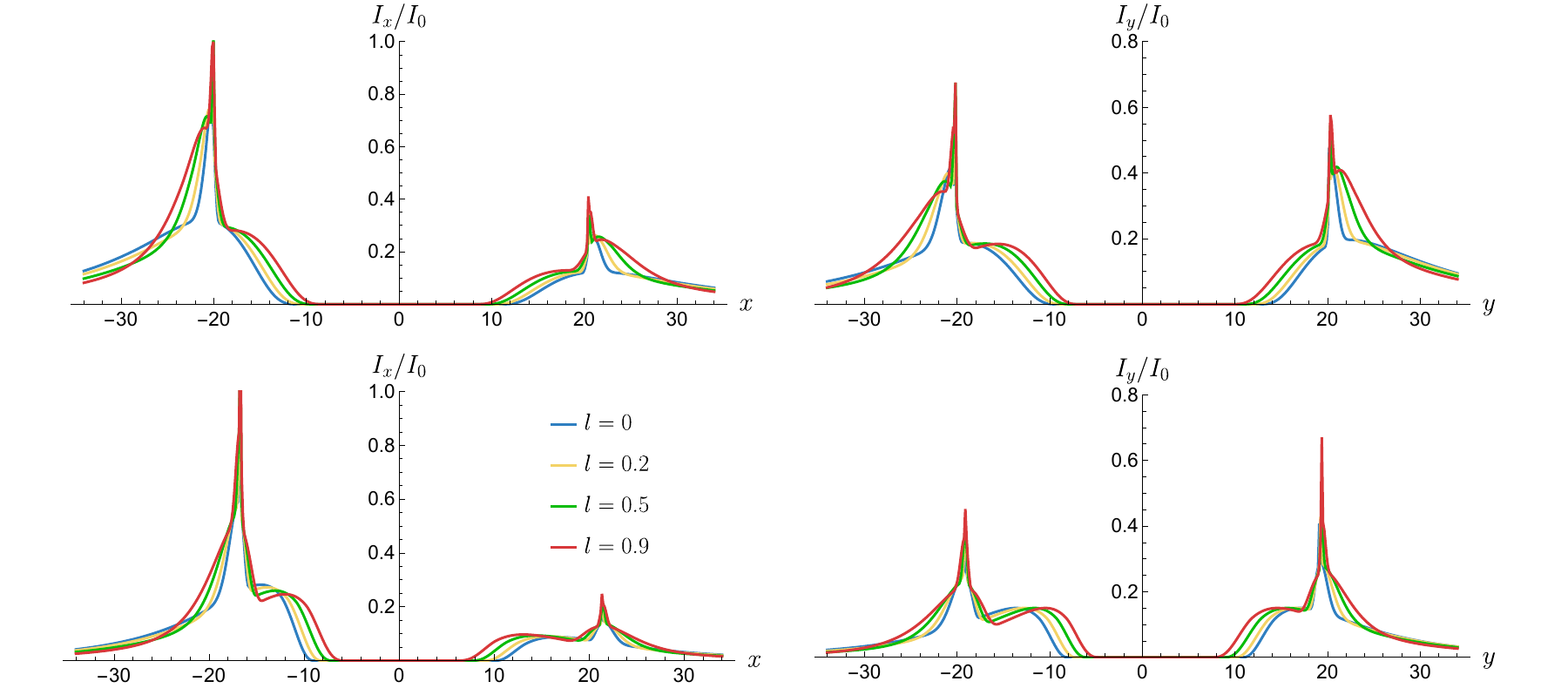}
	\caption{Imaging results at an observational inclination angle $\theta_o=17^\circ$. \textbf{Top two rows:} Images of a rotating bumblebee spacetime illuminated by a geometrically thin accretion disk. The white dashed line indicates the boundary of the inner shadow, and the blue dashed line represents the critical curve. The horizontal and vertical coordinates are in units of $\mu$as. \textbf{Bottom two rows:} Intensity profiles along the $x$ and $y$‑axes for different parameter sets. The horizontal coordinate is in $\mu$as, and the vertical coordinate has been normalized to the peak intensity $I_0$ for each spin case.}
	\label{fig17}
\end{figure*}

Utilizing the numerical ray-tracing method introduced earlier, we image the accretion disk light source around the rotating bumblebee spacetime and scale all images using M87* as a template, so that their angular gravitational size is $M/D=3.78\,\mu$as \cite{EventHorizonTelescope:2019ggy}.

\begin{figure*}[htbp]
	\centering
	\includegraphics[width=6in]{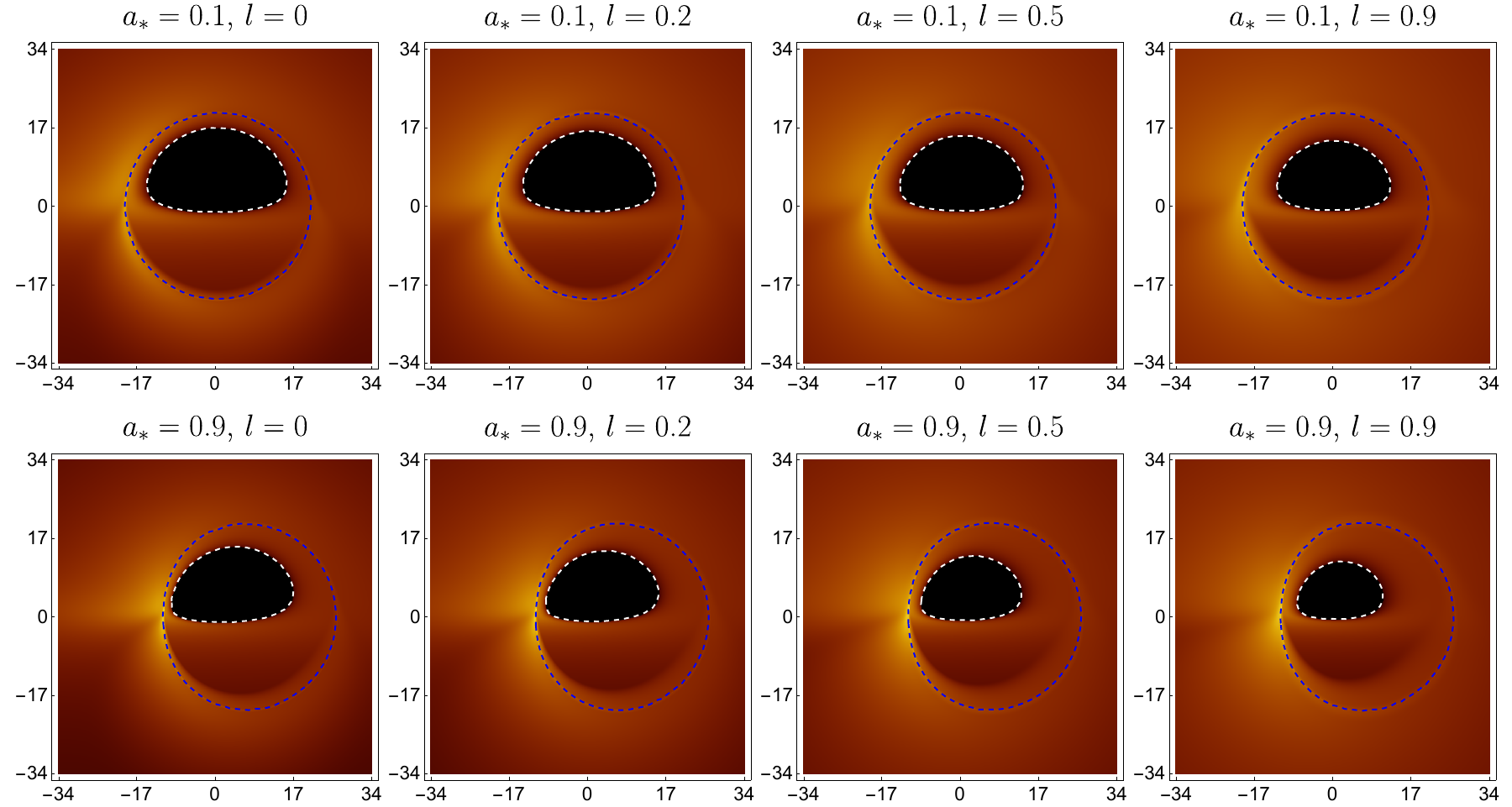}\\
	\includegraphics[width=6in]{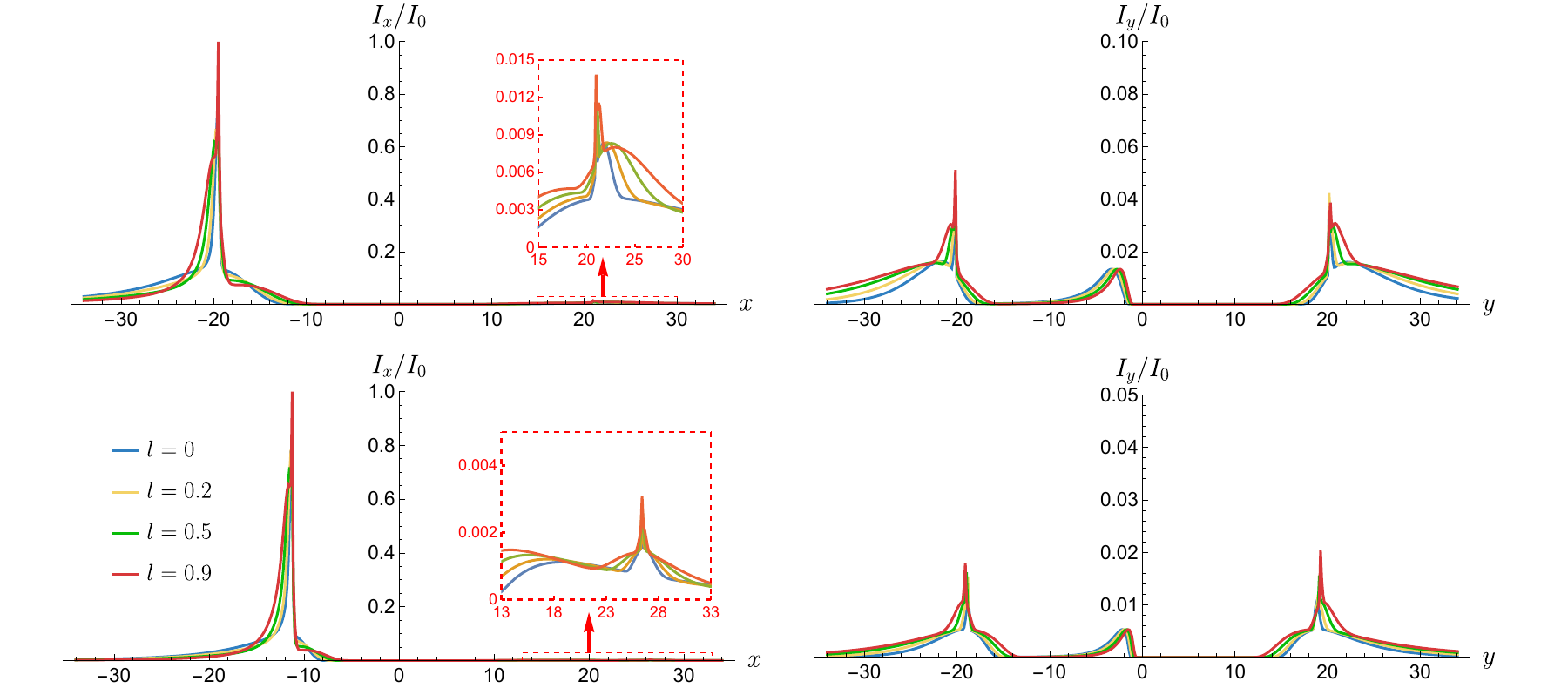}
	\caption{Imaging results at an observational inclination angle $\theta_o=80^\circ$. \textbf{Top two rows:} Images of a rotating bumblebee spacetime illuminated by a geometrically thin accretion disk. The white dashed line indicates the boundary of the inner shadow, and the blue dashed line represents the critical curve. The horizontal and vertical coordinates are in units of $\mu$as. \textbf{Bottom two rows:} Intensity profiles along the $x$ and $y$‑axes for different parameter sets. The horizontal coordinate is in $\mu$as, and the vertical coordinate has been normalized to the peak intensity $I_0$ for each spin case.}
	\label{fig80}
\end{figure*}

Fig.~\ref{fig17} and Fig.~\ref{fig80} show the optical appearance of the rotating bumblebee spacetime illuminated by the geometrically thin accretion disk for observer inclination angles $\theta_o=17^\circ$ and $\theta_o=80^\circ$, respectively. 
In each figure, the first and second rows correspond to spin parameters $a_*=0.1$ and $a_*=0.9$, with the Lorentz violation parameter $l$ increasing from left to right. 
The blue and white dashed curves denote the critical curve and inner shadow \footnote{In this work, the term “inner shadow” is retained only as an operational label for the lensed contour associated with rays terminated at the adopted absorbing inner boundary. For configurations with $al\neq0$, this terminology does not imply that $r=r_+$ is a regular horizon or that the inner surface is physically equivalent to one.}, respectively. 
For quantitative comparison, the third and fourth rows present intensity cut profiles along the $x$- and $y$-axes for $a_*=0.1$ and $a_*=0.9$; the intensity in each row has been normalized to the respective maximum value for the corresponding spin.

\begin{figure*}[htbp]
	\centering
	\includegraphics[width=6in]{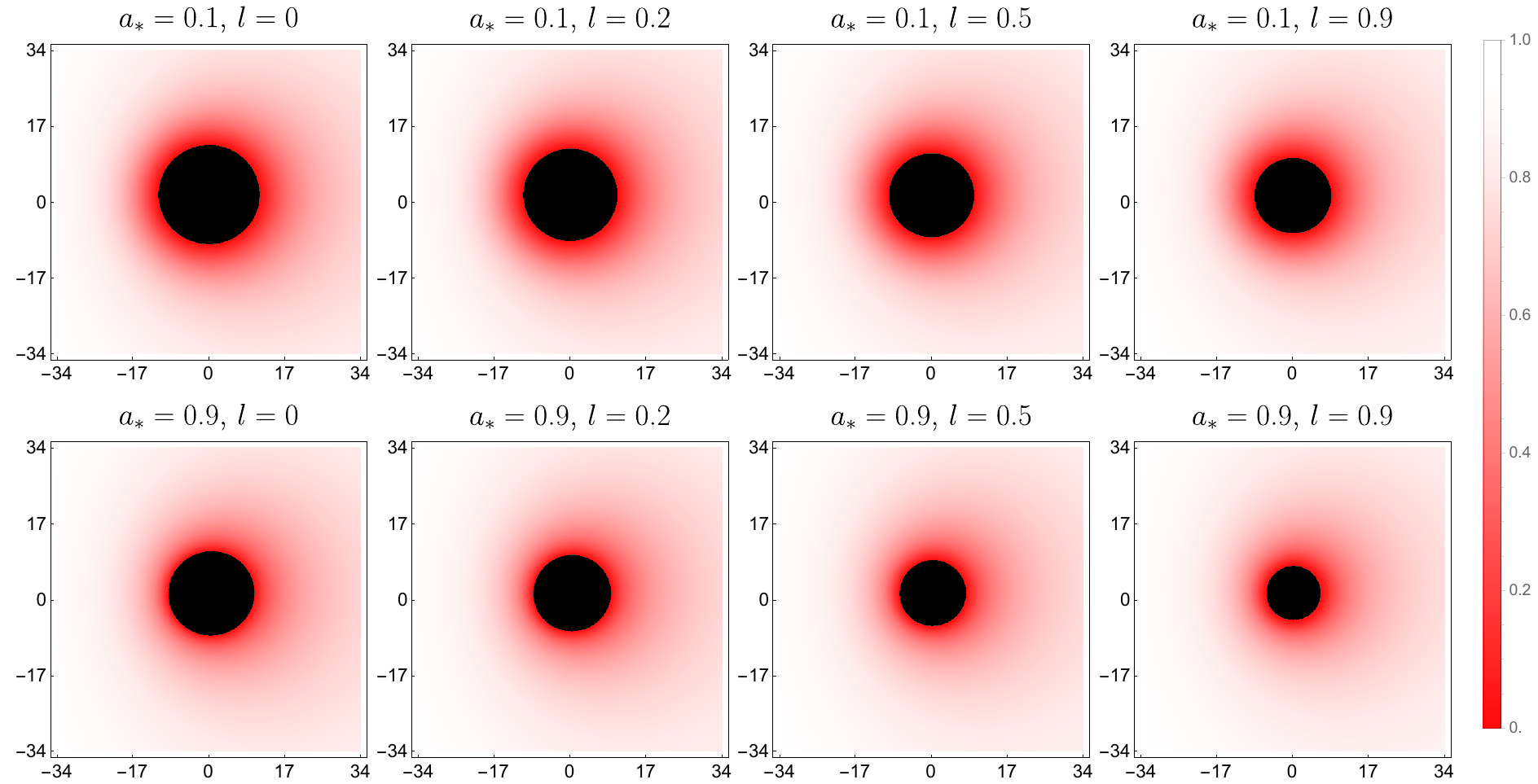}\\
	\includegraphics[width=6in]{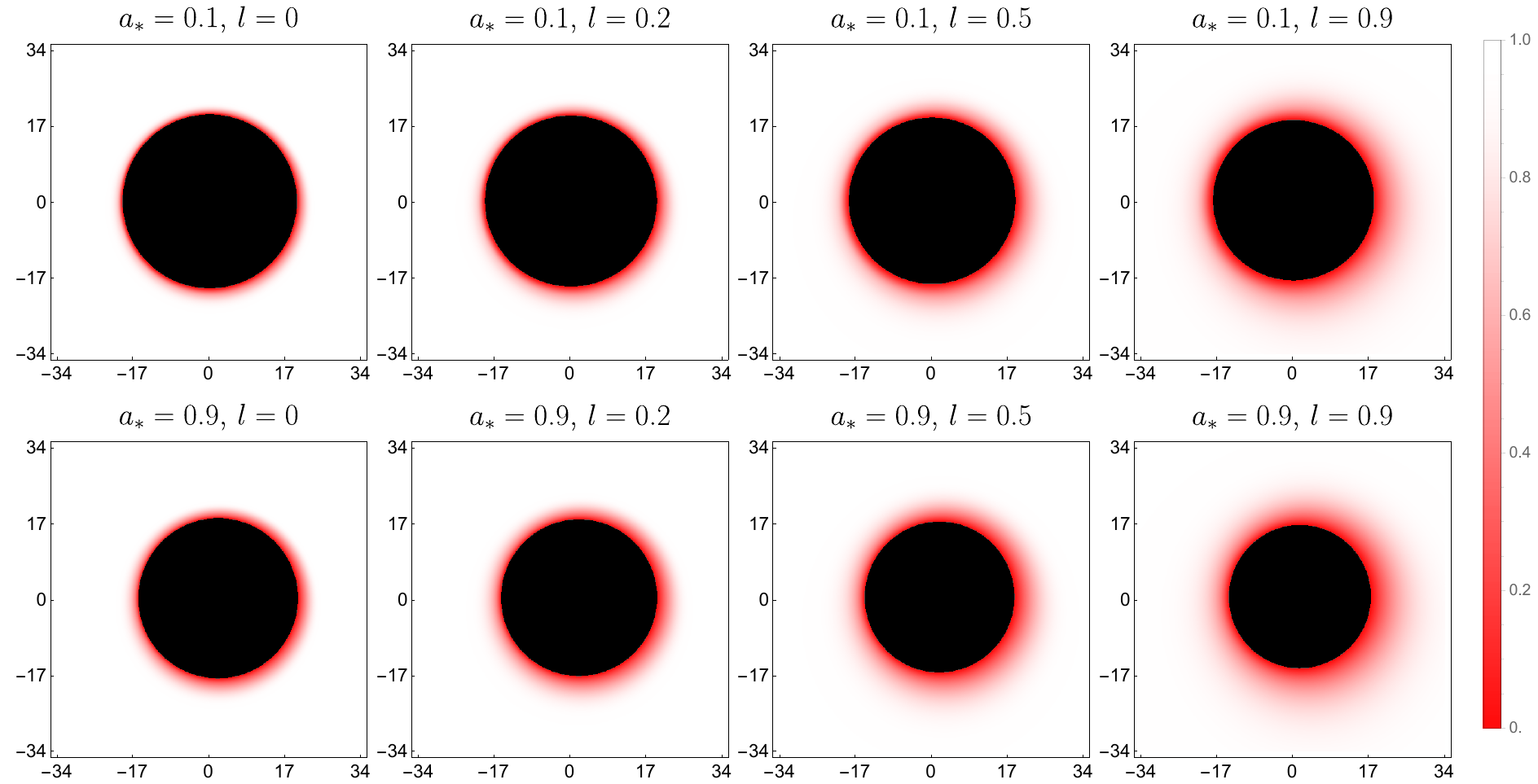}
	\caption{Redshift distribution of the direct (top two rows) and lensed (bottom two rows) images from the accretion disk at an observational inclination angle $\theta_o=17^\circ$. Red and blue colors denote redshift and blueshift, respectively. The boundaries of the black regions correspond to the inner shadow and the lensed images of the adopted absorbing inner boundary.}
	\label{fig17rs}
\end{figure*}

From Fig.~\ref{fig17}, a prominent feature in the images is a central brightness depression, which arises from a combination of the adopted perfectly absorbing inner-boundary prescription and the strong gravitational redshift of emission from the innermost region (detailed information can be extracted from Fig.~\ref{fig17rs}). 
It is worth noting that, for the geometrically thin accretion disk considered here, the outline of this central brightness depression nearly coincides with the inner shadow (white dashed curve in the figures). As an image feature determined jointly by the spacetime geometry, observational inclination, and emission model, the inner shadow serves as a potential probe of the spacetime geometry in the region near the inner surface $r=r_+$ \cite{Chael:2021rjo}. 
Comparing different panels in the same row reveals that, although the roots of $\Delta=0$ coincide with the Kerr horizon radii, the lensed projections of the adopted inner boundaries---the inner shadows in our operational terminology---differ significantly.
Specifically, the size of the inner shadow of a rotating bumblebee spacetime decreases noticeably as the parameter $l$ increases. 
By comparing vertical and horizontal panels, one can also see that the influence of parameter $l$ on the inner‑shadow size is more significant than that of the spin parameter $a_*$. 
Within the fixed emission model and absorbing-boundary prescription adopted here, the calculated inner-shadow morphology is sensitive to $l$ and may help distinguish the rotating bumblebee spacetime from the Kerr geometry. 

Besides the inner shadow, another prominent feature in the images is the photon ring structure surrounding the central brightness depression. 
This ring is formed by photons that orbit the central object multiple times, creating higher‑order images. 
As the number of photon orbits increases, the photon ring gradually converges toward the critical curve (represented by the blue dashed line in the figure). 
It is noteworthy that, unlike the inner shadow, the size and shape of the critical curve are almost unaffected by $l$. 
This can be seen more clearly in the intensity cuts along the $x$- and $y$-axes, where the peak positions of the intensity cuts show no significant shift with varying $l$. 
However, the size and intensity of the lensed ring are altered by the parameter $l$: increasing $l$ broadens the distribution and enhances the intensity of the lensed ring. 
Consequently, in a rotating bumblebee spacetime the lensed ring appears brighter and more pronounced than in the Kerr black hole case.

Fig.~\ref{fig80} shows the imaging results for $\theta_o=80^\circ$. 
Here, a clearly visible inner shadow and a bright ring structure are also present. 
Similar to the case at $\theta_o=17^\circ$, although the shape of the inner shadow changes from  nearly circular to an approximate semi-circle, its size also decreases with increasing $l$, while the critical curve size remains nearly unchanged, and the brightness and width of the lensed ring increase with larger $l$. 
Furthermore, the relative motion between the prograde accretion disk and the observer produces a Doppler blueshift effect, resulting in a crescent‑shaped bright region on the left side of the critical curve. 
In contrast, the intensity on the right side is significantly suppressed by Doppler redshift. 
As illustrated in the inset of Fig.~\ref{fig80}, the intensity of the left lensed ring is approximately 100 times larger than that of the right side.

\begin{figure*}[htbp]
	\centering
	\includegraphics[width=6in]{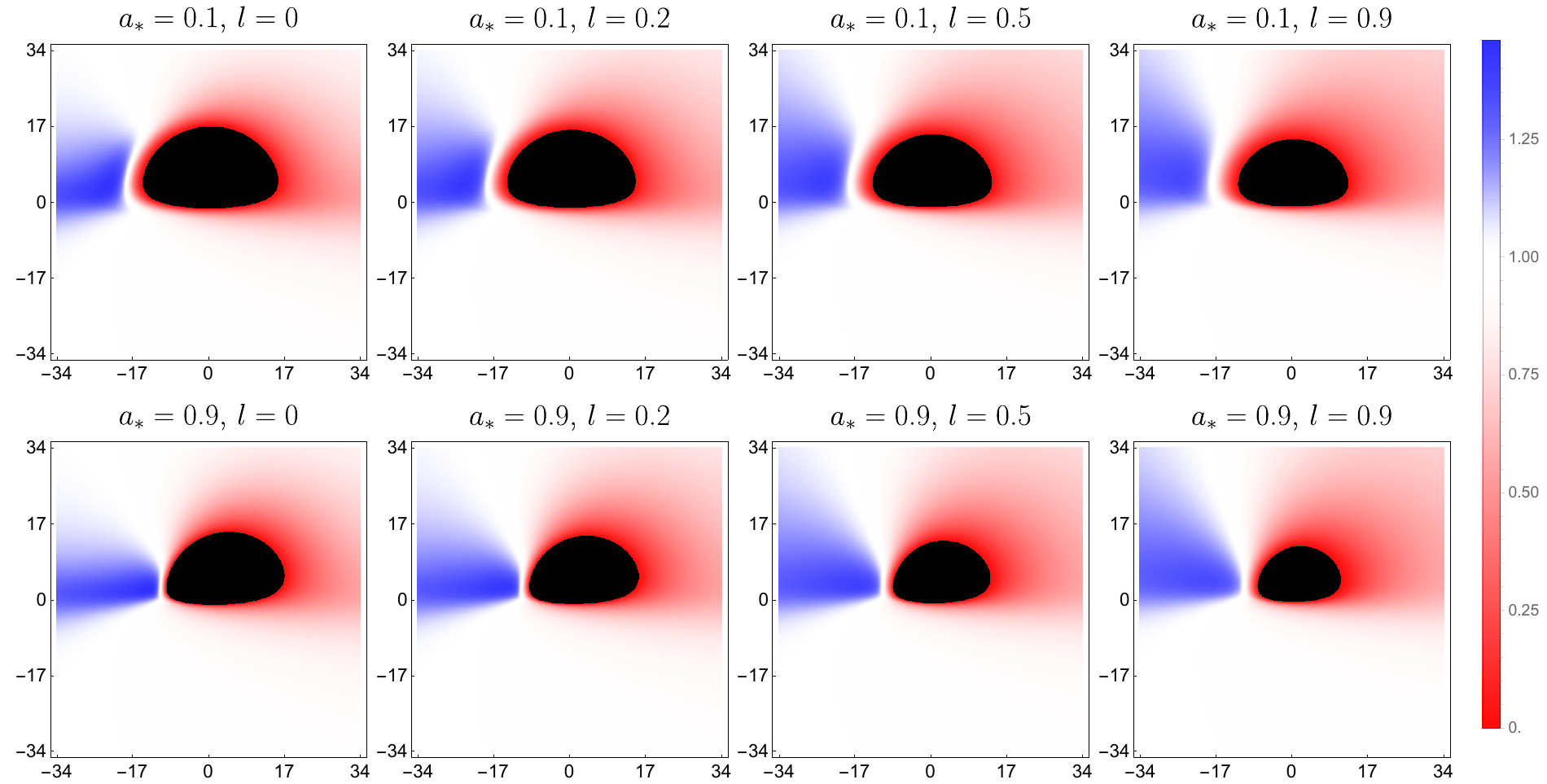}\\
	\includegraphics[width=6in]{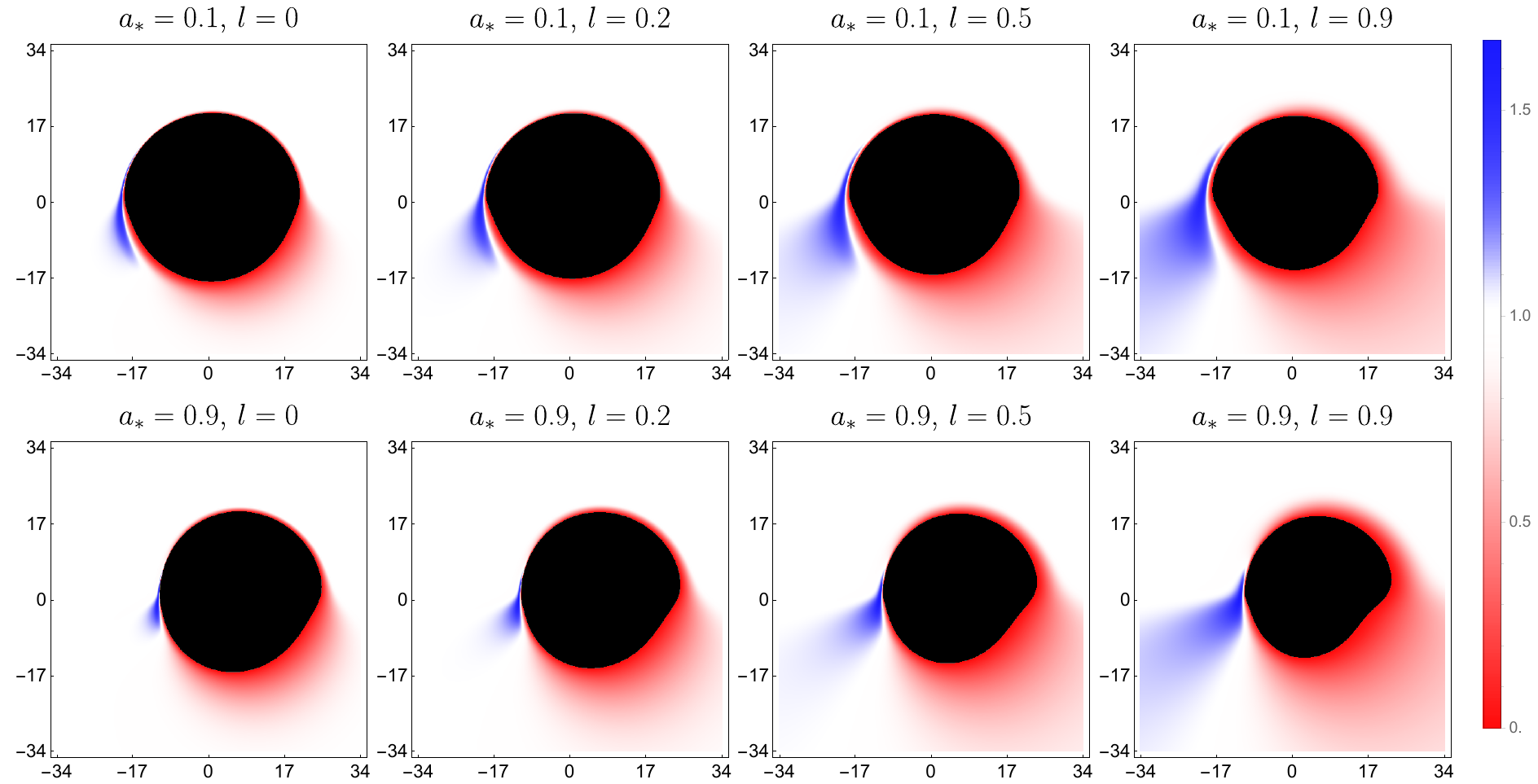}
	\caption{Redshift distribution of the direct (top two rows) and lensed (bottom two rows) images from the accretion disk at an observational inclination angle $\theta_o=80^\circ$. Red and blue colors denote redshift and blueshift, respectively. The boundaries of the black regions correspond to the inner shadow and the lensed images of the adopted absorbing inner boundary.}
	\label{fig80rs}
\end{figure*}

The strong gravitational redshift effects near the inner surface significantly alter the brightness of disk images. 
In Figs.~\ref{fig17rs} and \ref{fig80rs}, we show the redshift distributions of the direct emission and the lensed ring emission from the accretion disk at different observational inclination angles, where red and blue colors denote redshift and blueshift effects, respectively. 
The central black regions represent the inner shadow and the lensed images of the adopted absorbing inner boundary. 
For the convenience of comparing the influence of the Lorentz-violating parameter $l$ on the redshift, we also present in Appendix Fig.~\ref{figrsx} the distribution of the redshift factor along the $x$-axis for both the direct emission and the lensed ring emission.

From the figures, it can be seen that both the direct and lensed images exhibit varying degrees of redshift at an inclination angle of $\theta_o=17^\circ$, which is primarily contributed by gravitational redshift. 
The redshift effect becomes stronger closer to the inner surface. 
As the inclination angle increases to $\theta_o=80^\circ$, the relative motion of the accretion flow induces a pronounced Doppler blueshift in the left (equator-side) regions of both the direct emission and the lensed ring emission, while enhancing the redshift effect in the right-side regions. 
Furthermore, both the spin parameter $a_*$ and the Lorentz-violating parameter $l$ influence the redshift distribution. 
Compared with the Kerr black hole, an increase of the Lorentz-violating parameter $l$ enlarges both the region and the magnitude of the redshift distribution, and reduces the declining rate of the redshift factor near the inner surface. 
This effect is particularly prominent at high inclinations (specific details can be seen in Fig.~\ref{figrsx}), thereby making the overall image brighter. 
Similarly, increasing the spin parameter $a_*$ shifts the peak of the redshift-factor distribution on the $x$-axis inward, but its position is almost unaffected by the Lorentz-violating parameter---a conclusion consistent with the earlier intensity-cuts analysis.

\subsection{Quantitative analysis of image features}

\begin{figure*}[htbp]
	\centering
	\includegraphics[width=6in]{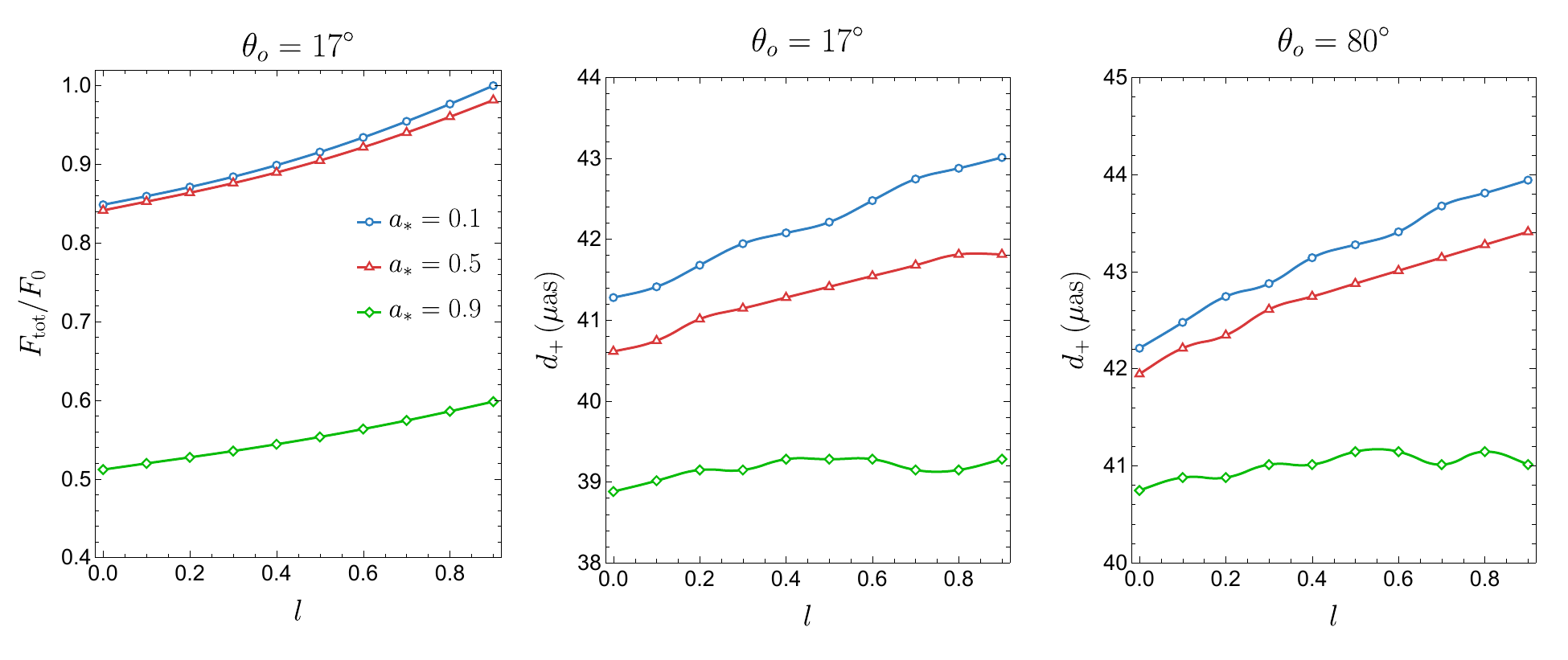}
	\caption{The variation of typical observational quantities in accretion disk images with the parameter $l$. The left panel shows the normalized total flux $F_\text{tot}/F_0$ as a function of $l$ at an observational inclination angle $\theta_o=17^\circ$, for different spin values, where $F_0$ is the total flux at $a_*=0.1$ and $l=0.9$. The middle ($\theta_o=17^\circ$) and right ($\theta_o=80^\circ$) panels depict the variation of the maximum interferometric diameter $d_+$ of the first photon ring with $l$.}
	\label{figfd}
\end{figure*}

The discussion in the previous subsection on the intensity distribution and redshift distribution in the images reveals that the Lorentz-violating parameter $l$ not only alters the overall morphology and brightness of the image, but also influences observable features such as the shape of the lensed ring and the size of the inner shadow. 
To gain a deeper understanding of these effects, we present a quantitative analysis of potential observables in the images in this subsection. 
For the geometrically thin, optically thin accretion disk model discussed in this work, the maximum number $N_\text{max}$ of times a photon crosses the equatorial plane can be used to classify the multiple-order images: $N_\text{max}=0,\,1,\,2,\,3,\,\cdots$ correspond, respectively, to the inner shadow, the direct emission, the lensed ring emission, and the photon ring \cite{Gralla:2019xty}, all of which represent potential observational features in the adopted imaging model.

We first focus on the direct emission ($N_\text{max}=1$), which contributes most to the total flux of the image \cite{Gralla:2019xty}. 
The left panel of Fig.~\ref{figfd} shows the behavior of the total flux $F_\text{tot}=r_o^{-2}\int I(x,y)\dd x\dd y$ as a function of $l$ for $\theta_o=17^\circ$. 
To facilitate comparison, the total flux for different parameter sets has been normalized to $F_\text{tot}$ obtained for $a_*=0.1$ and $l=0.9$. 
The figure reveals that the gravitational redshift induced by high spin significantly reduces the total flux, whereas the parameter $l$ increases the total flux of the image. 
As can be seen from Fig.~\ref{figrsx}, this is mainly because $l$ reduces the decreasing rate of the redshift factor $g$ in the direct emission near the inner surface.

We next consider the lensed ring ($N_\text{max}=2$), also called the first photon ring, which appears in the image as a bright ring structure with non-negligible width. 
Its geometry and brightness distribution encode key information about the strong gravitational field near the central object \cite{Cardenas-Avendano:2023dzo}. 
Moreover, future projects such as the Black Hole Explorer (BHEX) are expected to resolve the first photon ring directly in black hole images \cite{Johnson:2024ttr}. 
It is therefore valuable to assess the potential of using the geometric properties of the first photon ring to probe Lorentz-violating effects.

\begin{figure*}[htbp]
	\centering
	\includegraphics[width=6in]{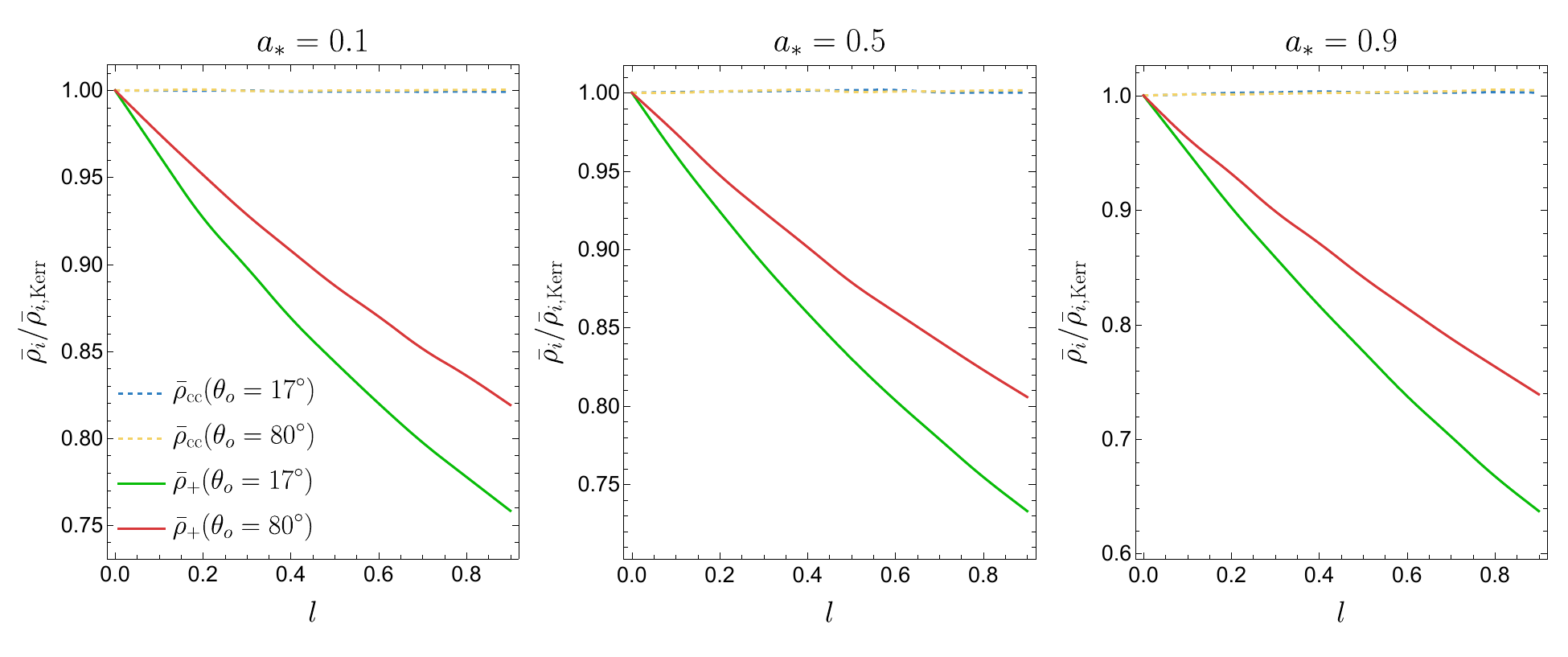}
	\caption{The variation of the average radius of the inner shadow $\rho_+$ and the average radius of the critical curve $\rho_\text{cc}$ with $l$ for different spin values. The dashed and solid lines represent the results for the critical curve and the boundary of the inner shadow, respectively.}
	\label{figrbar}
\end{figure*}

For the first photon ring, the spacing $d_\phi$ between intensity peaks at different azimuthal angles $\phi$ is a geometric observable. 
Denoting the width of the ring by $w$, when the baseline length $u$ satisfies $1/d_\phi\ll u\ll 1/w$, the Fourier component of the brightness distribution of the first photon ring in the sky, i.e., the visibility amplitude, can be approximated as \cite{Cardenas-Avendano:2023dzo}
\begin{equation}
	|V(u,\phi)|\approx\sqrt{(A^{L}_{\phi})^2+(A^{R}_{\phi})^{2}+2A^{L}_{\phi}A^{R}_{\phi}\sin(2\pi d_{\phi}u)},
\end{equation}
where $A_\phi^{L/R}(u)=(e_\text{up}(u)\pm e_\text{low}(u))/2$ with $e_\text{up}(u)$ and $e_\text{low}(u)$ denote the upper and lower envelopes of visibility amplitude. 
Hence \(d_\phi\) is also referred to as the interferometric diameter of the first photon ring, which can be extracted directly from the visibility amplitude. 
Existing studies have shown that \(d_\phi\) approximately follows a Gaussian distribution, and its maximum value \(d_+\) can be used to infer information about the background spacetime \cite{Farah:2025kpb, Wan:2025gbm}.

In Fig.~\ref{figfd}, we present the variation of the maximum interferometric diameter $d_+$ of the first photon ring with the parameter $l$ for different observational inclination angles. 
The figure shows that $d_+$ is indeed modified by $l$, but the magnitude of the change gradually decreases as the spin parameter $a$ increases. 
Taking $\theta_o=17^\circ$ as an example, when $a_*=0.1$, increasing $l$ from 0 to 0.9 alters $d_+$ by $1.73\,\mu$as, whereas for $a_*=0.9$ the change in $d_+$ reduces to only $0.4\,\mu$as. 
Consequently, at high spin the variation in $d_+$ is too small to be resolved directly by BHEX, and it could serve as a potential observable for testing Lorentz-violating effects only when the spin parameter is not too large. 
Moreover, this conclusion holds for different inclination angles, although the value of $d_+$ is slightly larger at high inclination angles.

Thus, for rotating bumblebee spacetimes, the maximum interferometric diameter $d_+$ of the first photon ring provides only very limited constraining power on the Lorentz-violating parameter $l$. 
It is therefore necessary to identify other observables in the images that are more sensitive to $l$. 
Previous qualitative discussions have shown that the size and shape of the inner shadow are modified by the parameter $l$, warranting a quantitative analysis of their dependence on $l$.

For closed convex curves such as the inner shadow contour and the critical curve, the geometric center is defined as $x_\text{c}=(x_\text{max}+x_\text{min})/2$ and $y_\text{c}=(y_\text{max}+y_\text{min})/2$, where the subscripts ``max'' and ``min'' denote the maximum and minimum values of the curve along the horizontal and vertical directions, respectively. 
Introducing a polar coordinate $(\rho,\alpha)$ with origin at the geometric center, the average radius of the curve can be defined as
\begin{equation}
	\bar{\rho}_i=\frac{1}{2\pi}\int_0^{2\pi}\rho_i(\alpha)\dd\alpha\,.
\end{equation}
In Fig.~\ref{figrbar} we display the variations of the average radius of the inner shadow (solid lines) and the critical curve (dashed lines) with the parameter $l$ for different spin values. 
To highlight the influence of the Lorentz‑violating parameter $l$ in rotating bumblebee spacetimes, the vertical axis has been normalized to the corresponding average radius of a Kerr black hole.

From the figure, it can be seen that the average radius of the critical curve remains almost unchanged with $l$ across different spins and observational inclination angles, consistent with the conclusion drawn earlier from the intensity profile analysis. 
In contrast, the average radius $\bar{\rho}_{+}$ of the inner shadow gradually decreases as $l$ increases, and this variation is significant for all spin values and inclination angles considered. 
Notably, when the inclination angle is small or the black‑hole spin is large---conditions that align with the inferred parameters for M87* from EHT data \cite{CraigWalker:2018vam}---the average radius of the inner shadow becomes even more sensitive to $l$. 
For instance, for $a_*=0.9$ and $\theta_o=17^\circ$, increasing $l$ from 0 to 0.9 reduces the average inner shadow radius from $9.54\,\mu$as to $6.08\,\mu$as, corresponding to a deviation of 36.3\% from the Kerr case. 
Therefore, future measurements of the central dark-region morphology could constrain Lorentz-violating parameter $l$, provided that the exterior spacetime model and the adopted absorbing-boundary prescription are applicable to M87*.

\section{CONCLUSIONS}
\label{sec6}

In this work, we investigated the kinematic and optical imprints of Lorentz symmetry violation in the strong-field region of a rotating bumblebee spacetime arising from a scalar-gradient bumblebee field \cite{Poulis:2021nqh}. We found that the Lorentz-violating parameter $l$ leaves distinct signatures on both timelike motion and null geodesics. 
We emphasize that $r=r_+$ is curvature singular for generic $a l\neq0$ in the nonextremal regime and is not a regular event horizon. The orbital and precession results characterize the exterior geometry, whereas the inner-shadow results are conditional on treating this surface as a perfectly absorbing boundary.

On the dynamical side, the periastron precession frequency $\Omega_{\text{peri}}$ of bound circular timelike orbits on the equatorial plane increases monotonically with $l$. Since both the ISCO and the nodal precession frequency remain unchanged from their Kerr values, $\Omega_{\text{peri}}$ emerges as a sensitive kinematic probe of Lorentz violation. 
Furthermore, the analysis of test gyroscope spin precession indicates that increasing $l$ suppresses the LT precession outside the ergosurface while enhancing the geodetic precession in the static limit. 
On the optical side, our imaging results for a geometrically thin accretion disk show that the critical curve is largely insensitive to $l$, whereas the average radius of the inner shadow decreases noticeably as $l$ increases. Within the imaging setup adopted here, Lorentz violation also makes the lensed ring broader and brighter than in the Kerr case. 

These results suggest that Lorentz violation leaves complementary imprints on timelike motion and null geodesics, so that combining dynamical and optical probes may help reduce parameter degeneracies in future tests of strong-field gravity. 
In particular, joint analyses involving optical imaging, stellar-orbit astrometry, and possibly X-ray timing observables could provide useful constraints on $l$. 
Looking ahead, it will be important to go beyond the scalar-gradient realization considered here, to incorporate more realistic accretion flow physics, and to investigate gravitational-wave signatures of extreme-mass-ratio inspirals in the exterior geometry. 
Such developments would provide a broader and more complete picture of Lorentz-violating effects in rotating strong-gravity systems. 
 
\begin{acknowledgments}
	We are grateful to Yehui Hou and Minyong Guo for helpful discussions. The work is in part supported by NSFC Grant No.12205104 and the startup funding of South China University of Technology.
\end{acknowledgments}

\bibliographystyle{apsrev4-2}
\bibliography{ref}

\begin{thebibliography}{119}%
\makeatletter
\providecommand \@ifxundefined [1]{%
 \@ifx{#1\undefined}
}%
\providecommand \@ifnum [1]{%
 \ifnum #1\expandafter \@firstoftwo
 \else \expandafter \@secondoftwo
 \fi
}%
\providecommand \@ifx [1]{%
 \ifx #1\expandafter \@firstoftwo
 \else \expandafter \@secondoftwo
 \fi
}%
\providecommand \natexlab [1]{#1}%
\providecommand \enquote  [1]{``#1''}%
\providecommand \bibnamefont  [1]{#1}%
\providecommand \bibfnamefont [1]{#1}%
\providecommand \citenamefont [1]{#1}%
\providecommand \href@noop [0]{\@secondoftwo}%
\providecommand \href [0]{\begingroup \@sanitize@url \@href}%
\providecommand \@href[1]{\@@startlink{#1}\@@href}%
\providecommand \@@href[1]{\endgroup#1\@@endlink}%
\providecommand \@sanitize@url [0]{\catcode `\\12\catcode `\$12\catcode
  `\&12\catcode `\#12\catcode `\^12\catcode `\_12\catcode `\%12\relax}%
\providecommand \@@startlink[1]{}%
\providecommand \@@endlink[0]{}%
\providecommand \url  [0]{\begingroup\@sanitize@url \@url }%
\providecommand \@url [1]{\endgroup\@href {#1}{\urlprefix }}%
\providecommand \urlprefix  [0]{URL }%
\providecommand \Eprint [0]{\href }%
\providecommand \doibase [0]{https://doi.org/}%
\providecommand \selectlanguage [0]{\@gobble}%
\providecommand \bibinfo  [0]{\@secondoftwo}%
\providecommand \bibfield  [0]{\@secondoftwo}%
\providecommand \translation [1]{[#1]}%
\providecommand \BibitemOpen [0]{}%
\providecommand \bibitemStop [0]{}%
\providecommand \bibitemNoStop [0]{.\EOS\space}%
\providecommand \EOS [0]{\spacefactor3000\relax}%
\providecommand \BibitemShut  [1]{\csname bibitem#1\endcsname}%
\let\auto@bib@innerbib\@empty
\bibitem [{\citenamefont {Will}(2001)}]{Will:2001mx}%
  \BibitemOpen
  \bibfield  {author} {\bibinfo {author} {\bibfnamefont {C.~M.}\ \bibnamefont
  {Will}},\ }\href {https://doi.org/10.12942/lrr-2001-4} {\bibfield  {journal}
  {\bibinfo  {journal} {Living Rev. Rel.}\ }\textbf {\bibinfo {volume} {4}},\
  \bibinfo {pages} {4} (\bibinfo {year} {2001})},\ \Eprint
  {https://arxiv.org/abs/gr-qc/0103036} {arXiv:gr-qc/0103036} \BibitemShut
  {NoStop}%
\bibitem [{\citenamefont {Turyshev}(2008)}]{Turyshev:2008dr}%
  \BibitemOpen
  \bibfield  {author} {\bibinfo {author} {\bibfnamefont {S.~G.}\ \bibnamefont
  {Turyshev}},\ }\href {https://doi.org/10.1146/annurev.nucl.58.020807.111839}
  {\bibfield  {journal} {\bibinfo  {journal} {Ann. Rev. Nucl. Part. Sci.}\
  }\textbf {\bibinfo {volume} {58}},\ \bibinfo {pages} {207} (\bibinfo {year}
  {2008})},\ \Eprint {https://arxiv.org/abs/0806.1731} {arXiv:0806.1731
  [gr-qc]} \BibitemShut {NoStop}%
\bibitem [{\citenamefont {Abbott}\ \emph {et~al.}(2016)\citenamefont {Abbott}
  \emph {et~al.}}]{LIGOScientific:2016aoc}%
  \BibitemOpen
  \bibfield  {author} {\bibinfo {author} {\bibfnamefont {B.~P.}\ \bibnamefont
  {Abbott}} \emph {et~al.} (\bibinfo {collaboration} {LIGO Scientific,
  Virgo}),\ }\href {https://doi.org/10.1103/PhysRevLett.116.061102} {\bibfield
  {journal} {\bibinfo  {journal} {Phys. Rev. Lett.}\ }\textbf {\bibinfo
  {volume} {116}},\ \bibinfo {pages} {061102} (\bibinfo {year} {2016})},\
  \Eprint {https://arxiv.org/abs/1602.03837} {arXiv:1602.03837 [gr-qc]}
  \BibitemShut {NoStop}%
\bibitem [{\citenamefont {Abbott}\ \emph {et~al.}(2019)\citenamefont {Abbott}
  \emph {et~al.}}]{LIGOScientific:2019fpa}%
  \BibitemOpen
  \bibfield  {author} {\bibinfo {author} {\bibfnamefont {B.~P.}\ \bibnamefont
  {Abbott}} \emph {et~al.} (\bibinfo {collaboration} {LIGO Scientific,
  Virgo}),\ }\href {https://doi.org/10.1103/PhysRevD.100.104036} {\bibfield
  {journal} {\bibinfo  {journal} {Phys. Rev. D}\ }\textbf {\bibinfo {volume}
  {100}},\ \bibinfo {pages} {104036} (\bibinfo {year} {2019})},\ \Eprint
  {https://arxiv.org/abs/1903.04467} {arXiv:1903.04467 [gr-qc]} \BibitemShut
  {NoStop}%
\bibitem [{\citenamefont {Crivellin}\ and\ \citenamefont
  {Mellado}(2024)}]{Crivellin:2023zui}%
  \BibitemOpen
  \bibfield  {author} {\bibinfo {author} {\bibfnamefont {A.}~\bibnamefont
  {Crivellin}}\ and\ \bibinfo {author} {\bibfnamefont {B.}~\bibnamefont
  {Mellado}},\ }\href {https://doi.org/10.1038/s42254-024-00703-6} {\bibfield
  {journal} {\bibinfo  {journal} {Nature Rev. Phys.}\ }\textbf {\bibinfo
  {volume} {6}},\ \bibinfo {pages} {294} (\bibinfo {year} {2024})},\ \Eprint
  {https://arxiv.org/abs/2309.03870} {arXiv:2309.03870 [hep-ph]} \BibitemShut
  {NoStop}%
\bibitem [{\citenamefont {Carlip}\ \emph {et~al.}(2015)\citenamefont {Carlip},
  \citenamefont {Chiou}, \citenamefont {Ni},\ and\ \citenamefont
  {Woodard}}]{Carlip:2015asa}%
  \BibitemOpen
  \bibfield  {author} {\bibinfo {author} {\bibfnamefont {S.}~\bibnamefont
  {Carlip}}, \bibinfo {author} {\bibfnamefont {D.-W.}\ \bibnamefont {Chiou}},
  \bibinfo {author} {\bibfnamefont {W.-T.}\ \bibnamefont {Ni}},\ and\ \bibinfo
  {author} {\bibfnamefont {R.}~\bibnamefont {Woodard}},\ }\href
  {https://doi.org/10.1142/S0218271815300281} {\bibfield  {journal} {\bibinfo
  {journal} {Int. J. Mod. Phys. D}\ }\textbf {\bibinfo {volume} {24}},\
  \bibinfo {pages} {1530028} (\bibinfo {year} {2015})},\ \Eprint
  {https://arxiv.org/abs/1507.08194} {arXiv:1507.08194 [gr-qc]} \BibitemShut
  {NoStop}%
\bibitem [{\citenamefont {Collins}\ \emph {et~al.}(2004)\citenamefont
  {Collins}, \citenamefont {Perez}, \citenamefont {Sudarsky}, \citenamefont
  {Urrutia},\ and\ \citenamefont {Vucetich}}]{Collins:2004bp}%
  \BibitemOpen
  \bibfield  {author} {\bibinfo {author} {\bibfnamefont {J.}~\bibnamefont
  {Collins}}, \bibinfo {author} {\bibfnamefont {A.}~\bibnamefont {Perez}},
  \bibinfo {author} {\bibfnamefont {D.}~\bibnamefont {Sudarsky}}, \bibinfo
  {author} {\bibfnamefont {L.}~\bibnamefont {Urrutia}},\ and\ \bibinfo {author}
  {\bibfnamefont {H.}~\bibnamefont {Vucetich}},\ }\href
  {https://doi.org/10.1103/PhysRevLett.93.191301} {\bibfield  {journal}
  {\bibinfo  {journal} {Phys. Rev. Lett.}\ }\textbf {\bibinfo {volume} {93}},\
  \bibinfo {pages} {191301} (\bibinfo {year} {2004})},\ \Eprint
  {https://arxiv.org/abs/gr-qc/0403053} {arXiv:gr-qc/0403053} \BibitemShut
  {NoStop}%
\bibitem [{\citenamefont {Kostelecky}\ and\ \citenamefont
  {Russell}(2011)}]{Kostelecky:2008ts}%
  \BibitemOpen
  \bibfield  {author} {\bibinfo {author} {\bibfnamefont {V.~A.}\ \bibnamefont
  {Kostelecky}}\ and\ \bibinfo {author} {\bibfnamefont {N.}~\bibnamefont
  {Russell}},\ }\href {https://doi.org/10.1103/RevModPhys.83.11} {\bibfield
  {journal} {\bibinfo  {journal} {Rev. Mod. Phys.}\ }\textbf {\bibinfo {volume}
  {83}},\ \bibinfo {pages} {11} (\bibinfo {year} {2011})},\ \Eprint
  {https://arxiv.org/abs/0801.0287} {arXiv:0801.0287 [hep-ph]} \BibitemShut
  {NoStop}%
\bibitem [{\citenamefont {Acciari}\ \emph {et~al.}(2020)\citenamefont {Acciari}
  \emph {et~al.}}]{MAGIC:2020egb}%
  \BibitemOpen
  \bibfield  {author} {\bibinfo {author} {\bibfnamefont {V.~A.}\ \bibnamefont
  {Acciari}} \emph {et~al.} (\bibinfo {collaboration} {MAGIC, Armenian
  Consortium: ICRANet-Armenia at NAS RA, A. Alikhanyan National Laboratory,
  Finnish MAGIC Consortium: Finnish Centre of Astronomy with ESO}),\ }\href
  {https://doi.org/10.1103/PhysRevLett.125.021301} {\bibfield  {journal}
  {\bibinfo  {journal} {Phys. Rev. Lett.}\ }\textbf {\bibinfo {volume} {125}},\
  \bibinfo {pages} {021301} (\bibinfo {year} {2020})},\ \Eprint
  {https://arxiv.org/abs/2001.09728} {arXiv:2001.09728 [astro-ph.HE]}
  \BibitemShut {NoStop}%
\bibitem [{\citenamefont {Kostelecky}\ and\ \citenamefont
  {Potting}(1995)}]{Kostelecky:1994rn}%
  \BibitemOpen
  \bibfield  {author} {\bibinfo {author} {\bibfnamefont {V.~A.}\ \bibnamefont
  {Kostelecky}}\ and\ \bibinfo {author} {\bibfnamefont {R.}~\bibnamefont
  {Potting}},\ }\href {https://doi.org/10.1103/PhysRevD.51.3923} {\bibfield
  {journal} {\bibinfo  {journal} {Phys. Rev. D}\ }\textbf {\bibinfo {volume}
  {51}},\ \bibinfo {pages} {3923} (\bibinfo {year} {1995})},\ \Eprint
  {https://arxiv.org/abs/hep-ph/9501341} {arXiv:hep-ph/9501341} \BibitemShut
  {NoStop}%
\bibitem [{\citenamefont {Colladay}\ and\ \citenamefont
  {Kostelecky}(1997)}]{Colladay:1996iz}%
  \BibitemOpen
  \bibfield  {author} {\bibinfo {author} {\bibfnamefont {D.}~\bibnamefont
  {Colladay}}\ and\ \bibinfo {author} {\bibfnamefont {V.~A.}\ \bibnamefont
  {Kostelecky}},\ }\href {https://doi.org/10.1103/PhysRevD.55.6760} {\bibfield
  {journal} {\bibinfo  {journal} {Phys. Rev. D}\ }\textbf {\bibinfo {volume}
  {55}},\ \bibinfo {pages} {6760} (\bibinfo {year} {1997})},\ \Eprint
  {https://arxiv.org/abs/hep-ph/9703464} {arXiv:hep-ph/9703464} \BibitemShut
  {NoStop}%
\bibitem [{\citenamefont {Colladay}\ and\ \citenamefont
  {Kostelecky}(1998)}]{Colladay:1998fq}%
  \BibitemOpen
  \bibfield  {author} {\bibinfo {author} {\bibfnamefont {D.}~\bibnamefont
  {Colladay}}\ and\ \bibinfo {author} {\bibfnamefont {V.~A.}\ \bibnamefont
  {Kostelecky}},\ }\href {https://doi.org/10.1103/PhysRevD.58.116002}
  {\bibfield  {journal} {\bibinfo  {journal} {Phys. Rev. D}\ }\textbf {\bibinfo
  {volume} {58}},\ \bibinfo {pages} {116002} (\bibinfo {year} {1998})},\
  \Eprint {https://arxiv.org/abs/hep-ph/9809521} {arXiv:hep-ph/9809521}
  \BibitemShut {NoStop}%
\bibitem [{\citenamefont {Kostelecky}(2004)}]{Kostelecky:2003fs}%
  \BibitemOpen
  \bibfield  {author} {\bibinfo {author} {\bibfnamefont {V.~A.}\ \bibnamefont
  {Kostelecky}},\ }\href {https://doi.org/10.1103/PhysRevD.69.105009}
  {\bibfield  {journal} {\bibinfo  {journal} {Phys. Rev. D}\ }\textbf {\bibinfo
  {volume} {69}},\ \bibinfo {pages} {105009} (\bibinfo {year} {2004})},\
  \Eprint {https://arxiv.org/abs/hep-th/0312310} {arXiv:hep-th/0312310}
  \BibitemShut {NoStop}%
\bibitem [{\citenamefont {Tasson}(2016)}]{Tasson:2016xib}%
  \BibitemOpen
  \bibfield  {author} {\bibinfo {author} {\bibfnamefont {J.~D.}\ \bibnamefont
  {Tasson}},\ }\href@noop {} {\bibfield  {journal} {\bibinfo  {journal}
  {Symmetry}\ }\textbf {\bibinfo {volume} {8}},\ \bibinfo {pages} {111}
  (\bibinfo {year} {2016})},\ \Eprint {https://arxiv.org/abs/1610.05357}
  {arXiv:1610.05357 [gr-qc]} \BibitemShut {NoStop}%
\bibitem [{\citenamefont {Bluhm}\ \emph {et~al.}(2019)\citenamefont {Bluhm},
  \citenamefont {Bossi},\ and\ \citenamefont {Wen}}]{Bluhm:2019ato}%
  \BibitemOpen
  \bibfield  {author} {\bibinfo {author} {\bibfnamefont {R.}~\bibnamefont
  {Bluhm}}, \bibinfo {author} {\bibfnamefont {H.}~\bibnamefont {Bossi}},\ and\
  \bibinfo {author} {\bibfnamefont {Y.}~\bibnamefont {Wen}},\ }\href
  {https://doi.org/10.1103/PhysRevD.100.084022} {\bibfield  {journal} {\bibinfo
   {journal} {Phys. Rev. D}\ }\textbf {\bibinfo {volume} {100}},\ \bibinfo
  {pages} {084022} (\bibinfo {year} {2019})},\ \Eprint
  {https://arxiv.org/abs/1907.13209} {arXiv:1907.13209 [gr-qc]} \BibitemShut
  {NoStop}%
\bibitem [{\citenamefont {Maluf}\ \emph {et~al.}(2014)\citenamefont {Maluf},
  \citenamefont {Almeida}, \citenamefont {Casana},\ and\ \citenamefont
  {Ferreira}}]{Maluf:2014dpa}%
  \BibitemOpen
  \bibfield  {author} {\bibinfo {author} {\bibfnamefont {R.~V.}\ \bibnamefont
  {Maluf}}, \bibinfo {author} {\bibfnamefont {C.~A.~S.}\ \bibnamefont
  {Almeida}}, \bibinfo {author} {\bibfnamefont {R.}~\bibnamefont {Casana}},\
  and\ \bibinfo {author} {\bibfnamefont {M.~M.}\ \bibnamefont {Ferreira},
  \bibfnamefont {Jr.}},\ }\href {https://doi.org/10.1103/PhysRevD.90.025007}
  {\bibfield  {journal} {\bibinfo  {journal} {Phys. Rev. D}\ }\textbf {\bibinfo
  {volume} {90}},\ \bibinfo {pages} {025007} (\bibinfo {year} {2014})},\
  \Eprint {https://arxiv.org/abs/1402.3554} {arXiv:1402.3554 [hep-th]}
  \BibitemShut {NoStop}%
\bibitem [{\citenamefont {Casana}\ \emph {et~al.}(2018)\citenamefont {Casana},
  \citenamefont {Cavalcante}, \citenamefont {Poulis},\ and\ \citenamefont
  {Santos}}]{Casana:2017jkc}%
  \BibitemOpen
  \bibfield  {author} {\bibinfo {author} {\bibfnamefont {R.}~\bibnamefont
  {Casana}}, \bibinfo {author} {\bibfnamefont {A.}~\bibnamefont {Cavalcante}},
  \bibinfo {author} {\bibfnamefont {F.~P.}\ \bibnamefont {Poulis}},\ and\
  \bibinfo {author} {\bibfnamefont {E.~B.}\ \bibnamefont {Santos}},\ }\href
  {https://doi.org/10.1103/PhysRevD.97.104001} {\bibfield  {journal} {\bibinfo
  {journal} {Phys. Rev. D}\ }\textbf {\bibinfo {volume} {97}},\ \bibinfo
  {pages} {104001} (\bibinfo {year} {2018})},\ \Eprint
  {https://arxiv.org/abs/1711.02273} {arXiv:1711.02273 [gr-qc]} \BibitemShut
  {NoStop}%
\bibitem [{\citenamefont {Bluhm}\ and\ \citenamefont
  {Kostelecky}(2005)}]{Bluhm:2004ep}%
  \BibitemOpen
  \bibfield  {author} {\bibinfo {author} {\bibfnamefont {R.}~\bibnamefont
  {Bluhm}}\ and\ \bibinfo {author} {\bibfnamefont {V.~A.}\ \bibnamefont
  {Kostelecky}},\ }\href {https://doi.org/10.1103/PhysRevD.71.065008}
  {\bibfield  {journal} {\bibinfo  {journal} {Phys. Rev. D}\ }\textbf {\bibinfo
  {volume} {71}},\ \bibinfo {pages} {065008} (\bibinfo {year} {2005})},\
  \Eprint {https://arxiv.org/abs/hep-th/0412320} {arXiv:hep-th/0412320}
  \BibitemShut {NoStop}%
\bibitem [{\citenamefont {Bluhm}(2007)}]{Bluhm:2007xzd}%
  \BibitemOpen
  \bibfield  {author} {\bibinfo {author} {\bibfnamefont {R.}~\bibnamefont
  {Bluhm}},\ }\href {https://doi.org/10.22323/1.043.0009} {\bibfield  {journal}
  {\bibinfo  {journal} {PoS}\ }\textbf {\bibinfo {volume} {QG-PH}},\ \bibinfo
  {pages} {009} (\bibinfo {year} {2007})},\ \Eprint
  {https://arxiv.org/abs/0801.0141} {arXiv:0801.0141 [gr-qc]} \BibitemShut
  {NoStop}%
\bibitem [{\citenamefont {Ding}\ \emph {et~al.}(2020)\citenamefont {Ding},
  \citenamefont {Liu}, \citenamefont {Casana},\ and\ \citenamefont
  {Cavalcante}}]{Ding:2019mal}%
  \BibitemOpen
  \bibfield  {author} {\bibinfo {author} {\bibfnamefont {C.}~\bibnamefont
  {Ding}}, \bibinfo {author} {\bibfnamefont {C.}~\bibnamefont {Liu}}, \bibinfo
  {author} {\bibfnamefont {R.}~\bibnamefont {Casana}},\ and\ \bibinfo {author}
  {\bibfnamefont {A.}~\bibnamefont {Cavalcante}},\ }\href
  {https://doi.org/10.1140/epjc/s10052-020-7743-y} {\bibfield  {journal}
  {\bibinfo  {journal} {Eur. Phys. J. C}\ }\textbf {\bibinfo {volume} {80}},\
  \bibinfo {pages} {178} (\bibinfo {year} {2020})},\ \Eprint
  {https://arxiv.org/abs/1910.02674} {arXiv:1910.02674 [gr-qc]} \BibitemShut
  {NoStop}%
\bibitem [{\citenamefont {Maluf}\ and\ \citenamefont
  {Neves}(2021)}]{Maluf:2020kgf}%
  \BibitemOpen
  \bibfield  {author} {\bibinfo {author} {\bibfnamefont {R.~V.}\ \bibnamefont
  {Maluf}}\ and\ \bibinfo {author} {\bibfnamefont {J.~C.~S.}\ \bibnamefont
  {Neves}},\ }\href {https://doi.org/10.1103/PhysRevD.103.044002} {\bibfield
  {journal} {\bibinfo  {journal} {Phys. Rev. D}\ }\textbf {\bibinfo {volume}
  {103}},\ \bibinfo {pages} {044002} (\bibinfo {year} {2021})},\ \Eprint
  {https://arxiv.org/abs/2011.12841} {arXiv:2011.12841 [gr-qc]} \BibitemShut
  {NoStop}%
\bibitem [{\citenamefont {Xu}\ \emph {et~al.}(2023)\citenamefont {Xu},
  \citenamefont {Liang},\ and\ \citenamefont {Shao}}]{Xu:2022frb}%
  \BibitemOpen
  \bibfield  {author} {\bibinfo {author} {\bibfnamefont {R.}~\bibnamefont
  {Xu}}, \bibinfo {author} {\bibfnamefont {D.}~\bibnamefont {Liang}},\ and\
  \bibinfo {author} {\bibfnamefont {L.}~\bibnamefont {Shao}},\ }\href
  {https://doi.org/10.1103/PhysRevD.107.024011} {\bibfield  {journal} {\bibinfo
   {journal} {Phys. Rev. D}\ }\textbf {\bibinfo {volume} {107}},\ \bibinfo
  {pages} {024011} (\bibinfo {year} {2023})},\ \Eprint
  {https://arxiv.org/abs/2209.02209} {arXiv:2209.02209 [gr-qc]} \BibitemShut
  {NoStop}%
\bibitem [{\citenamefont {Filho}\ \emph {et~al.}(2023)\citenamefont {Filho},
  \citenamefont {Nascimento}, \citenamefont {Petrov},\ and\ \citenamefont
  {Porf{\'\i}rio}}]{Filho:2022yrk}%
  \BibitemOpen
  \bibfield  {author} {\bibinfo {author} {\bibfnamefont {A.~A.~A.}\
  \bibnamefont {Filho}}, \bibinfo {author} {\bibfnamefont {J.~R.}\ \bibnamefont
  {Nascimento}}, \bibinfo {author} {\bibfnamefont {A.~Y.}\ \bibnamefont
  {Petrov}},\ and\ \bibinfo {author} {\bibfnamefont {P.~J.}\ \bibnamefont
  {Porf{\'\i}rio}},\ }\href {https://doi.org/10.1103/PhysRevD.108.085010}
  {\bibfield  {journal} {\bibinfo  {journal} {Phys. Rev. D}\ }\textbf {\bibinfo
  {volume} {108}},\ \bibinfo {pages} {085010} (\bibinfo {year} {2023})},\
  \Eprint {https://arxiv.org/abs/2211.11821} {arXiv:2211.11821 [gr-qc]}
  \BibitemShut {NoStop}%
\bibitem [{\citenamefont {Liu}\ \emph {et~al.}(2025{\natexlab{a}})\citenamefont
  {Liu}, \citenamefont {Guo}, \citenamefont {Wei},\ and\ \citenamefont
  {Liu}}]{Liu:2024axg}%
  \BibitemOpen
  \bibfield  {author} {\bibinfo {author} {\bibfnamefont {J.-Z.}\ \bibnamefont
  {Liu}}, \bibinfo {author} {\bibfnamefont {W.-D.}\ \bibnamefont {Guo}},
  \bibinfo {author} {\bibfnamefont {S.-W.}\ \bibnamefont {Wei}},\ and\ \bibinfo
  {author} {\bibfnamefont {Y.-X.}\ \bibnamefont {Liu}},\ }\href
  {https://doi.org/10.1140/epjc/s10052-025-13859-x} {\bibfield  {journal}
  {\bibinfo  {journal} {Eur. Phys. J. C}\ }\textbf {\bibinfo {volume} {85}},\
  \bibinfo {pages} {145} (\bibinfo {year} {2025}{\natexlab{a}})},\ \Eprint
  {https://arxiv.org/abs/2407.08396} {arXiv:2407.08396 [gr-qc]} \BibitemShut
  {NoStop}%
\bibitem [{\citenamefont {Ara{\'u}jo~Filho}\ \emph {et~al.}(2024)\citenamefont
  {Ara{\'u}jo~Filho}, \citenamefont {Nascimento}, \citenamefont {Petrov},\ and\
  \citenamefont {Porf{\'\i}rio}}]{AraujoFilho:2024ykw}%
  \BibitemOpen
  \bibfield  {author} {\bibinfo {author} {\bibfnamefont {A.~A.}\ \bibnamefont
  {Ara{\'u}jo~Filho}}, \bibinfo {author} {\bibfnamefont {J.~R.}\ \bibnamefont
  {Nascimento}}, \bibinfo {author} {\bibfnamefont {A.~Y.}\ \bibnamefont
  {Petrov}},\ and\ \bibinfo {author} {\bibfnamefont {P.~J.}\ \bibnamefont
  {Porf{\'\i}rio}},\ }\href {https://doi.org/10.1088/1475-7516/2024/07/004}
  {\bibfield  {journal} {\bibinfo  {journal} {JCAP}\ }\textbf {\bibinfo
  {volume} {07}},\ \bibinfo {pages} {004}},\ \Eprint
  {https://arxiv.org/abs/2402.13014} {arXiv:2402.13014 [gr-qc]} \BibitemShut
  {NoStop}%
\bibitem [{\citenamefont {Li}\ and\ \citenamefont {Zhu}(2026)}]{Li:2025rjv}%
  \BibitemOpen
  \bibfield  {author} {\bibinfo {author} {\bibfnamefont {H.}~\bibnamefont
  {Li}}\ and\ \bibinfo {author} {\bibfnamefont {J.}~\bibnamefont {Zhu}},\
  }\href {https://doi.org/10.1140/epjc/s10052-025-15229-z} {\bibfield
  {journal} {\bibinfo  {journal} {Eur. Phys. J. C}\ }\textbf {\bibinfo {volume}
  {86}},\ \bibinfo {pages} {2} (\bibinfo {year} {2026})},\ \Eprint
  {https://arxiv.org/abs/2506.17957} {arXiv:2506.17957 [gr-qc]} \BibitemShut
  {NoStop}%
\bibitem [{\citenamefont {Zhu}\ and\ \citenamefont {Li}(2025)}]{Zhu:2025fiy}%
  \BibitemOpen
  \bibfield  {author} {\bibinfo {author} {\bibfnamefont {J.}~\bibnamefont
  {Zhu}}\ and\ \bibinfo {author} {\bibfnamefont {H.}~\bibnamefont {Li}},\
  }\href@noop {} {\  (\bibinfo {year} {2025})},\ \Eprint
  {https://arxiv.org/abs/2511.03231} {arXiv:2511.03231 [gr-qc]} \BibitemShut
  {NoStop}%
\bibitem [{\citenamefont {Liu}\ \emph {et~al.}(2025{\natexlab{b}})\citenamefont
  {Liu}, \citenamefont {Wu}, \citenamefont {Wei},\ and\ \citenamefont
  {Liu}}]{Liu:2025oho}%
  \BibitemOpen
  \bibfield  {author} {\bibinfo {author} {\bibfnamefont {J.-Z.}\ \bibnamefont
  {Liu}}, \bibinfo {author} {\bibfnamefont {S.-P.}\ \bibnamefont {Wu}},
  \bibinfo {author} {\bibfnamefont {S.-W.}\ \bibnamefont {Wei}},\ and\ \bibinfo
  {author} {\bibfnamefont {Y.-X.}\ \bibnamefont {Liu}},\ }\href@noop {} {\
  (\bibinfo {year} {2025}{\natexlab{b}})},\ \Eprint
  {https://arxiv.org/abs/2510.16731} {arXiv:2510.16731 [gr-qc]} \BibitemShut
  {NoStop}%
\bibitem [{\citenamefont {Ovcharenko}(2026)}]{Ovcharenko:2026rvj}%
  \BibitemOpen
  \bibfield  {author} {\bibinfo {author} {\bibfnamefont {H.}~\bibnamefont
  {Ovcharenko}},\ }\href@noop {} {\  (\bibinfo {year} {2026})},\ \Eprint
  {https://arxiv.org/abs/2601.16037} {arXiv:2601.16037 [gr-qc]} \BibitemShut
  {NoStop}%
\bibitem [{\citenamefont {Xu}\ \emph {et~al.}(2026)\citenamefont {Xu},
  \citenamefont {Mai},\ and\ \citenamefont {Liang}}]{Xu:2026zgd}%
  \BibitemOpen
  \bibfield  {author} {\bibinfo {author} {\bibfnamefont {R.}~\bibnamefont
  {Xu}}, \bibinfo {author} {\bibfnamefont {Z.-F.}\ \bibnamefont {Mai}},\ and\
  \bibinfo {author} {\bibfnamefont {D.}~\bibnamefont {Liang}},\ }\href@noop {}
  {\  (\bibinfo {year} {2026})},\ \Eprint {https://arxiv.org/abs/2601.18809}
  {arXiv:2601.18809 [gr-qc]} \BibitemShut {NoStop}%
\bibitem [{\citenamefont {Mai}\ \emph {et~al.}(2023)\citenamefont {Mai},
  \citenamefont {Xu}, \citenamefont {Liang},\ and\ \citenamefont
  {Shao}}]{Mai:2023ggs}%
  \BibitemOpen
  \bibfield  {author} {\bibinfo {author} {\bibfnamefont {Z.-F.}\ \bibnamefont
  {Mai}}, \bibinfo {author} {\bibfnamefont {R.}~\bibnamefont {Xu}}, \bibinfo
  {author} {\bibfnamefont {D.}~\bibnamefont {Liang}},\ and\ \bibinfo {author}
  {\bibfnamefont {L.}~\bibnamefont {Shao}},\ }\href
  {https://doi.org/10.1103/PhysRevD.108.024004} {\bibfield  {journal} {\bibinfo
   {journal} {Phys. Rev. D}\ }\textbf {\bibinfo {volume} {108}},\ \bibinfo
  {pages} {024004} (\bibinfo {year} {2023})},\ \Eprint
  {https://arxiv.org/abs/2304.08030} {arXiv:2304.08030 [gr-qc]} \BibitemShut
  {NoStop}%
\bibitem [{\citenamefont {Ara{\'u}jo~Filho}\ \emph {et~al.}(2025)\citenamefont
  {Ara{\'u}jo~Filho}, \citenamefont {Reis},\ and\ \citenamefont
  {{\"O}vg{\"u}n}}]{AraujoFilho:2024iox}%
  \BibitemOpen
  \bibfield  {author} {\bibinfo {author} {\bibfnamefont {A.~A.}\ \bibnamefont
  {Ara{\'u}jo~Filho}}, \bibinfo {author} {\bibfnamefont {J.~A. A.~S.}\
  \bibnamefont {Reis}},\ and\ \bibinfo {author} {\bibfnamefont
  {A.}~\bibnamefont {{\"O}vg{\"u}n}},\ }\href
  {https://doi.org/10.1140/epjc/s10052-025-13789-8} {\bibfield  {journal}
  {\bibinfo  {journal} {Eur. Phys. J. C}\ }\textbf {\bibinfo {volume} {85}},\
  \bibinfo {pages} {83} (\bibinfo {year} {2025})},\ \Eprint
  {https://arxiv.org/abs/2409.17351} {arXiv:2409.17351 [gr-qc]} \BibitemShut
  {NoStop}%
\bibitem [{\citenamefont {Liu}\ \emph {et~al.}(2025{\natexlab{c}})\citenamefont
  {Liu}, \citenamefont {Liu}, \citenamefont {Liu},\ and\ \citenamefont
  {Wang}}]{Liu:2025bpp}%
  \BibitemOpen
  \bibfield  {author} {\bibinfo {author} {\bibfnamefont {X.}~\bibnamefont
  {Liu}}, \bibinfo {author} {\bibfnamefont {W.}~\bibnamefont {Liu}}, \bibinfo
  {author} {\bibfnamefont {Z.}~\bibnamefont {Liu}},\ and\ \bibinfo {author}
  {\bibfnamefont {J.}~\bibnamefont {Wang}},\ }\href
  {https://doi.org/10.1007/JHEP08(2025)094} {\bibfield  {journal} {\bibinfo
  {journal} {JHEP}\ }\textbf {\bibinfo {volume} {08}},\ \bibinfo {pages}
  {094}},\ \Eprint {https://arxiv.org/abs/2503.06404} {arXiv:2503.06404
  [gr-qc]} \BibitemShut {NoStop}%
\bibitem [{\citenamefont {Ara{\'u}jo~Filho}(2025)}]{AraujoFilho:2025hkm}%
  \BibitemOpen
  \bibfield  {author} {\bibinfo {author} {\bibfnamefont {A.~A.}\ \bibnamefont
  {Ara{\'u}jo~Filho}},\ }\href {https://doi.org/10.1088/1475-7516/2025/06/026}
  {\bibfield  {journal} {\bibinfo  {journal} {JCAP}\ }\textbf {\bibinfo
  {volume} {06}},\ \bibinfo {pages} {026}},\ \bibinfo {note} {[Erratum: JCAP
  02, E01 (2026)]},\ \Eprint {https://arxiv.org/abs/2501.00927}
  {arXiv:2501.00927 [gr-qc]} \BibitemShut {NoStop}%
\bibitem [{\citenamefont {Jha}\ and\ \citenamefont
  {Rahaman}(2021)}]{2011.14916}%
  \BibitemOpen
  \bibfield  {author} {\bibinfo {author} {\bibfnamefont {S.~K.}\ \bibnamefont
  {Jha}}\ and\ \bibinfo {author} {\bibfnamefont {A.}~\bibnamefont {Rahaman}},\
  }\href {https://doi.org/10.1140/epjc/s10052-021-09132-6} {\bibfield
  {journal} {\bibinfo  {journal} {Eur. Phys. J. C}\ }\textbf {\bibinfo {volume}
  {81}},\ \bibinfo {pages} {345} (\bibinfo {year} {2021})},\ \Eprint
  {https://arxiv.org/abs/2011.14916} {arXiv:2011.14916 [gr-qc]} \BibitemShut
  {NoStop}%
\bibitem [{\citenamefont {Wang}\ and\ \citenamefont {Wei}(2022)}]{2106.14602}%
  \BibitemOpen
  \bibfield  {author} {\bibinfo {author} {\bibfnamefont {H.-M.}\ \bibnamefont
  {Wang}}\ and\ \bibinfo {author} {\bibfnamefont {S.-W.}\ \bibnamefont {Wei}},\
  }\href {https://doi.org/10.1140/epjp/s13360-022-02785-6} {\bibfield
  {journal} {\bibinfo  {journal} {Eur. Phys. J. Plus}\ }\textbf {\bibinfo
  {volume} {137}},\ \bibinfo {pages} {571} (\bibinfo {year} {2022})},\ \Eprint
  {https://arxiv.org/abs/2106.14602} {arXiv:2106.14602 [gr-qc]} \BibitemShut
  {NoStop}%
\bibitem [{\citenamefont {Izmailov}\ and\ \citenamefont
  {Nandi}(2022)}]{Izmailov:2022jon}%
  \BibitemOpen
  \bibfield  {author} {\bibinfo {author} {\bibfnamefont {R.~N.}\ \bibnamefont
  {Izmailov}}\ and\ \bibinfo {author} {\bibfnamefont {K.~K.}\ \bibnamefont
  {Nandi}},\ }\href {https://doi.org/10.1088/1361-6382/ac8fda} {\bibfield
  {journal} {\bibinfo  {journal} {Class. Quant. Grav.}\ }\textbf {\bibinfo
  {volume} {39}},\ \bibinfo {pages} {215006} (\bibinfo {year}
  {2022})}\BibitemShut {NoStop}%
\bibitem [{\citenamefont {Islam}\ \emph {et~al.}(2024)\citenamefont {Islam},
  \citenamefont {Ghosh},\ and\ \citenamefont {Maharaj}}]{Islam:2024sph}%
  \BibitemOpen
  \bibfield  {author} {\bibinfo {author} {\bibfnamefont {S.~U.}\ \bibnamefont
  {Islam}}, \bibinfo {author} {\bibfnamefont {S.~G.}\ \bibnamefont {Ghosh}},\
  and\ \bibinfo {author} {\bibfnamefont {S.~D.}\ \bibnamefont {Maharaj}},\
  }\href {https://doi.org/10.1088/1475-7516/2024/12/047} {\bibfield  {journal}
  {\bibinfo  {journal} {JCAP}\ }\textbf {\bibinfo {volume} {12}},\ \bibinfo
  {pages} {047}},\ \Eprint {https://arxiv.org/abs/2410.05395} {arXiv:2410.05395
  [gr-qc]} \BibitemShut {NoStop}%
\bibitem [{\citenamefont {Kumar}\ \emph {et~al.}(2025)\citenamefont {Kumar},
  \citenamefont {Islam},\ and\ \citenamefont {Ghosh}}]{Kumar:2025bim}%
  \BibitemOpen
  \bibfield  {author} {\bibinfo {author} {\bibfnamefont {A.}~\bibnamefont
  {Kumar}}, \bibinfo {author} {\bibfnamefont {S.~U.}\ \bibnamefont {Islam}},\
  and\ \bibinfo {author} {\bibfnamefont {S.~G.}\ \bibnamefont {Ghosh}},\
  }\href@noop {} {\  (\bibinfo {year} {2025})},\ \Eprint
  {https://arxiv.org/abs/2509.00127} {arXiv:2509.00127 [gr-qc]} \BibitemShut
  {NoStop}%
\bibitem [{\citenamefont {Ara{\'u}jo~Filho}\ \emph {et~al.}(2026)\citenamefont
  {Ara{\'u}jo~Filho}, \citenamefont {Heidari}, \citenamefont {Lobo},\ and\
  \citenamefont {Bezerra}}]{AraujoFilho:2025zaj}%
  \BibitemOpen
  \bibfield  {author} {\bibinfo {author} {\bibfnamefont {A.~A.}\ \bibnamefont
  {Ara{\'u}jo~Filho}}, \bibinfo {author} {\bibfnamefont {N.}~\bibnamefont
  {Heidari}}, \bibinfo {author} {\bibfnamefont {I.~P.}\ \bibnamefont {Lobo}},\
  and\ \bibinfo {author} {\bibfnamefont {V.~B.}\ \bibnamefont {Bezerra}},\
  }\href {https://doi.org/10.1016/j.aop.2026.170469} {\bibfield  {journal}
  {\bibinfo  {journal} {Annals Phys.}\ }\textbf {\bibinfo {volume} {489}},\
  \bibinfo {pages} {170469} (\bibinfo {year} {2026})},\ \Eprint
  {https://arxiv.org/abs/2511.12839} {arXiv:2511.12839 [gr-qc]} \BibitemShut
  {NoStop}%
\bibitem [{\citenamefont {Sui}\ \emph {et~al.}(2026)\citenamefont {Sui},
  \citenamefont {Meng},\ and\ \citenamefont {Wang}}]{Sui:2026pyf}%
  \BibitemOpen
  \bibfield  {author} {\bibinfo {author} {\bibfnamefont {T.-T.}\ \bibnamefont
  {Sui}}, \bibinfo {author} {\bibfnamefont {X.-C.}\ \bibnamefont {Meng}},\ and\
  \bibinfo {author} {\bibfnamefont {X.-Y.}\ \bibnamefont {Wang}},\ }\href@noop
  {} {\  (\bibinfo {year} {2026})},\ \Eprint {https://arxiv.org/abs/2602.20586}
  {arXiv:2602.20586 [gr-qc]} \BibitemShut {NoStop}%
\bibitem [{\citenamefont {Qi}\ \emph {et~al.}(2026)\citenamefont {Qi},
  \citenamefont {Sang},\ and\ \citenamefont {Kuang}}]{QiQi:2026zwp}%
  \BibitemOpen
  \bibfield  {author} {\bibinfo {author} {\bibfnamefont {Q.}~\bibnamefont
  {Qi}}, \bibinfo {author} {\bibfnamefont {Y.}~\bibnamefont {Sang}},\ and\
  \bibinfo {author} {\bibfnamefont {X.-M.}\ \bibnamefont {Kuang}},\ }\href@noop
  {} {\  (\bibinfo {year} {2026})},\ \Eprint {https://arxiv.org/abs/2601.06491}
  {arXiv:2601.06491 [gr-qc]} \BibitemShut {NoStop}%
\bibitem [{\citenamefont {Kanzi}\ and\ \citenamefont
  {Sakall{\i}}(2021)}]{2102.06303}%
  \BibitemOpen
  \bibfield  {author} {\bibinfo {author} {\bibfnamefont {S.}~\bibnamefont
  {Kanzi}}\ and\ \bibinfo {author} {\bibfnamefont {{\.I}.}~\bibnamefont
  {Sakall{\i}}},\ }\href {https://doi.org/10.1140/epjc/s10052-021-09299-y}
  {\bibfield  {journal} {\bibinfo  {journal} {Eur. Phys. J. C}\ }\textbf
  {\bibinfo {volume} {81}},\ \bibinfo {pages} {501} (\bibinfo {year} {2021})},\
  \Eprint {https://arxiv.org/abs/2102.06303} {arXiv:2102.06303 [hep-th]}
  \BibitemShut {NoStop}%
\bibitem [{\citenamefont {Liu}\ \emph {et~al.}(2023)\citenamefont {Liu},
  \citenamefont {Fang}, \citenamefont {Jing},\ and\ \citenamefont
  {Wang}}]{Liu:2022dcn}%
  \BibitemOpen
  \bibfield  {author} {\bibinfo {author} {\bibfnamefont {W.}~\bibnamefont
  {Liu}}, \bibinfo {author} {\bibfnamefont {X.}~\bibnamefont {Fang}}, \bibinfo
  {author} {\bibfnamefont {J.}~\bibnamefont {Jing}},\ and\ \bibinfo {author}
  {\bibfnamefont {J.}~\bibnamefont {Wang}},\ }\href
  {https://doi.org/10.1140/epjc/s10052-023-11231-5} {\bibfield  {journal}
  {\bibinfo  {journal} {Eur. Phys. J. C}\ }\textbf {\bibinfo {volume} {83}},\
  \bibinfo {pages} {83} (\bibinfo {year} {2023})},\ \Eprint
  {https://arxiv.org/abs/2211.03156} {arXiv:2211.03156 [gr-qc]} \BibitemShut
  {NoStop}%
\bibitem [{\citenamefont {Liu}\ \emph {et~al.}(2024)\citenamefont {Liu},
  \citenamefont {Fang}, \citenamefont {Jing},\ and\ \citenamefont
  {Wang}}]{Liu:2024oeq}%
  \BibitemOpen
  \bibfield  {author} {\bibinfo {author} {\bibfnamefont {W.}~\bibnamefont
  {Liu}}, \bibinfo {author} {\bibfnamefont {X.}~\bibnamefont {Fang}}, \bibinfo
  {author} {\bibfnamefont {J.}~\bibnamefont {Jing}},\ and\ \bibinfo {author}
  {\bibfnamefont {J.}~\bibnamefont {Wang}},\ }\href
  {https://doi.org/10.1007/s11433-024-2405-y} {\bibfield  {journal} {\bibinfo
  {journal} {Sci. China Phys. Mech. Astron.}\ }\textbf {\bibinfo {volume}
  {67}},\ \bibinfo {pages} {280413} (\bibinfo {year} {2024})},\ \Eprint
  {https://arxiv.org/abs/2402.09686} {arXiv:2402.09686 [gr-qc]} \BibitemShut
  {NoStop}%
\bibitem [{\citenamefont {Mai}\ \emph {et~al.}(2024)\citenamefont {Mai},
  \citenamefont {Xu}, \citenamefont {Liang},\ and\ \citenamefont
  {Shao}}]{Mai:2024lgk}%
  \BibitemOpen
  \bibfield  {author} {\bibinfo {author} {\bibfnamefont {Z.-F.}\ \bibnamefont
  {Mai}}, \bibinfo {author} {\bibfnamefont {R.}~\bibnamefont {Xu}}, \bibinfo
  {author} {\bibfnamefont {D.}~\bibnamefont {Liang}},\ and\ \bibinfo {author}
  {\bibfnamefont {L.}~\bibnamefont {Shao}},\ }\href
  {https://doi.org/10.1103/PhysRevD.109.084076} {\bibfield  {journal} {\bibinfo
   {journal} {Phys. Rev. D}\ }\textbf {\bibinfo {volume} {109}},\ \bibinfo
  {pages} {084076} (\bibinfo {year} {2024})},\ \Eprint
  {https://arxiv.org/abs/2401.07757} {arXiv:2401.07757 [gr-qc]} \BibitemShut
  {NoStop}%
\bibitem [{\citenamefont {de~Oliveira}\ \emph {et~al.}(2025)\citenamefont
  {de~Oliveira}, \citenamefont {Pavan}, \citenamefont {Lin},\ and\
  \citenamefont {Cui}}]{deOliveira:2025yeo}%
  \BibitemOpen
  \bibfield  {author} {\bibinfo {author} {\bibfnamefont {J.}~\bibnamefont
  {de~Oliveira}}, \bibinfo {author} {\bibfnamefont {A.~B.}\ \bibnamefont
  {Pavan}}, \bibinfo {author} {\bibfnamefont {K.}~\bibnamefont {Lin}},\ and\
  \bibinfo {author} {\bibfnamefont {Y.-H.}\ \bibnamefont {Cui}},\ }\href
  {https://doi.org/10.1088/1361-6382/ae2378} {\bibfield  {journal} {\bibinfo
  {journal} {Class. Quant. Grav.}\ }\textbf {\bibinfo {volume} {42}},\ \bibinfo
  {pages} {235018} (\bibinfo {year} {2025})}\BibitemShut {NoStop}%
\bibitem [{\citenamefont {Chen}\ \emph {et~al.}(2023)\citenamefont {Chen},
  \citenamefont {Pan},\ and\ \citenamefont {Jing}}]{Chen:2023cjd}%
  \BibitemOpen
  \bibfield  {author} {\bibinfo {author} {\bibfnamefont {C.}~\bibnamefont
  {Chen}}, \bibinfo {author} {\bibfnamefont {Q.}~\bibnamefont {Pan}},\ and\
  \bibinfo {author} {\bibfnamefont {J.}~\bibnamefont {Jing}},\ }\href
  {https://doi.org/10.1016/j.physletb.2023.138186} {\bibfield  {journal}
  {\bibinfo  {journal} {Phys. Lett. B}\ }\textbf {\bibinfo {volume} {846}},\
  \bibinfo {pages} {138186} (\bibinfo {year} {2023})},\ \Eprint
  {https://arxiv.org/abs/2302.05861} {arXiv:2302.05861 [gr-qc]} \BibitemShut
  {NoStop}%
\bibitem [{\citenamefont {Li}\ \emph {et~al.}(2025{\natexlab{a}})\citenamefont
  {Li}, \citenamefont {Liu}, \citenamefont {Guo},\ and\ \citenamefont
  {Liu}}]{Li:2025itp}%
  \BibitemOpen
  \bibfield  {author} {\bibinfo {author} {\bibfnamefont {B.-R.}\ \bibnamefont
  {Li}}, \bibinfo {author} {\bibfnamefont {J.-Z.}\ \bibnamefont {Liu}},
  \bibinfo {author} {\bibfnamefont {W.-D.}\ \bibnamefont {Guo}},\ and\ \bibinfo
  {author} {\bibfnamefont {Y.-X.}\ \bibnamefont {Liu}},\ }\href@noop {} {\
  (\bibinfo {year} {2025}{\natexlab{a}})},\ \Eprint
  {https://arxiv.org/abs/2510.20503} {arXiv:2510.20503 [gr-qc]} \BibitemShut
  {NoStop}%
\bibitem [{\citenamefont {Liu}\ \emph {et~al.}(2025{\natexlab{d}})\citenamefont
  {Liu}, \citenamefont {Huang}, \citenamefont {Wang},\ and\ \citenamefont
  {Yang}}]{Liu:2025swi}%
  \BibitemOpen
  \bibfield  {author} {\bibinfo {author} {\bibfnamefont {M.-J.}\ \bibnamefont
  {Liu}}, \bibinfo {author} {\bibfnamefont {L.-X.}\ \bibnamefont {Huang}},
  \bibinfo {author} {\bibfnamefont {Y.-Q.}\ \bibnamefont {Wang}},\ and\
  \bibinfo {author} {\bibfnamefont {K.}~\bibnamefont {Yang}},\ }\href@noop {}
  {\  (\bibinfo {year} {2025}{\natexlab{d}})},\ \Eprint
  {https://arxiv.org/abs/2512.19581} {arXiv:2512.19581 [gr-qc]} \BibitemShut
  {NoStop}%
\bibitem [{\citenamefont {Ara{\'u}jo~Filho}\ and\ \citenamefont
  {Liu}(2025)}]{AraujoFilho:2025nmc}%
  \BibitemOpen
  \bibfield  {author} {\bibinfo {author} {\bibfnamefont {A.~A.}\ \bibnamefont
  {Ara{\'u}jo~Filho}}\ and\ \bibinfo {author} {\bibfnamefont {W.}~\bibnamefont
  {Liu}},\ }\href@noop {} {\  (\bibinfo {year} {2025})},\ \Eprint
  {https://arxiv.org/abs/2512.17567} {arXiv:2512.17567 [gr-qc]} \BibitemShut
  {NoStop}%
\bibitem [{\citenamefont {Shi}\ and\ \citenamefont
  {Ara{\'u}jo~Filho}(2025)}]{Shi:2025plr}%
  \BibitemOpen
  \bibfield  {author} {\bibinfo {author} {\bibfnamefont {Y.}~\bibnamefont
  {Shi}}\ and\ \bibinfo {author} {\bibfnamefont {A.~A.}\ \bibnamefont
  {Ara{\'u}jo~Filho}},\ }\href {https://doi.org/10.1088/1475-7516/2025/11/045}
  {\bibfield  {journal} {\bibinfo  {journal} {JCAP}\ }\textbf {\bibinfo
  {volume} {11}},\ \bibinfo {pages} {045}},\ \Eprint
  {https://arxiv.org/abs/2505.02290} {arXiv:2505.02290 [gr-qc]} \BibitemShut
  {NoStop}%
\bibitem [{\citenamefont {Shi}\ \emph {et~al.}(2026)\citenamefont {Shi},
  \citenamefont {Zhang},\ and\ \citenamefont {Liu}}]{Shi:2026zxx}%
  \BibitemOpen
  \bibfield  {author} {\bibinfo {author} {\bibfnamefont {Z.}~\bibnamefont
  {Shi}}, \bibinfo {author} {\bibfnamefont {X.}~\bibnamefont {Zhang}},\ and\
  \bibinfo {author} {\bibfnamefont {Y.}~\bibnamefont {Liu}},\ }\href@noop {} {\
   (\bibinfo {year} {2026})},\ \Eprint {https://arxiv.org/abs/2603.14413}
  {arXiv:2603.14413 [gr-qc]} \BibitemShut {NoStop}%
\bibitem [{\citenamefont {Poulis}\ and\ \citenamefont
  {Soares}(2022)}]{Poulis:2021nqh}%
  \BibitemOpen
  \bibfield  {author} {\bibinfo {author} {\bibfnamefont {F.~P.}\ \bibnamefont
  {Poulis}}\ and\ \bibinfo {author} {\bibfnamefont {M.~A.~C.}\ \bibnamefont
  {Soares}},\ }\href {https://doi.org/10.1140/epjc/s10052-022-10547-y}
  {\bibfield  {journal} {\bibinfo  {journal} {Eur. Phys. J. C}\ }\textbf
  {\bibinfo {volume} {82}},\ \bibinfo {pages} {613} (\bibinfo {year} {2022})},\
  \Eprint {https://arxiv.org/abs/2112.04040} {arXiv:2112.04040 [gr-qc]}
  \BibitemShut {NoStop}%
\bibitem [{\citenamefont {Guo}\ and\ \citenamefont {Fan}(2026)}]{Guo:2026sfg}%
  \BibitemOpen
  \bibfield  {author} {\bibinfo {author} {\bibfnamefont {M.}~\bibnamefont
  {Guo}}\ and\ \bibinfo {author} {\bibfnamefont {Z.-Y.}\ \bibnamefont {Fan}},\
  }\href@noop {} {\  (\bibinfo {year} {2026})},\ \Eprint
  {https://arxiv.org/abs/2607.17223} {arXiv:2607.17223 [gr-qc]} \BibitemShut
  {NoStop}%
\bibitem [{\citenamefont {Akiyama}\ \emph
  {et~al.}(2019{\natexlab{a}})\citenamefont {Akiyama} \emph
  {et~al.}}]{EventHorizonTelescope:2019dse}%
  \BibitemOpen
  \bibfield  {author} {\bibinfo {author} {\bibfnamefont {K.}~\bibnamefont
  {Akiyama}} \emph {et~al.} (\bibinfo {collaboration} {Event Horizon
  Telescope}),\ }\href {https://doi.org/10.3847/2041-8213/ab0ec7} {\bibfield
  {journal} {\bibinfo  {journal} {Astrophys. J. Lett.}\ }\textbf {\bibinfo
  {volume} {875}},\ \bibinfo {pages} {L1} (\bibinfo {year}
  {2019}{\natexlab{a}})},\ \Eprint {https://arxiv.org/abs/1906.11238}
  {arXiv:1906.11238 [astro-ph.GA]} \BibitemShut {NoStop}%
\bibitem [{\citenamefont {Akiyama}\ \emph
  {et~al.}(2019{\natexlab{b}})\citenamefont {Akiyama} \emph
  {et~al.}}]{EventHorizonTelescope:2019ggy}%
  \BibitemOpen
  \bibfield  {author} {\bibinfo {author} {\bibfnamefont {K.}~\bibnamefont
  {Akiyama}} \emph {et~al.} (\bibinfo {collaboration} {Event Horizon
  Telescope}),\ }\href {https://doi.org/10.3847/2041-8213/ab1141} {\bibfield
  {journal} {\bibinfo  {journal} {Astrophys. J. Lett.}\ }\textbf {\bibinfo
  {volume} {875}},\ \bibinfo {pages} {L6} (\bibinfo {year}
  {2019}{\natexlab{b}})},\ \Eprint {https://arxiv.org/abs/1906.11243}
  {arXiv:1906.11243 [astro-ph.GA]} \BibitemShut {NoStop}%
\bibitem [{\citenamefont {Psaltis}\ \emph {et~al.}(2020)\citenamefont {Psaltis}
  \emph {et~al.}}]{EventHorizonTelescope:2020qrl}%
  \BibitemOpen
  \bibfield  {author} {\bibinfo {author} {\bibfnamefont {D.}~\bibnamefont
  {Psaltis}} \emph {et~al.} (\bibinfo {collaboration} {Event Horizon
  Telescope}),\ }\href {https://doi.org/10.1103/PhysRevLett.125.141104}
  {\bibfield  {journal} {\bibinfo  {journal} {Phys. Rev. Lett.}\ }\textbf
  {\bibinfo {volume} {125}},\ \bibinfo {pages} {141104} (\bibinfo {year}
  {2020})},\ \Eprint {https://arxiv.org/abs/2010.01055} {arXiv:2010.01055
  [gr-qc]} \BibitemShut {NoStop}%
\bibitem [{\citenamefont {Akiyama}\ \emph {et~al.}(2022)\citenamefont {Akiyama}
  \emph {et~al.}}]{EventHorizonTelescope:2022wkp}%
  \BibitemOpen
  \bibfield  {author} {\bibinfo {author} {\bibfnamefont {K.}~\bibnamefont
  {Akiyama}} \emph {et~al.} (\bibinfo {collaboration} {Event Horizon
  Telescope}),\ }\href {https://doi.org/10.3847/2041-8213/ac6674} {\bibfield
  {journal} {\bibinfo  {journal} {Astrophys. J. Lett.}\ }\textbf {\bibinfo
  {volume} {930}},\ \bibinfo {pages} {L12} (\bibinfo {year} {2022})},\ \Eprint
  {https://arxiv.org/abs/2311.08680} {arXiv:2311.08680 [astro-ph.HE]}
  \BibitemShut {NoStop}%
\bibitem [{\citenamefont {Genzel}\ \emph {et~al.}(2024)\citenamefont {Genzel},
  \citenamefont {Eisenhauer},\ and\ \citenamefont
  {Gillessen}}]{Genzel:2024vou}%
  \BibitemOpen
  \bibfield  {author} {\bibinfo {author} {\bibfnamefont {R.}~\bibnamefont
  {Genzel}}, \bibinfo {author} {\bibfnamefont {F.}~\bibnamefont {Eisenhauer}},\
  and\ \bibinfo {author} {\bibfnamefont {S.}~\bibnamefont {Gillessen}},\ }\href
  {https://doi.org/10.1007/s00159-024-00154-z} {\bibfield  {journal} {\bibinfo
  {journal} {Astron. Astrophys. Rev.}\ }\textbf {\bibinfo {volume} {32}},\
  \bibinfo {pages} {3} (\bibinfo {year} {2024})},\ \Eprint
  {https://arxiv.org/abs/2404.03522} {arXiv:2404.03522 [astro-ph.GA]}
  \BibitemShut {NoStop}%
\bibitem [{\citenamefont {Islam}\ and\ \citenamefont
  {Ghosh}(2021)}]{Islam:2021dyk}%
  \BibitemOpen
  \bibfield  {author} {\bibinfo {author} {\bibfnamefont {S.~U.}\ \bibnamefont
  {Islam}}\ and\ \bibinfo {author} {\bibfnamefont {S.~G.}\ \bibnamefont
  {Ghosh}},\ }\href {https://doi.org/10.1103/PhysRevD.103.124052} {\bibfield
  {journal} {\bibinfo  {journal} {Phys. Rev. D}\ }\textbf {\bibinfo {volume}
  {103}},\ \bibinfo {pages} {124052} (\bibinfo {year} {2021})},\ \Eprint
  {https://arxiv.org/abs/2102.08289} {arXiv:2102.08289 [gr-qc]} \BibitemShut
  {NoStop}%
\bibitem [{\citenamefont {Cunha}\ \emph {et~al.}(2018)\citenamefont {Cunha},
  \citenamefont {Herdeiro},\ and\ \citenamefont {Rodriguez}}]{Cunha:2018gql}%
  \BibitemOpen
  \bibfield  {author} {\bibinfo {author} {\bibfnamefont {P.~V.~P.}\
  \bibnamefont {Cunha}}, \bibinfo {author} {\bibfnamefont {C.~A.~R.}\
  \bibnamefont {Herdeiro}},\ and\ \bibinfo {author} {\bibfnamefont {M.~J.}\
  \bibnamefont {Rodriguez}},\ }\href
  {https://doi.org/10.1103/PhysRevD.97.084020} {\bibfield  {journal} {\bibinfo
  {journal} {Phys. Rev. D}\ }\textbf {\bibinfo {volume} {97}},\ \bibinfo
  {pages} {084020} (\bibinfo {year} {2018})},\ \Eprint
  {https://arxiv.org/abs/1802.02675} {arXiv:1802.02675 [gr-qc]} \BibitemShut
  {NoStop}%
\bibitem [{\citenamefont {Wu}\ \emph {et~al.}(2025)\citenamefont {Wu},
  \citenamefont {Guo},\ and\ \citenamefont {Kuang}}]{2509.26270}%
  \BibitemOpen
  \bibfield  {author} {\bibinfo {author} {\bibfnamefont {M.-H.}\ \bibnamefont
  {Wu}}, \bibinfo {author} {\bibfnamefont {H.}~\bibnamefont {Guo}},\ and\
  \bibinfo {author} {\bibfnamefont {X.-M.}\ \bibnamefont {Kuang}},\ }\href@noop
  {} {\  (\bibinfo {year} {2025})},\ \Eprint {https://arxiv.org/abs/2509.26270}
  {arXiv:2509.26270 [gr-qc]} \BibitemShut {NoStop}%
\bibitem [{\citenamefont {Wang}\ \emph {et~al.}(2022)\citenamefont {Wang},
  \citenamefont {Chen},\ and\ \citenamefont {Jing}}]{Wang:2021gtd}%
  \BibitemOpen
  \bibfield  {author} {\bibinfo {author} {\bibfnamefont {Z.}~\bibnamefont
  {Wang}}, \bibinfo {author} {\bibfnamefont {S.}~\bibnamefont {Chen}},\ and\
  \bibinfo {author} {\bibfnamefont {J.}~\bibnamefont {Jing}},\ }\href
  {https://doi.org/10.1140/epjc/s10052-022-10475-x} {\bibfield  {journal}
  {\bibinfo  {journal} {Eur. Phys. J. C}\ }\textbf {\bibinfo {volume} {82}},\
  \bibinfo {pages} {528} (\bibinfo {year} {2022})},\ \Eprint
  {https://arxiv.org/abs/2112.02895} {arXiv:2112.02895 [gr-qc]} \BibitemShut
  {NoStop}%
\bibitem [{\citenamefont {Zhen}\ \emph {et~al.}(2025)\citenamefont {Zhen},
  \citenamefont {Guo}, \citenamefont {Wu},\ and\ \citenamefont
  {Kuang}}]{Zhen:2025nah}%
  \BibitemOpen
  \bibfield  {author} {\bibinfo {author} {\bibfnamefont {W.-Q.}\ \bibnamefont
  {Zhen}}, \bibinfo {author} {\bibfnamefont {H.}~\bibnamefont {Guo}}, \bibinfo
  {author} {\bibfnamefont {M.-H.}\ \bibnamefont {Wu}},\ and\ \bibinfo {author}
  {\bibfnamefont {X.-M.}\ \bibnamefont {Kuang}},\ }\href
  {https://doi.org/10.1016/j.physletb.2025.139307} {\bibfield  {journal}
  {\bibinfo  {journal} {Phys. Lett. B}\ }\textbf {\bibinfo {volume} {862}},\
  \bibinfo {pages} {139307} (\bibinfo {year} {2025})}\BibitemShut {NoStop}%
\bibitem [{\citenamefont {Qi}\ \emph {et~al.}(2024)\citenamefont {Qi},
  \citenamefont {Kuang}, \citenamefont {Li},\ and\ \citenamefont
  {Sang}}]{QiQi:2024dwc}%
  \BibitemOpen
  \bibfield  {author} {\bibinfo {author} {\bibfnamefont {Q.}~\bibnamefont
  {Qi}}, \bibinfo {author} {\bibfnamefont {X.-M.}\ \bibnamefont {Kuang}},
  \bibinfo {author} {\bibfnamefont {Y.-Z.}\ \bibnamefont {Li}},\ and\ \bibinfo
  {author} {\bibfnamefont {Y.}~\bibnamefont {Sang}},\ }\href
  {https://doi.org/10.1140/epjc/s10052-024-12989-y} {\bibfield  {journal}
  {\bibinfo  {journal} {Eur. Phys. J. C}\ }\textbf {\bibinfo {volume} {84}},\
  \bibinfo {pages} {645} (\bibinfo {year} {2024})},\ \Eprint
  {https://arxiv.org/abs/2407.01958} {arXiv:2407.01958 [gr-qc]} \BibitemShut
  {NoStop}%
\bibitem [{\citenamefont {Wu}\ \emph {et~al.}(2023)\citenamefont {Wu},
  \citenamefont {Guo},\ and\ \citenamefont {Kuang}}]{Wu:2023wld}%
  \BibitemOpen
  \bibfield  {author} {\bibinfo {author} {\bibfnamefont {M.-H.}\ \bibnamefont
  {Wu}}, \bibinfo {author} {\bibfnamefont {H.}~\bibnamefont {Guo}},\ and\
  \bibinfo {author} {\bibfnamefont {X.-M.}\ \bibnamefont {Kuang}},\ }\href
  {https://doi.org/10.1103/PhysRevD.107.064033} {\bibfield  {journal} {\bibinfo
   {journal} {Phys. Rev. D}\ }\textbf {\bibinfo {volume} {107}},\ \bibinfo
  {pages} {064033} (\bibinfo {year} {2023})},\ \Eprint
  {https://arxiv.org/abs/2306.10467} {arXiv:2306.10467 [gr-qc]} \BibitemShut
  {NoStop}%
\bibitem [{\citenamefont {Li}\ and\ \citenamefont {Kuang}(2023)}]{Li:2023djs}%
  \BibitemOpen
  \bibfield  {author} {\bibinfo {author} {\bibfnamefont {Y.-Z.}\ \bibnamefont
  {Li}}\ and\ \bibinfo {author} {\bibfnamefont {X.-M.}\ \bibnamefont {Kuang}},\
  }\href {https://doi.org/10.1103/PhysRevD.107.064052} {\bibfield  {journal}
  {\bibinfo  {journal} {Phys. Rev. D}\ }\textbf {\bibinfo {volume} {107}},\
  \bibinfo {pages} {064052} (\bibinfo {year} {2023})}\BibitemShut {NoStop}%
\bibitem [{\citenamefont {Bambhaniya}\ \emph {et~al.}(2023)\citenamefont
  {Bambhaniya}, \citenamefont {Trivedi}, \citenamefont {Dey}, \citenamefont
  {Joshi},\ and\ \citenamefont {Joshi}}]{Bambhaniya:2021jum}%
  \BibitemOpen
  \bibfield  {author} {\bibinfo {author} {\bibfnamefont {P.}~\bibnamefont
  {Bambhaniya}}, \bibinfo {author} {\bibfnamefont {J.~V.}\ \bibnamefont
  {Trivedi}}, \bibinfo {author} {\bibfnamefont {D.}~\bibnamefont {Dey}},
  \bibinfo {author} {\bibfnamefont {P.~S.}\ \bibnamefont {Joshi}},\ and\
  \bibinfo {author} {\bibfnamefont {A.~B.}\ \bibnamefont {Joshi}},\ }\href
  {https://doi.org/10.1016/j.dark.2023.101215} {\bibfield  {journal} {\bibinfo
  {journal} {Phys. Dark Univ.}\ }\textbf {\bibinfo {volume} {40}},\ \bibinfo
  {pages} {101215} (\bibinfo {year} {2023})},\ \Eprint
  {https://arxiv.org/abs/2109.11137} {arXiv:2109.11137 [gr-qc]} \BibitemShut
  {NoStop}%
\bibitem [{\citenamefont {Wang}(2025)}]{Wang:2025zis}%
  \BibitemOpen
  \bibfield  {author} {\bibinfo {author} {\bibfnamefont {H.-M.}\ \bibnamefont
  {Wang}},\ }\href {https://doi.org/10.1088/1475-7516/2025/11/087} {\bibfield
  {journal} {\bibinfo  {journal} {JCAP}\ }\textbf {\bibinfo {volume} {11}},\
  \bibinfo {pages} {087}},\ \Eprint {https://arxiv.org/abs/2507.03443}
  {arXiv:2507.03443 [gr-qc]} \BibitemShut {NoStop}%
\bibitem [{\citenamefont {Wang}\ \emph {et~al.}(2025)\citenamefont {Wang},
  \citenamefont {Liao},\ and\ \citenamefont {Wei}}]{2505.20736}%
  \BibitemOpen
  \bibfield  {author} {\bibinfo {author} {\bibfnamefont {H.-M.}\ \bibnamefont
  {Wang}}, \bibinfo {author} {\bibfnamefont {K.}~\bibnamefont {Liao}},\ and\
  \bibinfo {author} {\bibfnamefont {S.-W.}\ \bibnamefont {Wei}},\ }\href
  {https://doi.org/10.1140/epjc/s10052-025-14626-8} {\bibfield  {journal}
  {\bibinfo  {journal} {Eur. Phys. J. C}\ }\textbf {\bibinfo {volume} {85}},\
  \bibinfo {pages} {933} (\bibinfo {year} {2025})},\ \Eprint
  {https://arxiv.org/abs/2505.20736} {arXiv:2505.20736 [gr-qc]} \BibitemShut
  {NoStop}%
\bibitem [{\citenamefont {Merloni}\ \emph {et~al.}(1999)\citenamefont
  {Merloni}, \citenamefont {Vietri}, \citenamefont {Stella},\ and\
  \citenamefont {Bini}}]{astro-ph/9811198}%
  \BibitemOpen
  \bibfield  {author} {\bibinfo {author} {\bibfnamefont {A.}~\bibnamefont
  {Merloni}}, \bibinfo {author} {\bibfnamefont {M.}~\bibnamefont {Vietri}},
  \bibinfo {author} {\bibfnamefont {L.}~\bibnamefont {Stella}},\ and\ \bibinfo
  {author} {\bibfnamefont {D.}~\bibnamefont {Bini}},\ }\href
  {https://doi.org/10.1046/j.1365-8711.1999.02307.x} {\bibfield  {journal}
  {\bibinfo  {journal} {Mon. Not. Roy. Astron. Soc.}\ }\textbf {\bibinfo
  {volume} {304}},\ \bibinfo {pages} {155} (\bibinfo {year} {1999})},\ \Eprint
  {https://arxiv.org/abs/astro-ph/9811198} {arXiv:astro-ph/9811198}
  \BibitemShut {NoStop}%
\bibitem [{\citenamefont {Karmakar}\ and\ \citenamefont
  {Sarkar}(2018)}]{1709.08935}%
  \BibitemOpen
  \bibfield  {author} {\bibinfo {author} {\bibfnamefont {T.}~\bibnamefont
  {Karmakar}}\ and\ \bibinfo {author} {\bibfnamefont {T.}~\bibnamefont
  {Sarkar}},\ }\href {https://doi.org/10.1007/s10714-018-2408-y} {\bibfield
  {journal} {\bibinfo  {journal} {Gen. Rel. Grav.}\ }\textbf {\bibinfo {volume}
  {50}},\ \bibinfo {pages} {85} (\bibinfo {year} {2018})},\ \Eprint
  {https://arxiv.org/abs/1709.08935} {arXiv:1709.08935 [gr-qc]} \BibitemShut
  {NoStop}%
\bibitem [{\citenamefont {Kraniotis}(2005)}]{Kraniotis:2005zm}%
  \BibitemOpen
  \bibfield  {author} {\bibinfo {author} {\bibfnamefont {G.~V.}\ \bibnamefont
  {Kraniotis}},\ }\href {https://doi.org/10.1088/0264-9381/22/21/001}
  {\bibfield  {journal} {\bibinfo  {journal} {Class. Quant. Grav.}\ }\textbf
  {\bibinfo {volume} {22}},\ \bibinfo {pages} {4391} (\bibinfo {year}
  {2005})},\ \Eprint {https://arxiv.org/abs/gr-qc/0507056}
  {arXiv:gr-qc/0507056} \BibitemShut {NoStop}%
\bibitem [{\citenamefont {Paul}\ \emph {et~al.}(2024)\citenamefont {Paul},
  \citenamefont {Bhattacharjee},\ and\ \citenamefont {Kalita}}]{2402.01192}%
  \BibitemOpen
  \bibfield  {author} {\bibinfo {author} {\bibfnamefont {D.}~\bibnamefont
  {Paul}}, \bibinfo {author} {\bibfnamefont {P.}~\bibnamefont
  {Bhattacharjee}},\ and\ \bibinfo {author} {\bibfnamefont {S.}~\bibnamefont
  {Kalita}},\ }\href {https://doi.org/10.3847/1538-4357/ad24f0} {\bibfield
  {journal} {\bibinfo  {journal} {Astrophys. J.}\ }\textbf {\bibinfo {volume}
  {964}},\ \bibinfo {pages} {127} (\bibinfo {year} {2024})},\ \Eprint
  {https://arxiv.org/abs/2402.01192} {arXiv:2402.01192 [gr-qc]} \BibitemShut
  {NoStop}%
\bibitem [{\citenamefont {Pantig}\ \emph {et~al.}(2024)\citenamefont {Pantig},
  \citenamefont {Kala}, \citenamefont {{\"O}vg{\"u}n},\ and\ \citenamefont
  {Lobos}}]{2410.13661}%
  \BibitemOpen
  \bibfield  {author} {\bibinfo {author} {\bibfnamefont {R.~C.}\ \bibnamefont
  {Pantig}}, \bibinfo {author} {\bibfnamefont {S.}~\bibnamefont {Kala}},
  \bibinfo {author} {\bibfnamefont {A.}~\bibnamefont {{\"O}vg{\"u}n}},\ and\
  \bibinfo {author} {\bibfnamefont {N.~J. L.~S.}\ \bibnamefont {Lobos}}\ }\href
  {https://doi.org/10.1142/S0219887825502408} {10.1142/S0219887825502408}
  (\bibinfo {year} {2024}),\ \Eprint {https://arxiv.org/abs/2410.13661}
  {arXiv:2410.13661 [gr-qc]} \BibitemShut {NoStop}%
\bibitem [{\citenamefont {Jha}\ and\ \citenamefont
  {Rahaman}(2022)}]{Jha:2022bpv}%
  \BibitemOpen
  \bibfield  {author} {\bibinfo {author} {\bibfnamefont {S.~K.}\ \bibnamefont
  {Jha}}\ and\ \bibinfo {author} {\bibfnamefont {A.}~\bibnamefont {Rahaman}},\
  }\href {https://doi.org/10.1140/epjc/s10052-022-10617-1} {\bibfield
  {journal} {\bibinfo  {journal} {Eur. Phys. J. C}\ }\textbf {\bibinfo {volume}
  {82}},\ \bibinfo {pages} {728} (\bibinfo {year} {2022})}\BibitemShut
  {NoStop}%
\bibitem [{\citenamefont {Kuang}\ and\ \citenamefont
  {{\"O}vg{\"u}n}(2022)}]{Kuang:2022xjp}%
  \BibitemOpen
  \bibfield  {author} {\bibinfo {author} {\bibfnamefont {X.-M.}\ \bibnamefont
  {Kuang}}\ and\ \bibinfo {author} {\bibfnamefont {A.}~\bibnamefont
  {{\"O}vg{\"u}n}},\ }\href {https://doi.org/10.1016/j.aop.2022.169147}
  {\bibfield  {journal} {\bibinfo  {journal} {Annals Phys.}\ }\textbf {\bibinfo
  {volume} {447}},\ \bibinfo {pages} {169147} (\bibinfo {year} {2022})},\
  \Eprint {https://arxiv.org/abs/2205.11003} {arXiv:2205.11003 [gr-qc]}
  \BibitemShut {NoStop}%
\bibitem [{\citenamefont {Li}\ \emph {et~al.}(2025{\natexlab{b}})\citenamefont
  {Li}, \citenamefont {Luo},\ and\ \citenamefont {Feng}}]{2507.03981}%
  \BibitemOpen
  \bibfield  {author} {\bibinfo {author} {\bibfnamefont {S.-Y.}\ \bibnamefont
  {Li}}, \bibinfo {author} {\bibfnamefont {S.-S.}\ \bibnamefont {Luo}},\ and\
  \bibinfo {author} {\bibfnamefont {Z.-W.}\ \bibnamefont {Feng}},\ }\href@noop
  {} {\  (\bibinfo {year} {2025}{\natexlab{b}})},\ \Eprint
  {https://arxiv.org/abs/2507.03981} {arXiv:2507.03981 [gr-qc]} \BibitemShut
  {NoStop}%
\bibitem [{\citenamefont {Ayzenberg}\ and\ \citenamefont
  {Yunes}(2018)}]{Ayzenberg:2018jip}%
  \BibitemOpen
  \bibfield  {author} {\bibinfo {author} {\bibfnamefont {D.}~\bibnamefont
  {Ayzenberg}}\ and\ \bibinfo {author} {\bibfnamefont {N.}~\bibnamefont
  {Yunes}},\ }\href {https://doi.org/10.1088/1361-6382/aae87b} {\bibfield
  {journal} {\bibinfo  {journal} {Class. Quant. Grav.}\ }\textbf {\bibinfo
  {volume} {35}},\ \bibinfo {pages} {235002} (\bibinfo {year} {2018})},\
  \Eprint {https://arxiv.org/abs/1807.08422} {arXiv:1807.08422 [gr-qc]}
  \BibitemShut {NoStop}%
\bibitem [{\citenamefont {Alhamzawi}(2017)}]{Alhamzawi:2017iyn}%
  \BibitemOpen
  \bibfield  {author} {\bibinfo {author} {\bibfnamefont {A.}~\bibnamefont
  {Alhamzawi}},\ }\href {https://doi.org/10.1142/S0218271817501565} {\bibfield
  {journal} {\bibinfo  {journal} {Int. J. Mod. Phys. D}\ }\textbf {\bibinfo
  {volume} {26}},\ \bibinfo {pages} {1750156} (\bibinfo {year}
  {2017})}\BibitemShut {NoStop}%
\bibitem [{\citenamefont {Chen}\ \emph {et~al.}(2020)\citenamefont {Chen},
  \citenamefont {Wang},\ and\ \citenamefont {Jing}}]{2004.08857}%
  \BibitemOpen
  \bibfield  {author} {\bibinfo {author} {\bibfnamefont {S.}~\bibnamefont
  {Chen}}, \bibinfo {author} {\bibfnamefont {M.}~\bibnamefont {Wang}},\ and\
  \bibinfo {author} {\bibfnamefont {J.}~\bibnamefont {Jing}},\ }\href
  {https://doi.org/10.1007/JHEP07(2020)054} {\bibfield  {journal} {\bibinfo
  {journal} {JHEP}\ }\textbf {\bibinfo {volume} {07}},\ \bibinfo {pages}
  {054}},\ \Eprint {https://arxiv.org/abs/2004.08857} {arXiv:2004.08857
  [gr-qc]} \BibitemShut {NoStop}%
\bibitem [{\citenamefont {Hou}\ \emph {et~al.}(2022{\natexlab{a}})\citenamefont
  {Hou}, \citenamefont {Liu}, \citenamefont {Guo}, \citenamefont {Yan},\ and\
  \citenamefont {Chen}}]{Hou:2022gge}%
  \BibitemOpen
  \bibfield  {author} {\bibinfo {author} {\bibfnamefont {Y.}~\bibnamefont
  {Hou}}, \bibinfo {author} {\bibfnamefont {P.}~\bibnamefont {Liu}}, \bibinfo
  {author} {\bibfnamefont {M.}~\bibnamefont {Guo}}, \bibinfo {author}
  {\bibfnamefont {H.}~\bibnamefont {Yan}},\ and\ \bibinfo {author}
  {\bibfnamefont {B.}~\bibnamefont {Chen}},\ }\href
  {https://doi.org/10.1088/1361-6382/ac8860} {\bibfield  {journal} {\bibinfo
  {journal} {Class. Quant. Grav.}\ }\textbf {\bibinfo {volume} {39}},\ \bibinfo
  {pages} {194001} (\bibinfo {year} {2022}{\natexlab{a}})},\ \Eprint
  {https://arxiv.org/abs/2203.02755} {arXiv:2203.02755 [gr-qc]} \BibitemShut
  {NoStop}%
\bibitem [{\citenamefont {Kuang}\ \emph {et~al.}(2022)\citenamefont {Kuang},
  \citenamefont {Tang}, \citenamefont {Wang},\ and\ \citenamefont
  {Wang}}]{2206.05878}%
  \BibitemOpen
  \bibfield  {author} {\bibinfo {author} {\bibfnamefont {X.-M.}\ \bibnamefont
  {Kuang}}, \bibinfo {author} {\bibfnamefont {Z.-Y.}\ \bibnamefont {Tang}},
  \bibinfo {author} {\bibfnamefont {B.}~\bibnamefont {Wang}},\ and\ \bibinfo
  {author} {\bibfnamefont {A.}~\bibnamefont {Wang}},\ }\href
  {https://doi.org/10.1103/PhysRevD.106.064012} {\bibfield  {journal} {\bibinfo
   {journal} {Phys. Rev. D}\ }\textbf {\bibinfo {volume} {106}},\ \bibinfo
  {pages} {064012} (\bibinfo {year} {2022})},\ \Eprint
  {https://arxiv.org/abs/2206.05878} {arXiv:2206.05878 [gr-qc]} \BibitemShut
  {NoStop}%
\bibitem [{\citenamefont {Vagnozzi}\ \emph {et~al.}(2023)\citenamefont
  {Vagnozzi} \emph {et~al.}}]{Vagnozzi:2022moj}%
  \BibitemOpen
  \bibfield  {author} {\bibinfo {author} {\bibfnamefont {S.}~\bibnamefont
  {Vagnozzi}} \emph {et~al.},\ }\href
  {https://doi.org/10.1088/1361-6382/acd97b} {\bibfield  {journal} {\bibinfo
  {journal} {Class. Quant. Grav.}\ }\textbf {\bibinfo {volume} {40}},\ \bibinfo
  {pages} {165007} (\bibinfo {year} {2023})},\ \Eprint
  {https://arxiv.org/abs/2205.07787} {arXiv:2205.07787 [gr-qc]} \BibitemShut
  {NoStop}%
\bibitem [{\citenamefont {Zhang}\ \emph
  {et~al.}(2024{\natexlab{a}})\citenamefont {Zhang}, \citenamefont {Hou},
  \citenamefont {Hu}, \citenamefont {Guo},\ and\ \citenamefont
  {Chen}}]{Zhang:2023cuw}%
  \BibitemOpen
  \bibfield  {author} {\bibinfo {author} {\bibfnamefont {Z.}~\bibnamefont
  {Zhang}}, \bibinfo {author} {\bibfnamefont {Y.}~\bibnamefont {Hou}}, \bibinfo
  {author} {\bibfnamefont {Z.}~\bibnamefont {Hu}}, \bibinfo {author}
  {\bibfnamefont {M.}~\bibnamefont {Guo}},\ and\ \bibinfo {author}
  {\bibfnamefont {B.}~\bibnamefont {Chen}},\ }\href
  {https://doi.org/10.1088/1475-7516/2024/03/013} {\bibfield  {journal}
  {\bibinfo  {journal} {JCAP}\ }\textbf {\bibinfo {volume} {03}},\ \bibinfo
  {pages} {013}},\ \Eprint {https://arxiv.org/abs/2304.03642} {arXiv:2304.03642
  [gr-qc]} \BibitemShut {NoStop}%
\bibitem [{\citenamefont {Zhang}\ \emph
  {et~al.}(2024{\natexlab{b}})\citenamefont {Zhang}, \citenamefont {Hou},
  \citenamefont {Guo},\ and\ \citenamefont {Chen}}]{Zhang:2024lsf}%
  \BibitemOpen
  \bibfield  {author} {\bibinfo {author} {\bibfnamefont {Z.}~\bibnamefont
  {Zhang}}, \bibinfo {author} {\bibfnamefont {Y.}~\bibnamefont {Hou}}, \bibinfo
  {author} {\bibfnamefont {M.}~\bibnamefont {Guo}},\ and\ \bibinfo {author}
  {\bibfnamefont {B.}~\bibnamefont {Chen}},\ }\href
  {https://doi.org/10.1088/1475-7516/2024/05/032} {\bibfield  {journal}
  {\bibinfo  {journal} {JCAP}\ }\textbf {\bibinfo {volume} {05}},\ \bibinfo
  {pages} {032}},\ \Eprint {https://arxiv.org/abs/2401.14794} {arXiv:2401.14794
  [astro-ph.HE]} \BibitemShut {NoStop}%
\bibitem [{\citenamefont {Afrin}\ \emph {et~al.}(2024)\citenamefont {Afrin},
  \citenamefont {Ghosh},\ and\ \citenamefont {Wang}}]{2409.06218}%
  \BibitemOpen
  \bibfield  {author} {\bibinfo {author} {\bibfnamefont {M.}~\bibnamefont
  {Afrin}}, \bibinfo {author} {\bibfnamefont {S.~G.}\ \bibnamefont {Ghosh}},\
  and\ \bibinfo {author} {\bibfnamefont {A.}~\bibnamefont {Wang}},\ }\href
  {https://doi.org/10.1016/j.dark.2024.101642} {\bibfield  {journal} {\bibinfo
  {journal} {Phys. Dark Univ.}\ }\textbf {\bibinfo {volume} {46}},\ \bibinfo
  {pages} {101642} (\bibinfo {year} {2024})},\ \Eprint
  {https://arxiv.org/abs/2409.06218} {arXiv:2409.06218 [gr-qc]} \BibitemShut
  {NoStop}%
\bibitem [{\citenamefont {Zhang}\ \emph {et~al.}(2025)\citenamefont {Zhang},
  \citenamefont {Hou}, \citenamefont {Guo}, \citenamefont {Mizuno},\ and\
  \citenamefont {Chen}}]{Zhang:2025vyx}%
  \BibitemOpen
  \bibfield  {author} {\bibinfo {author} {\bibfnamefont {Z.}~\bibnamefont
  {Zhang}}, \bibinfo {author} {\bibfnamefont {Y.}~\bibnamefont {Hou}}, \bibinfo
  {author} {\bibfnamefont {M.}~\bibnamefont {Guo}}, \bibinfo {author}
  {\bibfnamefont {Y.}~\bibnamefont {Mizuno}},\ and\ \bibinfo {author}
  {\bibfnamefont {B.}~\bibnamefont {Chen}},\ }\href
  {https://doi.org/10.1103/zmnz-p2rs} {\bibfield  {journal} {\bibinfo
  {journal} {Phys. Rev. D}\ }\textbf {\bibinfo {volume} {112}},\ \bibinfo
  {pages} {083024} (\bibinfo {year} {2025})},\ \Eprint
  {https://arxiv.org/abs/2503.17200} {arXiv:2503.17200 [astro-ph.HE]}
  \BibitemShut {NoStop}%
\bibitem [{\citenamefont {Ingram}\ and\ \citenamefont
  {Motta}(2019)}]{Ingram:2019mna}%
  \BibitemOpen
  \bibfield  {author} {\bibinfo {author} {\bibfnamefont {A.}~\bibnamefont
  {Ingram}}\ and\ \bibinfo {author} {\bibfnamefont {S.}~\bibnamefont {Motta}},\
  }\href {https://doi.org/10.1016/j.newar.2020.101524} {\bibfield  {journal}
  {\bibinfo  {journal} {New Astron. Rev.}\ }\textbf {\bibinfo {volume} {85}},\
  \bibinfo {pages} {101524} (\bibinfo {year} {2019})},\ \Eprint
  {https://arxiv.org/abs/2001.08758} {arXiv:2001.08758 [astro-ph.HE]}
  \BibitemShut {NoStop}%
\bibitem [{\citenamefont {Motta}(2017)}]{Motta:2016vwf}%
  \BibitemOpen
  \bibfield  {author} {\bibinfo {author} {\bibfnamefont {S.~E.}\ \bibnamefont
  {Motta}},\ }\href {https://doi.org/10.1002/asna.201612320} {\bibfield
  {journal} {\bibinfo  {journal} {Astron. Nachr.}\ }\textbf {\bibinfo {volume}
  {337}},\ \bibinfo {pages} {398} (\bibinfo {year} {2017})},\ \Eprint
  {https://arxiv.org/abs/1603.07885} {arXiv:1603.07885 [astro-ph.HE]}
  \BibitemShut {NoStop}%
\bibitem [{\citenamefont {van~der Klis}(2000)}]{vanderKlis:2000ca}%
  \BibitemOpen
  \bibfield  {author} {\bibinfo {author} {\bibfnamefont {M.}~\bibnamefont
  {van~der Klis}},\ }\href {https://doi.org/10.1146/annurev.astro.38.1.717}
  {\bibfield  {journal} {\bibinfo  {journal} {Ann. Rev. Astron. Astrophys.}\
  }\textbf {\bibinfo {volume} {38}},\ \bibinfo {pages} {717} (\bibinfo {year}
  {2000})},\ \Eprint {https://arxiv.org/abs/astro-ph/0001167}
  {arXiv:astro-ph/0001167} \BibitemShut {NoStop}%
\bibitem [{\citenamefont {Bambi}(2015)}]{Bambi:2013fea}%
  \BibitemOpen
  \bibfield  {author} {\bibinfo {author} {\bibfnamefont {C.}~\bibnamefont
  {Bambi}},\ }\href {https://doi.org/10.1140/epjc/s10052-015-3396-7} {\bibfield
   {journal} {\bibinfo  {journal} {Eur. Phys. J. C}\ }\textbf {\bibinfo
  {volume} {75}},\ \bibinfo {pages} {162} (\bibinfo {year} {2015})},\ \Eprint
  {https://arxiv.org/abs/1312.2228} {arXiv:1312.2228 [gr-qc]} \BibitemShut
  {NoStop}%
\bibitem [{\citenamefont {Bardeen}\ \emph {et~al.}(1972)\citenamefont
  {Bardeen}, \citenamefont {Press},\ and\ \citenamefont
  {Teukolsky}}]{Bardeen:1972fi}%
  \BibitemOpen
  \bibfield  {author} {\bibinfo {author} {\bibfnamefont {J.~M.}\ \bibnamefont
  {Bardeen}}, \bibinfo {author} {\bibfnamefont {W.~H.}\ \bibnamefont {Press}},\
  and\ \bibinfo {author} {\bibfnamefont {S.~A.}\ \bibnamefont {Teukolsky}},\
  }\href {https://doi.org/10.1086/151796} {\bibfield  {journal} {\bibinfo
  {journal} {Astrophys. J.}\ }\textbf {\bibinfo {volume} {178}},\ \bibinfo
  {pages} {347} (\bibinfo {year} {1972})}\BibitemShut {NoStop}%
\bibitem [{\citenamefont {Ryan}(1995)}]{PhysRevD.52.5707}%
  \BibitemOpen
  \bibfield  {author} {\bibinfo {author} {\bibfnamefont {F.~D.}\ \bibnamefont
  {Ryan}},\ }\href {https://doi.org/10.1103/PhysRevD.52.5707} {\bibfield
  {journal} {\bibinfo  {journal} {Phys. Rev. D}\ }\textbf {\bibinfo {volume}
  {52}},\ \bibinfo {pages} {5707} (\bibinfo {year} {1995})}\BibitemShut
  {NoStop}%
\bibitem [{\citenamefont {Doneva}\ \emph {et~al.}(2014)\citenamefont {Doneva},
  \citenamefont {Yazadjiev}, \citenamefont {Stergioulas}, \citenamefont
  {Kokkotas},\ and\ \citenamefont {Athanasiadis}}]{PhysRevD.90.044004}%
  \BibitemOpen
  \bibfield  {author} {\bibinfo {author} {\bibfnamefont {D.~D.}\ \bibnamefont
  {Doneva}}, \bibinfo {author} {\bibfnamefont {S.~S.}\ \bibnamefont
  {Yazadjiev}}, \bibinfo {author} {\bibfnamefont {N.}~\bibnamefont
  {Stergioulas}}, \bibinfo {author} {\bibfnamefont {K.~D.}\ \bibnamefont
  {Kokkotas}},\ and\ \bibinfo {author} {\bibfnamefont {T.~M.}\ \bibnamefont
  {Athanasiadis}},\ }\href {https://doi.org/10.1103/PhysRevD.90.044004}
  {\bibfield  {journal} {\bibinfo  {journal} {Phys. Rev. D}\ }\textbf {\bibinfo
  {volume} {90}},\ \bibinfo {pages} {044004} (\bibinfo {year}
  {2014})}\BibitemShut {NoStop}%
\bibitem [{\citenamefont {Motta}\ \emph {et~al.}(2014)\citenamefont {Motta},
  \citenamefont {Belloni}, \citenamefont {Stella}, \citenamefont
  {Mu{\~n}oz-Darias},\ and\ \citenamefont {Fender}}]{Motta:2013wga}%
  \BibitemOpen
  \bibfield  {author} {\bibinfo {author} {\bibfnamefont {S.~E.}\ \bibnamefont
  {Motta}}, \bibinfo {author} {\bibfnamefont {T.~M.}\ \bibnamefont {Belloni}},
  \bibinfo {author} {\bibfnamefont {L.}~\bibnamefont {Stella}}, \bibinfo
  {author} {\bibfnamefont {T.}~\bibnamefont {Mu{\~n}oz-Darias}},\ and\ \bibinfo
  {author} {\bibfnamefont {R.}~\bibnamefont {Fender}},\ }\href
  {https://doi.org/10.1093/mnras/stt2068} {\bibfield  {journal} {\bibinfo
  {journal} {Mon. Not. Roy. Astron. Soc.}\ }\textbf {\bibinfo {volume} {437}},\
  \bibinfo {pages} {2554} (\bibinfo {year} {2014})},\ \Eprint
  {https://arxiv.org/abs/1309.3652} {arXiv:1309.3652 [astro-ph.HE]}
  \BibitemShut {NoStop}%
\bibitem [{\citenamefont {Straumann}(2004)}]{straumann2004general}%
  \BibitemOpen
  \bibfield  {author} {\bibinfo {author} {\bibfnamefont {N.}~\bibnamefont
  {Straumann}},\ }\href@noop {} {\emph {\bibinfo {title} {{General Relativity}
  with Applications to {Astrophysics}}}},\ \bibinfo {edition} {1st}\ ed.\
  (\bibinfo  {publisher} {Springer},\ \bibinfo {address} {Berlin, Heidelberg},\
  \bibinfo {year} {2004})\BibitemShut {NoStop}%
\bibitem [{\citenamefont {Chakraborty}\ \emph
  {et~al.}(2017{\natexlab{a}})\citenamefont {Chakraborty}, \citenamefont
  {Patil}, \citenamefont {Kocherlakota}, \citenamefont {Bhattacharyya},
  \citenamefont {Joshi},\ and\ \citenamefont
  {Kr{\'o}lak}}]{Chakraborty:2016mhx}%
  \BibitemOpen
  \bibfield  {author} {\bibinfo {author} {\bibfnamefont {C.}~\bibnamefont
  {Chakraborty}}, \bibinfo {author} {\bibfnamefont {M.}~\bibnamefont {Patil}},
  \bibinfo {author} {\bibfnamefont {P.}~\bibnamefont {Kocherlakota}}, \bibinfo
  {author} {\bibfnamefont {S.}~\bibnamefont {Bhattacharyya}}, \bibinfo {author}
  {\bibfnamefont {P.~S.}\ \bibnamefont {Joshi}},\ and\ \bibinfo {author}
  {\bibfnamefont {A.}~\bibnamefont {Kr{\'o}lak}},\ }\href
  {https://doi.org/10.1103/PhysRevD.95.084024} {\bibfield  {journal} {\bibinfo
  {journal} {Phys. Rev. D}\ }\textbf {\bibinfo {volume} {95}},\ \bibinfo
  {pages} {084024} (\bibinfo {year} {2017}{\natexlab{a}})},\ \Eprint
  {https://arxiv.org/abs/1611.08808} {arXiv:1611.08808 [gr-qc]} \BibitemShut
  {NoStop}%
\bibitem [{\citenamefont {Chakraborty}\ \emph
  {et~al.}(2017{\natexlab{b}})\citenamefont {Chakraborty}, \citenamefont
  {Kocherlakota},\ and\ \citenamefont {Joshi}}]{Chakraborty:2016ipk}%
  \BibitemOpen
  \bibfield  {author} {\bibinfo {author} {\bibfnamefont {C.}~\bibnamefont
  {Chakraborty}}, \bibinfo {author} {\bibfnamefont {P.}~\bibnamefont
  {Kocherlakota}},\ and\ \bibinfo {author} {\bibfnamefont {P.~S.}\ \bibnamefont
  {Joshi}},\ }\href {https://doi.org/10.1103/PhysRevD.95.044006} {\bibfield
  {journal} {\bibinfo  {journal} {Phys. Rev. D}\ }\textbf {\bibinfo {volume}
  {95}},\ \bibinfo {pages} {044006} (\bibinfo {year} {2017}{\natexlab{b}})},\
  \Eprint {https://arxiv.org/abs/1605.00600} {arXiv:1605.00600 [gr-qc]}
  \BibitemShut {NoStop}%
\bibitem [{\citenamefont {Sakina}\ and\ \citenamefont
  {Chiba}(1979)}]{PhysRevD.19.2280}%
  \BibitemOpen
  \bibfield  {author} {\bibinfo {author} {\bibfnamefont {K.-i.}\ \bibnamefont
  {Sakina}}\ and\ \bibinfo {author} {\bibfnamefont {J.}~\bibnamefont {Chiba}},\
  }\href {https://doi.org/10.1103/PhysRevD.19.2280} {\bibfield  {journal}
  {\bibinfo  {journal} {Phys. Rev. D}\ }\textbf {\bibinfo {volume} {19}},\
  \bibinfo {pages} {2280} (\bibinfo {year} {1979})}\BibitemShut {NoStop}%
\bibitem [{\citenamefont {Hartle}(2021)}]{10.1119/1.1604390}%
  \BibitemOpen
  \bibfield  {author} {\bibinfo {author} {\bibfnamefont {J.}~\bibnamefont
  {Hartle}},\ }\href {https://doi.org/10.1017/9781009042604} {\emph {\bibinfo
  {title} {Gravity: An Introduction to Einstein's General Relativity}}}\
  (\bibinfo {year} {2021})\BibitemShut {NoStop}%
\bibitem [{\citenamefont {Gralla}\ \emph {et~al.}(2020)\citenamefont {Gralla},
  \citenamefont {Lupsasca},\ and\ \citenamefont {Marrone}}]{Gralla:2020srx}%
  \BibitemOpen
  \bibfield  {author} {\bibinfo {author} {\bibfnamefont {S.~E.}\ \bibnamefont
  {Gralla}}, \bibinfo {author} {\bibfnamefont {A.}~\bibnamefont {Lupsasca}},\
  and\ \bibinfo {author} {\bibfnamefont {D.~P.}\ \bibnamefont {Marrone}},\
  }\href {https://doi.org/10.1103/PhysRevD.102.124004} {\bibfield  {journal}
  {\bibinfo  {journal} {Phys. Rev. D}\ }\textbf {\bibinfo {volume} {102}},\
  \bibinfo {pages} {124004} (\bibinfo {year} {2020})},\ \Eprint
  {https://arxiv.org/abs/2008.03879} {arXiv:2008.03879 [gr-qc]} \BibitemShut
  {NoStop}%
\bibitem [{\citenamefont {Hou}\ \emph {et~al.}(2022{\natexlab{b}})\citenamefont
  {Hou}, \citenamefont {Zhang}, \citenamefont {Yan}, \citenamefont {Guo},\ and\
  \citenamefont {Chen}}]{Hou:2022eev}%
  \BibitemOpen
  \bibfield  {author} {\bibinfo {author} {\bibfnamefont {Y.}~\bibnamefont
  {Hou}}, \bibinfo {author} {\bibfnamefont {Z.}~\bibnamefont {Zhang}}, \bibinfo
  {author} {\bibfnamefont {H.}~\bibnamefont {Yan}}, \bibinfo {author}
  {\bibfnamefont {M.}~\bibnamefont {Guo}},\ and\ \bibinfo {author}
  {\bibfnamefont {B.}~\bibnamefont {Chen}},\ }\href
  {https://doi.org/10.1103/PhysRevD.106.064058} {\bibfield  {journal} {\bibinfo
   {journal} {Phys. Rev. D}\ }\textbf {\bibinfo {volume} {106}},\ \bibinfo
  {pages} {064058} (\bibinfo {year} {2022}{\natexlab{b}})},\ \Eprint
  {https://arxiv.org/abs/2206.13744} {arXiv:2206.13744 [gr-qc]} \BibitemShut
  {NoStop}%
\bibitem [{\citenamefont {Zhang}\ \emph
  {et~al.}(2024{\natexlab{c}})\citenamefont {Zhang}, \citenamefont {Hou},\ and\
  \citenamefont {Guo}}]{Zhang:2023bzv}%
  \BibitemOpen
  \bibfield  {author} {\bibinfo {author} {\bibfnamefont {Z.}~\bibnamefont
  {Zhang}}, \bibinfo {author} {\bibfnamefont {Y.}~\bibnamefont {Hou}},\ and\
  \bibinfo {author} {\bibfnamefont {M.}~\bibnamefont {Guo}},\ }\href
  {https://doi.org/10.1088/1674-1137/ad432b} {\bibfield  {journal} {\bibinfo
  {journal} {Chin. Phys. C}\ }\textbf {\bibinfo {volume} {48}},\ \bibinfo
  {pages} {085106} (\bibinfo {year} {2024}{\natexlab{c}})},\ \Eprint
  {https://arxiv.org/abs/2305.14924} {arXiv:2305.14924 [gr-qc]} \BibitemShut
  {NoStop}%
\bibitem [{\citenamefont {Zeng}\ \emph {et~al.}(2026)\citenamefont {Zeng},
  \citenamefont {Yang}, \citenamefont {Aslam},\ and\ \citenamefont
  {Saleem}}]{Zeng:2025pch}%
  \BibitemOpen
  \bibfield  {author} {\bibinfo {author} {\bibfnamefont {X.-X.}\ \bibnamefont
  {Zeng}}, \bibinfo {author} {\bibfnamefont {C.-Y.}\ \bibnamefont {Yang}},
  \bibinfo {author} {\bibfnamefont {M.~I.}\ \bibnamefont {Aslam}},\ and\
  \bibinfo {author} {\bibfnamefont {R.}~\bibnamefont {Saleem}},\ }\href
  {https://doi.org/10.1016/j.jheap.2025.100540} {\bibfield  {journal} {\bibinfo
   {journal} {JHEAp}\ }\textbf {\bibinfo {volume} {51}},\ \bibinfo {pages}
  {100540} (\bibinfo {year} {2026})},\ \Eprint
  {https://arxiv.org/abs/2509.05803} {arXiv:2509.05803 [gr-qc]} \BibitemShut
  {NoStop}%
\bibitem [{\citenamefont {Zeng}\ \emph
  {et~al.}(2025{\natexlab{a}})\citenamefont {Zeng}, \citenamefont {Yang},\ and\
  \citenamefont {Yu}}]{Zeng:2025tji}%
  \BibitemOpen
  \bibfield  {author} {\bibinfo {author} {\bibfnamefont {X.-X.}\ \bibnamefont
  {Zeng}}, \bibinfo {author} {\bibfnamefont {C.-Y.}\ \bibnamefont {Yang}},\
  and\ \bibinfo {author} {\bibfnamefont {H.}~\bibnamefont {Yu}},\ }\href
  {https://doi.org/10.1140/epjc/s10052-025-14989-y} {\bibfield  {journal}
  {\bibinfo  {journal} {Eur. Phys. J. C}\ }\textbf {\bibinfo {volume} {85}},\
  \bibinfo {pages} {1242} (\bibinfo {year} {2025}{\natexlab{a}})},\ \Eprint
  {https://arxiv.org/abs/2508.03020} {arXiv:2508.03020 [gr-qc]} \BibitemShut
  {NoStop}%
\bibitem [{\citenamefont {Zeng}\ \emph
  {et~al.}(2025{\natexlab{b}})\citenamefont {Zeng}, \citenamefont {Yang},
  \citenamefont {Aslam}, \citenamefont {Saleem},\ and\ \citenamefont
  {Aslam}}]{Zeng:2025kqw}%
  \BibitemOpen
  \bibfield  {author} {\bibinfo {author} {\bibfnamefont {X.-X.}\ \bibnamefont
  {Zeng}}, \bibinfo {author} {\bibfnamefont {C.-Y.}\ \bibnamefont {Yang}},
  \bibinfo {author} {\bibfnamefont {M.~I.}\ \bibnamefont {Aslam}}, \bibinfo
  {author} {\bibfnamefont {R.}~\bibnamefont {Saleem}},\ and\ \bibinfo {author}
  {\bibfnamefont {S.}~\bibnamefont {Aslam}},\ }\href
  {https://doi.org/10.1088/1475-7516/2025/08/066} {\bibfield  {journal}
  {\bibinfo  {journal} {JCAP}\ }\textbf {\bibinfo {volume} {08}},\ \bibinfo
  {pages} {066}},\ \Eprint {https://arxiv.org/abs/2505.07063} {arXiv:2505.07063
  [gr-qc]} \BibitemShut {NoStop}%
\bibitem [{\citenamefont {Meng}\ \emph {et~al.}(2025)\citenamefont {Meng},
  \citenamefont {Wang}, \citenamefont {Li},\ and\ \citenamefont
  {Kuang}}]{Meng:2025ivb}%
  \BibitemOpen
  \bibfield  {author} {\bibinfo {author} {\bibfnamefont {Y.}~\bibnamefont
  {Meng}}, \bibinfo {author} {\bibfnamefont {X.-J.}\ \bibnamefont {Wang}},
  \bibinfo {author} {\bibfnamefont {Y.-Z.}\ \bibnamefont {Li}},\ and\ \bibinfo
  {author} {\bibfnamefont {X.-M.}\ \bibnamefont {Kuang}},\ }\href
  {https://doi.org/10.1140/epjc/s10052-025-14346-z} {\bibfield  {journal}
  {\bibinfo  {journal} {Eur. Phys. J. C}\ }\textbf {\bibinfo {volume} {85}},\
  \bibinfo {pages} {627} (\bibinfo {year} {2025})},\ \Eprint
  {https://arxiv.org/abs/2501.02496} {arXiv:2501.02496 [gr-qc]} \BibitemShut
  {NoStop}%
\bibitem [{\citenamefont {Wan}\ \emph {et~al.}(2026{\natexlab{a}})\citenamefont
  {Wan}, \citenamefont {Hou}, \citenamefont {Huang}, \citenamefont {Li},
  \citenamefont {Guo},\ and\ \citenamefont {Chen}}]{Wan:2026xzs}%
  \BibitemOpen
  \bibfield  {author} {\bibinfo {author} {\bibfnamefont {Q.}~\bibnamefont
  {Wan}}, \bibinfo {author} {\bibfnamefont {Y.}~\bibnamefont {Hou}}, \bibinfo
  {author} {\bibfnamefont {Y.}~\bibnamefont {Huang}}, \bibinfo {author}
  {\bibfnamefont {P.-C.}\ \bibnamefont {Li}}, \bibinfo {author} {\bibfnamefont
  {M.}~\bibnamefont {Guo}},\ and\ \bibinfo {author} {\bibfnamefont
  {B.}~\bibnamefont {Chen}},\ }\href@noop {} {\  (\bibinfo {year}
  {2026}{\natexlab{a}})},\ \Eprint {https://arxiv.org/abs/2605.28376}
  {arXiv:2605.28376 [gr-qc]} \BibitemShut {NoStop}%
\bibitem [{\citenamefont {Hu}\ \emph {et~al.}(2021)\citenamefont {Hu},
  \citenamefont {Zhong}, \citenamefont {Li}, \citenamefont {Guo},\ and\
  \citenamefont {Chen}}]{Hu:2020usx}%
  \BibitemOpen
  \bibfield  {author} {\bibinfo {author} {\bibfnamefont {Z.}~\bibnamefont
  {Hu}}, \bibinfo {author} {\bibfnamefont {Z.}~\bibnamefont {Zhong}}, \bibinfo
  {author} {\bibfnamefont {P.-C.}\ \bibnamefont {Li}}, \bibinfo {author}
  {\bibfnamefont {M.}~\bibnamefont {Guo}},\ and\ \bibinfo {author}
  {\bibfnamefont {B.}~\bibnamefont {Chen}},\ }\href
  {https://doi.org/10.1103/PhysRevD.103.044057} {\bibfield  {journal} {\bibinfo
   {journal} {Phys. Rev. D}\ }\textbf {\bibinfo {volume} {103}},\ \bibinfo
  {pages} {044057} (\bibinfo {year} {2021})},\ \Eprint
  {https://arxiv.org/abs/2012.07022} {arXiv:2012.07022 [gr-qc]} \BibitemShut
  {NoStop}%
\bibitem [{\citenamefont {Lindquist}(1966)}]{LINDQUIST1966487}%
  \BibitemOpen
  \bibfield  {author} {\bibinfo {author} {\bibfnamefont {R.~W.}\ \bibnamefont
  {Lindquist}},\ }\href
  {https://doi.org/https://doi.org/10.1016/0003-4916(66)90207-7} {\bibfield
  {journal} {\bibinfo  {journal} {Annals of Physics}\ }\textbf {\bibinfo
  {volume} {37}},\ \bibinfo {pages} {487} (\bibinfo {year} {1966})}\BibitemShut
  {NoStop}%
\bibitem [{\citenamefont {Chael}\ \emph {et~al.}(2021)\citenamefont {Chael},
  \citenamefont {Johnson},\ and\ \citenamefont {Lupsasca}}]{Chael:2021rjo}%
  \BibitemOpen
  \bibfield  {author} {\bibinfo {author} {\bibfnamefont {A.}~\bibnamefont
  {Chael}}, \bibinfo {author} {\bibfnamefont {M.~D.}\ \bibnamefont {Johnson}},\
  and\ \bibinfo {author} {\bibfnamefont {A.}~\bibnamefont {Lupsasca}},\ }\href
  {https://doi.org/10.3847/1538-4357/ac09ee} {\bibfield  {journal} {\bibinfo
  {journal} {Astrophys. J.}\ }\textbf {\bibinfo {volume} {918}},\ \bibinfo
  {pages} {6} (\bibinfo {year} {2021})},\ \Eprint
  {https://arxiv.org/abs/2106.00683} {arXiv:2106.00683 [astro-ph.HE]}
  \BibitemShut {NoStop}%
\bibitem [{\citenamefont {Gralla}\ \emph {et~al.}(2019)\citenamefont {Gralla},
  \citenamefont {Holz},\ and\ \citenamefont {Wald}}]{Gralla:2019xty}%
  \BibitemOpen
  \bibfield  {author} {\bibinfo {author} {\bibfnamefont {S.~E.}\ \bibnamefont
  {Gralla}}, \bibinfo {author} {\bibfnamefont {D.~E.}\ \bibnamefont {Holz}},\
  and\ \bibinfo {author} {\bibfnamefont {R.~M.}\ \bibnamefont {Wald}},\ }\href
  {https://doi.org/10.1103/PhysRevD.100.024018} {\bibfield  {journal} {\bibinfo
   {journal} {Phys. Rev. D}\ }\textbf {\bibinfo {volume} {100}},\ \bibinfo
  {pages} {024018} (\bibinfo {year} {2019})},\ \Eprint
  {https://arxiv.org/abs/1906.00873} {arXiv:1906.00873 [astro-ph.HE]}
  \BibitemShut {NoStop}%
\bibitem [{\citenamefont {C{\'a}rdenas-Avenda{\~n}o}\ and\ \citenamefont
  {Lupsasca}(2023)}]{Cardenas-Avendano:2023dzo}%
  \BibitemOpen
  \bibfield  {author} {\bibinfo {author} {\bibfnamefont {A.}~\bibnamefont
  {C{\'a}rdenas-Avenda{\~n}o}}\ and\ \bibinfo {author} {\bibfnamefont
  {A.}~\bibnamefont {Lupsasca}},\ }\href
  {https://doi.org/10.1103/PhysRevD.108.064043} {\bibfield  {journal} {\bibinfo
   {journal} {Phys. Rev. D}\ }\textbf {\bibinfo {volume} {108}},\ \bibinfo
  {pages} {064043} (\bibinfo {year} {2023})},\ \Eprint
  {https://arxiv.org/abs/2305.12956} {arXiv:2305.12956 [gr-qc]} \BibitemShut
  {NoStop}%
\bibitem [{\citenamefont {Johnson}\ \emph {et~al.}(2024)\citenamefont {Johnson}
  \emph {et~al.}}]{Johnson:2024ttr}%
  \BibitemOpen
  \bibfield  {author} {\bibinfo {author} {\bibfnamefont {M.~D.}\ \bibnamefont
  {Johnson}} \emph {et~al.},\ }\href {https://doi.org/10.1117/12.3019835}
  {\bibfield  {journal} {\bibinfo  {journal} {Proc. SPIE Int. Soc. Opt. Eng.}\
  }\textbf {\bibinfo {volume} {13092}},\ \bibinfo {pages} {130922D} (\bibinfo
  {year} {2024})},\ \Eprint {https://arxiv.org/abs/2406.12917}
  {arXiv:2406.12917 [astro-ph.IM]} \BibitemShut {NoStop}%
\bibitem [{\citenamefont {Farah}\ \emph {et~al.}(2025)\citenamefont {Farah},
  \citenamefont {Lupsasca}, \citenamefont {Quataert},\ and\ \citenamefont
  {Johnson}}]{Farah:2025kpb}%
  \BibitemOpen
  \bibfield  {author} {\bibinfo {author} {\bibfnamefont {J.~R.}\ \bibnamefont
  {Farah}}, \bibinfo {author} {\bibfnamefont {A.}~\bibnamefont {Lupsasca}},
  \bibinfo {author} {\bibfnamefont {E.}~\bibnamefont {Quataert}},\ and\
  \bibinfo {author} {\bibfnamefont {M.~D.}\ \bibnamefont {Johnson}},\
  }\href@noop {} {\  (\bibinfo {year} {2025})},\ \Eprint
  {https://arxiv.org/abs/2509.23628} {arXiv:2509.23628 [astro-ph.HE]}
  \BibitemShut {NoStop}%
\bibitem [{\citenamefont {Wan}\ \emph {et~al.}(2026{\natexlab{b}})\citenamefont
  {Wan}, \citenamefont {Hou},\ and\ \citenamefont {Guo}}]{Wan:2025gbm}%
  \BibitemOpen
  \bibfield  {author} {\bibinfo {author} {\bibfnamefont {Q.}~\bibnamefont
  {Wan}}, \bibinfo {author} {\bibfnamefont {Y.}~\bibnamefont {Hou}},\ and\
  \bibinfo {author} {\bibfnamefont {M.}~\bibnamefont {Guo}},\ }\href
  {https://doi.org/10.1103/zjtb-3dyz} {\bibfield  {journal} {\bibinfo
  {journal} {Phys. Rev. D}\ }\textbf {\bibinfo {volume} {113}},\ \bibinfo
  {pages} {083023} (\bibinfo {year} {2026}{\natexlab{b}})},\ \Eprint
  {https://arxiv.org/abs/2512.00917} {arXiv:2512.00917 [gr-qc]} \BibitemShut
  {NoStop}%
\bibitem [{\citenamefont {Craig~Walker}\ \emph {et~al.}(2018)\citenamefont
  {Craig~Walker}, \citenamefont {Hardee}, \citenamefont {Davies}, \citenamefont
  {Ly},\ and\ \citenamefont {Junor}}]{CraigWalker:2018vam}%
  \BibitemOpen
  \bibfield  {author} {\bibinfo {author} {\bibfnamefont {R.}~\bibnamefont
  {Craig~Walker}}, \bibinfo {author} {\bibfnamefont {P.~E.}\ \bibnamefont
  {Hardee}}, \bibinfo {author} {\bibfnamefont {F.~B.}\ \bibnamefont {Davies}},
  \bibinfo {author} {\bibfnamefont {C.}~\bibnamefont {Ly}},\ and\ \bibinfo
  {author} {\bibfnamefont {W.}~\bibnamefont {Junor}},\ }\href
  {https://doi.org/10.3847/1538-4357/aaafcc} {\bibfield  {journal} {\bibinfo
  {journal} {Astrophys. J.}\ }\textbf {\bibinfo {volume} {855}},\ \bibinfo
  {pages} {128} (\bibinfo {year} {2018})},\ \Eprint
  {https://arxiv.org/abs/1802.06166} {arXiv:1802.06166 [astro-ph.HE]}
  \BibitemShut {NoStop}%
\end{thebibliography}%

\appendix

\section{Ricci scalar and the regularity of the inner surface}
\label{app:ricci}

For the rotating bumblebee metric in Eq.~\eqref{metric}, the full four-dimensional Ricci scalar can be written as
\begin{equation}
	\label{eq:ricci-full}
	R=\frac{2l}{(1+l)\Sigma^3}\left[\frac{ar(r-M)}
	{\sqrt{\Delta}\sin\theta}\mathcal{F}(r,\theta)+\mathcal{Q}(r,\theta)+\frac{a^3\sqrt{\Delta}\cos^4\theta}{\sin\theta}\right]
\end{equation}
where
\begin{equation}
	\mathcal{F}(r,\theta)=\Sigma\cos^2\theta-r^2\sin^2\theta,\quad\mathcal{Q}(r,\theta)=r^4-3a^2Mr\cos^2\theta-a^4\cos^4\theta.
\end{equation}
For a nonextremal configuration, $0<a<M$, the leading behavior
near the inner surface is
\begin{equation}
	R =
	\frac{2lar_+(r_+-M)\mathcal{F}(r_+,\theta)}{(1+l)\Sigma_+^3\sin\theta}
	\frac{1}{\sqrt{\Delta}}
	+O(1),
	\label{eq:ricci-near}
\end{equation}
where $\Sigma_+=r_+^2+a^2\cos^2\theta$. 
Thus, for $al\neq0$, the Ricci scalar diverges as
$\Delta^{-1/2}$ at generic latitudes when $r\to r_+^+$.
Although the leading coefficient may vanish at isolated latitudes,
its divergence over generic portions of the surface is sufficient
to show that $r=r_+$ is not a regular horizon. 
We also note that Eq.~\eqref{eq:ricci-full} generically exhibits an
axial divergence proportional to $1/\sin\theta$ as
$\theta\to0,\,\pi$ for $al\neq0$. Accordingly, the regular exterior
domain considered in this work is understood to exclude the symmetry
axis.

The singular term disappears in the two limiting cases: $R=0$ in the Kerr limit $l=0$, while $R=2l/[(1+l)r^2]$ in the static limit $a=0$, which is
finite at $r=2M$. 
In the extremal case $a=M$, the factor $(r-M)/\sqrt{\Delta}$ remains finite when
approached from the exterior, so Eq.~\eqref{eq:ricci-near} does not
imply the same radial divergence. The extremal geometry would require
a separate regularity analysis.

\section{Distribution of the Redshift Factor along the $x$-axis}

In this Appendix, we present the distribution of the redshift factor $g_n$ along the $x$-axis for both the direct emission (solid curves) and the lensed ring emission (dashed curves). 
The plots correspond to different values of the spin parameter $a_*$ and observation inclination angles $\theta_o$, with the horizontal coordinate given in units of $\mu$as.

\begin{figure*}[htbp]
	\centering
	\includegraphics[width=6in]{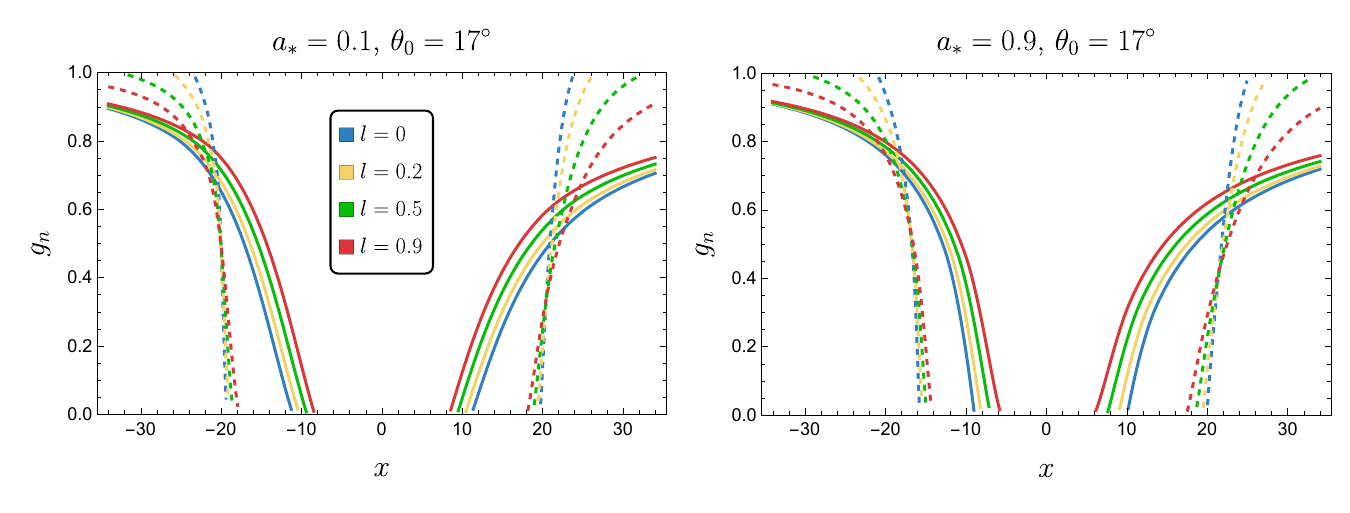}\\
	\includegraphics[width=6in]{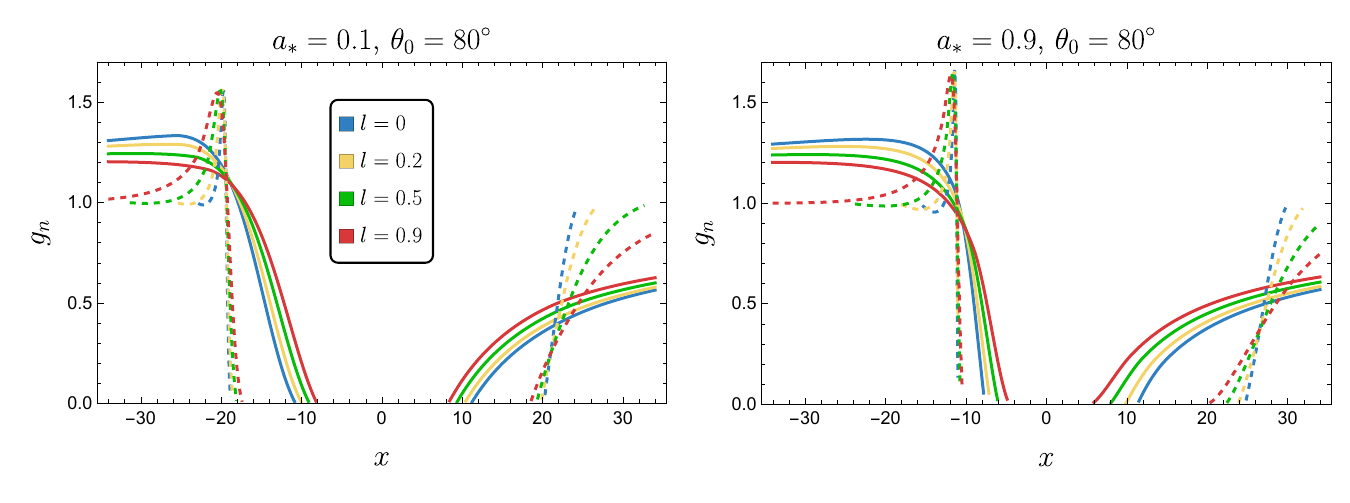}
	\caption{The distribution of redshift factor $g_n$ along the $x$-axis (in $\mu$as) for direct emission (solid curves) and lensed ring emission (dashed curves). Each row/column corresponds to a different combination of the spin parameter $a_*$ and observation inclination angle $\theta_o$.}
	\label{figrsx}
\end{figure*}

\end{document}